\documentclass[useAMS,usenatbib]{mn2e}

\usepackage{graphicx}
\usepackage{longtable}
\usepackage{lscape}
\usepackage{times}      % for font
\usepackage{epsfig}     % for figure inclusion

\usepackage[]{natbib}     % for bibliography
\title[Kinematics of BHB and RR Lyrae stars towards the  AC and the NGP]
{The Kinematic Properties of BHB and RR Lyrae stars towards the   
Anticentre and the North Galactic Pole: The Transition between the Inner
and the Outer Halo.}

\author[T.D. Kinman et al.]
{T.D. Kinman$^{1}$\thanks{e-mail: kinman@noao.edu}, C. Cacciari$^{2}$, 
A. Bragaglia$^{2}$, R. Smart$^{3},  $A. Spagna$^{3}$ \\
\medskip \\
$^{1}$ NOAO, P.O.\ Box 26732, Tucson, AZ 85726-6732, 
USA\thanks{NOAO is operated by AURA Inc.\ under contract with the 
National Science Foundation.}\\
$^{2}$ INAF, Osservatorio Astronomico di Bologna, Via Ranzani 1, I-40127 
Bologna, Italy\\
$^{3}$ INAF, Osservatorio Astronomico di Torino,  Via Osservatorio 20, I-10025 
Pino Torinese, Italy}

\begin{document}

\date{Accepted 2012 February 14. Received 2012 February 8; 
in original form 2011 November 29.}

\maketitle

\begin{abstract}

 We identify 51 blue horizontal branch (BHB) stars, 12 possible BHB stars
 and 58 RR Lyrae stars in Anticentre fields. Their selection does not depend 
 on their kinematics.  Light curves and ephemerides are
 given for 7  previously unknown RR Lyrae stars. All but
 4 of the RR Lyrae stars are of Oosterhoff type I. 
 
  Our selection criteria for BHB
 stars give results that agree with those used by Smith et al. (2010) and 
 Ruhland et al. (2011). We use 5 methods to determine distances for the
 BHB stars and 3 methods for the RR Lyrae stars to get distances on a 
 uniform scale. 
 Absolute proper motions (largely derived from the
 GSCII and SDSS (DR7) databases) are given for these stars; radial 
 velocities are given for 31 of the BHB stars and 37 of the RR Lyrae stars.
 
 Combining these data for BHB and RR Lyrae stars with those previously found  
in fields at the North Galactic Pole, we find that retrograde 
orbits dominate for galactocentric distances greater than 12.5  kpc.
 The majority of metal-poor stars in the solar neighbourhood are known to be
concentrated in a L$_{\perp}$ $vs.$ L$_{z}$ angular momentum plot. We 
 show that the ratio of the number of {\it outliers} to the number in the
{\it main concentration} increases with galactocentric distance. The
location of these {\it outliers} with L$_{\perp}$ and L$_{z}$ shows
that the halo BHB and RR Lyrae stars have more retrograde orbits and a 
more spherical distribution with increasing galactocentric distance.
Six RR Lyrae stars are identified in the H99 group of {\it outliers}; the
small spread in their [Fe/H] suggests that they could have come from a
single globular cluster. Another group of {\it outliers} contains two 
pairs of RR Lyrae stars; the stars in each pair have similar properties.

\end{abstract}


\section{Introduction}

 The blue horizontal branch (BHB) and RR Lyrae stars are well-established 
 tracers of the oldest stars in the galactic halo although their galactic
 distribution may  not coincide with other halo tracers such
 as the turn-off stars (Bell et al. 2010). This paper continues those on 
 other surveys for BHB and RR Lyrae stars (Kinman et al. 2007b and
 Kinman \& Brown 2011) and extends  previous work in the Anticentre direction
 (Kinman et al. 1994). Our BHB and RR Lyrae stars are in the range
 10$<V<$17 and so  range in distance from those in the solar 
 neighbourhood to those distant enough to be included 
 in the SDSS DR 7 survey (Abazajian et al. 2009).
 This allows our selection methods for BHB stars to be compared with other 
 methods for identifying halo SDSS stars ( Smith et al. 2010
 and Ruhland et al. 2011). 

 The globular-cluster halo is thought to consist of an Old Halo and a Young Halo
 in which the clusters have different horizontal branch (HB) morphologies that
 can be interpreted as an age difference (Zinn 1993). The shape of the field
 star halo changes with galactocentric distance (Schmidt 1956; Kinman et al. 1966) 
 which suggests that it may not be homogeneous. 
  It is now usual to postulate that the field stars belong to an ``inner halo"
 that was formed $in situ$ and an ``outer halo" that has largely been 
 accreted.  
 Earlier work on the field star halo has been summarized by Helmi (2008).
 Since then, there have been several more investigations of halo structure
 using various tracers. Surveys using BHB stars have been made by Smith et al.
 (2010), Xue et al. (2011), Ruhland et al. (2011) and Deason et al. (2011).
 Surveys using RR Lyrae stars have been made by Watkins et al. (2009), and
 Sesar et al. (2010). The stars in these surveys are mostly too distant to 
 have good proper motions from which space motions could be derived. They do,
 however, allow space densities to be determined as a function of 
 galactocentric distance. The surveys of Deason et al. (2011) for BHB stars 
 and of Sesar et al. (2011) for turn-off stars both show breaks in the slopes
 of their density distributions at 28 kpc from the galactic centre; this 
 suggests that the halo may have two components.

 Recent surveys for stars that are close enough for existing proper motions to
 allow a kinematic analysis include the survey for subdwarfs by Smith et al.
 (2009), but this survey does not extend far enough in galactocentric distance
 to sample the outer halo. Carollo et al. (2007, 2010) have analyzed the
 space motions of a large number of stars within 4 kpc and find that a 
 two-component halo is needed to account for the galactic rotation of these
 stars. They find that there is an outer halo that is more metal-poor and that
 has a more retrograde rotation than the inner halo. It should be noted that 
 two halos overlap spatially and the outer halo is less centrally concentrated 
 than the inner. The difference in metallicity between the two halos
    has been confirmed by De Jong et al. (2010). Recently 
 Carollo et al. (2012) have shown that the fraction of carbon-enhanced 
 metal-poor stars is twice as great in the outer halo as in the inner halo. 
 Sch\"{o}nrich, Asplund \& Casagrande (2011) re-analysed the Carollo et al. 
 (2010) data and failed to find any reliable evidence for an outer 
 counter-rotating halo. Inter-alia they critized the luminosity classification
 of the turn-off stars. In a rebuttal, Beers et al. (2012) re-analyzed their
 data (re-classifying the turn-off stars and criticizing the main-sequence 
 luminosity relation used by Sch\"{o}nrich et al.); they substantiate their
 original conclusion  that the inner halo is nearly non-rotating
  while the outer halo has ``a retrograde signature" with a transition
  at 15 to 20 kpc from the Sun. 

 Recent simulations of galaxy formation support the idea that the halos of
 galaxies like the Milky Way have a dual origin and have been formed both
 in-situ and by accretion (Zolotov et al., 2009, 2010, 2011), (Oser et al.,
 2010). (Font et al., 2011), (McCarthy et al., 2012). 
  Zolotov (2011) discusses a dual halo in which
 the role of accretion increases outwards from the Galactic centre and the 
 halo is formed solely by accretion if R$_{gal}$ $>$ 20 kpc.
 In these simulations,  the fraction of the halo that is accreted depends 
 upon the mass of the galaxy. The average of the 400 simulations given by
 Font et al. has the inner ``in situ" component dropping to
 20\% and the accreted component rising to 80\% at a galactocentric distance  
 of 20 kpc. McCarthy et al. (2012) find that the ``in situ" component has a
 flattened distribution and a rotation that is intermediate between that of
 the disc and the ``outer halo". 

   In this paper we examine halo stars in the Anticentre direction 
 because this is the best direction in which to study the transition from the
 ``in situ" to the accreted halo and the kinematic properties of the outer halo.
 We do not have sufficient data to discuss the abundance differences between the
 two halos. 
  Sec. 2 identifies the sources and their galactic distributions from  
 which our BHB candidates and our RR Lyrae stars are taken. In Sec. 3 we give
 the photometric data for the BHB stars and in Appendix A we describe the 
 techniques used to identify these stars and the methods used to estimate their
 distances. Sec. 4 gives the photometric data and periods for the RR Lyrae stars
 and shows that they mostly belong to Oosterhoff type 1; in Appendix B we give
 the ephemerides for the new RR Lyrae stars and the methods used to estimate
 their distances. Sec. 5 gives the adopted distances, proper motions and radial
 velocities of our program stars together with their galactic space motons.
 It shown in Sec. 5  that the galactic rotation velocity (V) becomes more 
 retrograde with increasing galactocentric distance. Appendix C gives details of
 the sources of the proper motions. Sec.6 introduces the angular momenta 
 L$_{z}$ and L$_{\perp}$ and their relation to the galactic rotation (V) and
 the maximum height above the plane of the star's orbit (z$_{max}$). It is
 shown the halo becomes more spherical with increasing galactocentric distance.
 Appendix D discusses the location of thick disc in the L$_{z}$ and L$_{\perp}$
 plot. Appendix E gives the angular momenta of the RR Lyrae and BHB stars near
 the North Galactic Pole, those of the local BHB stars and those of the 
 galactic globular clusters within 10 kpc. Appendix F discusses the properties 
 of the kinematic groups H99 and K07. The results of the paper are summarized 
 in Sec. 7. 

\section{Target Selection}\label{s:tsel}

 Our study of the Anticentre halo began with a search for RR Lyrae stars in
the fields 
 RR VI ($l$ = 180$^{\circ}$, $b$ = +26$^{\circ}$) and RR VII 
($l$ = 183$^{\circ}$, $b$ = +37$^{\circ}$); each field covers an area of 30 
deg$^{2}$ (Kinman et al. 1982). Later Sanduleak provided BHB star candidates
for the RR VII field  and these were discussed by Kinman et al. (1994).
 These samples of BHB and RR Lyrae stars have been enlarged for the present 
 paper. New BHB candidates were taken from the objective prism surveys of 
 Pesch \& Sanduleak (1989; Case A-F stars) and of Beers et al.(1996; BPS BS stars).
   We are indebted to Dr Peter Pesch (1996, private communication) for 
  sending us unpublished candidate stars from the Case survey.
   These are given the prefix P in column 2 of Table \ref{t:bhb}  
   unless there is a previous identification in the literature. 

 Our methods of 
 selecting  BHB stars from these candidates and the calculation of their
 distances are described in Appendix A.  Apart from the RR Lyrae stars in 
 fields RR VI and RR VII, the additional RR Lyrae stars have mostly been
  found among BHB candidates that were found to be variable.
  Seven of these RR Lyrae stars have not been previously identified and
 their light curves and ephemerides are given in Appendix B.
 All our program stars  are listed in Tables \ref{t:bhb} and \ref{t:rrl} for the BHB
 and RR Lyrae stars respectively. These tables give positions and photometric 
 data for the BHB stars and, also,  metallicities and periods for the RR
 Lyrae stars; sources of these data are given in notes to these tables.  

\begin{figure*}
\includegraphics[width=13.0cm, bb=17 208 490 610, clip=true]{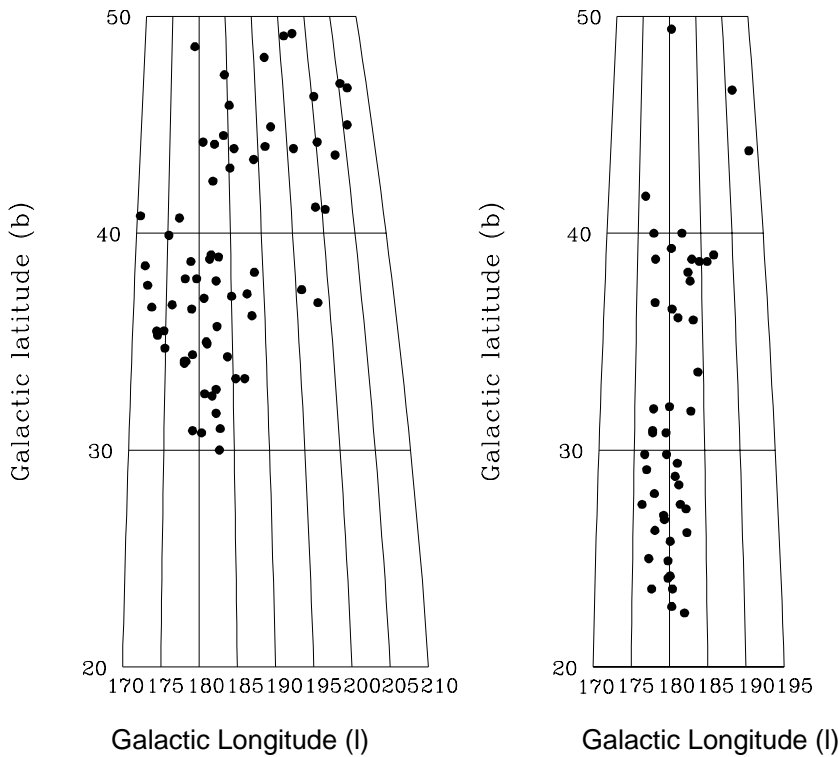}
\caption{The distribution in galactic coordinates (expressed in degrees) of 
 (left) the BHB
 candidates from Table \ref{t:bhb} and, (right) our program RR Lyrae stars 
 (Table \ref{t:rrl}).
 Seven RR Lyrae stars with b $>$ 50$^{\circ}$  are omitted from this Figure.
 \label{f:bl}
}
%\label{f:1}
\end{figure*}

 The galactic distributions of 
 our program stars are shown separately for the BHB and RR Lyrae stars in
 Fig. \ref{f:bl}. Not only do our BHB and RR Lyrae stars cover somewhat different
 areas of the Anticentre sky, but they also cover different magnitude ranges 
 so that the volumes of space that they occupy only partially overlap.
 Also, our selection of RR Lyrae stars favours the bluer (Bailey type $c$)
 and may miss some of the redder (Bailey type $ab$) 
 RR Lyrae variables and so our sample may not be complete. This must be
 taken into account in comparing the properties of our two samples.

\begin{figure}
\includegraphics[width=8.5cm, bb=95 210 490 610, clip=true]{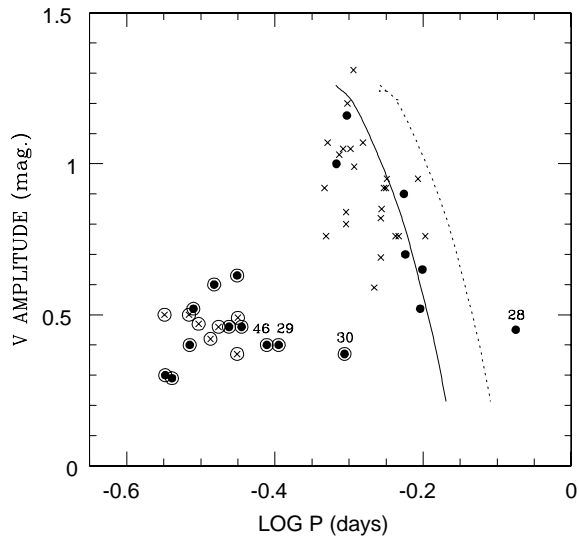}
\caption{The $V$ amplitude {\it vs.} $\log$ Period plot for our RR Lyrae
 sample with galactic latitudes less than 50$^{\circ}$.
  Filled circles indicate stars for which a $V$ amplitude was available.
 Crosses indicate stars where the $V$ amplitude was obtained by dividing the
 $B$ amplitude by 1.31. Encircled filled circles and crosses 
 show Bailey type $c$ variables. The solid and dotted curves show the 
 expected loci for Oo type I and Oo II variables of Bailey type $ab$
 respectively and were 
 taken from Cacciari et al. (2005). The four likely Oo II variables are 
 given their numbers in Table \ref{t:rrl}. 
 \label{f:pa} 
}
%\label{f:2}
%\end{figure*}
\end{figure}

 The period-amplitude distribution of our RR Lyrae sample is shown 
 in Fig. \ref{f:pa} 
 for the variables with galactic latitudes less than 50$^{\circ}$.
 The solid and dotted curves show the loci of the Oosterhoff type I and II variables 
 respectively (these were taken from Cacciari et al. 2005).                
 Most of our RR Lyrae stars lie close and to the left of the Oo I curve; this suggests 
 that the majority are Oo I variables. 
 The four stars that are most likely to be Oo II variables are
 indicated by their numbers in Fig. \ref{f:pa}. This preponderance of Oo I variables in
 the Anticenter is compatible with the discovery bu Miceli et al. (2008) that
 the Oo II variables are more concentrated towards the Galactic centre than
 the Oo I variables. The ratio of Oo I to Oo II variables should therefore 
 increase with galactocentric distance and so the Oo I variables should 
 predominate towards the Anticentre.

 \section{The BHB Stars.} 

 Our BHB stars were chosen from {\it candidates} in the sources given in
  Section \ref{s:tsel}. 
  Table \ref{t:bhb}  gives the equatorial and galactic coordinates of these stars,
  together with  photometric data (using the system used in Kinman
  et al. 1994)). Details of the photometric observing are given in
  Kinman et al. (1994) for photoelectric observations and Kinman \& Brown 
  (2011) for CCD observations. 
  Table \ref{t:bhb} also gives the $GALEX$ $NUV$ magnitude
  (effective wavelength 2267 \AA) that was taken from $MAST$.\footnote{The Multimission 
  Archive at the STSci, http://archive.stsci.edu/.} 
  We assumed that stars with $V<$12.5 had saturated $GALEX$ $NUV$ magnitudes 
  and should not be used. We also give the $2MASS$ $K$ magnitude that was
  taken from the $2MASS$ Point Source Catalog using the $Vizier$ access tool.

% Figure 1 \ref{t:bhb}

%\twocolumn
%\clearpage

 The selection of BHB stars from the candidates is described in Appendix A. 
 We show there how a weight W (Table \ref{t:bhb}, column 13) was assigned to each
  star that depends  on the probability that it is a BHB star. The stars
  were given a type (Table \ref{t:bhb}, column 15) that depended on this weight. 
  Stars with a high probability of being BHB stars were classified as $BHB$,
  those with a high probability of not being BHB were classified as $A$ while
  intermediate types were classified as $bhb$. A comparison of our 
  classifications with those obtained by methods based on SDSS photometry
  suggests that stars with both $BHB$ and $bhb$ classifications have a high
  probability of being BHB stars.

  Appendix A also contains a discussion of five different methods of getting
  the absolute magnitudes (and hence distances) of BHB stars. These distances
  and the adopted distances are given in  Table \ref{t:A1}.
  The reddenings of both BHB stars and RR Lyrae stars were taken from 
  Schlegel et al. (1998).

% Figure 2 
%\begin{figure*}

%\newpage

\section{The RR Lyrae Stars.} 

Most of the RR Lyrae stars in Table \ref{t:rrl}  are listed in the General 
Catalogue of Variable Stars (GCVS, Kholopov et al., 1985) and subsequent 
Name Lists and have the traditional identification by constellation;
those without GCVS names  are taken from Pier et al. (2003), Kinman et al.
 (2004) and Kinman \& Brown (2010). Seven of the stars in Table \ref{t:rrl} have not
 been previously identified as RR Lyrae stars; their light curves and 
 ephemerides are given in Appendix B. 
 The mean $K$ magnitudes ($\langle K \rangle$) in Table \ref{t:rrl} were derived from 
 the 2MASS $K$ magnitudes using the method given in Feast et al. (2008).
 We follow the methods given in Kinman et al. (2007b) to derive distances for
 the RR Lyraes. These are briefly restated in Appendix B where we give 
 distances by three separate methods together with adopted distances (D) that 
 are adjusted to be on the same scale as those adopted for the BHB stars 
 (Appendix A).

\subsection{Oosterhoff types of the RR Lyrae stars.}

 The globular clusters with Oosterhoff type I RR Lyrae stars are known to
 have different kinematics (more retrograde orbits) than those containing
 Oosterhoff type II RR Lyrae stars (Lee \& Carney, 1999; van den Bergh 
 1993).\footnote{The simple division into two Oosterhoff types does not cover all
 cases (Smith et al. 2011) but is a good approximation for our own Galaxy.}
 The Oosterhoff type is determined from the period-amplitude diagram
 and this is shown in Fig. \ref{f:pa} for our RR Lyrae sample with 
 b$\leq$50$^{\circ}$. In this figure the loci for the 
 Oosterhoff type I and II variables are shown by solid and dotted curves 
 respectively (these curves were taken from Cacciari et al. 2005). 
   Most of our RR Lyrae stars lie close and to the left of the Oo I curve.
  Type $ab$ stars that lie to the left of the Oo I curve may either be 
  metal-rich or have smaller mean amplitudes because of Blazhko effect. 
  This suggests that the majority of our stars
 are Oo I variables. Four stars that are most likely to be Oo II variables  
 are indicated by their numbers in Fig. \ref{f:pa}. 
 Stars 29, 30 and 46 are type $c$ variables while 28 is type $ab$. 
 These four stars have $\langle$Z$\rangle$ = 3.9~kpc and
 $\langle$R$_{gal}$$\rangle$ = 14.3~kpc compared with $\langle$Z$\rangle$ =  
 6.6~kpc and $\langle$R$_{gal}$$\rangle$ = 19.6~kpc for the whole sample.
  The preponderance of Oo I variables in our Anticenter fields and the smaller
 $\langle$R$_{gal}$$\rangle$ of our Oo II variables is to be expected if 
 the Oo II variables are more concentrated towards the Galactic centre than
 the Oo I variables (Miceli et al. 2008). The smaller $\langle$Z$\rangle$ of
 our Oo II variables also agrees with the preponderance of Oo I variables
 at high Z that was found by De Lee (2008). This suggests that
  the Oo II variables are not only more concentrated to the Galactic
 centre but also form a more flattened system than the Oo I variables.

\section{The Galactic Space Motions of the Program Stars.}

 Tables \ref{t:bhb2} and \ref{t:rrl2} give the parallaxes, proper motions, 
 radial velocities and the
 Galactic Space Motions (U,V,W) with respect to the Local Standard of Rest 
 (LSR) for the BHB and bhb stars and the RR Lyrae stars respectively. The 
 parallaxes are derived from the adopted distances (D) given 
 in Tables \ref{t:A1} and
 \ref{t:B2} in Appendices A and B respectively.  
 The error in the parallax is derived from the rms scatter given in the last
 columns of Tables \ref{t:A1} and \ref{t:B2} and does not include any systematic
 error in the case of the BHB stars. In the case of the RR Lyrae stars, a small
 distance-dependent error has been added in quadrature to the rms scatter to
 derive the error of the parallax as explained in the appendix Sec. B2.
  Heliocentric space-velocity components U, V, and W were derived from 
 the  data listed in  these tables.     We used the program by
 Johnson \& Soderblom (1987) (updated for the J2000 
  reference frame and further updated with the transformation 
  matrix derived from the Vol. 1 of the Hipparcos data catalogue).  
This program gives a right-handed system for U, V and W in which these vectors
are positive towards the directions of the Galactic centre, but we here use 
the left-handed system so as to be comparable with most other recent work.
 These heliocentric velocities were then corrected to velocities relative to 
 the LSR using the solar motion relative to the LSR                   
   ${\rm (U,V,W)_\odot=(10.0,\,5.25,\, 7.17)}$~km\,s$^{-1}$ 
  (Dehnen \& Binney 1998).   

\subsection{Radial velocities and Proper Motions}

The sources of our radial velocities are given in column (8) (S$_{RV}$) of
Tables \ref{t:bhb2} and \ref{t:rrl2}. 
The Bologna velocities were derived from spectra taken with the 3.5m-LRS (TNG) 
spectrograph. The Kitt Peak  velocities were derived from spectra taken with 
 RC spectrograph on the 4m Kitt Peak telescope and kindly made available to us 
 by Nick Suntzeff (1997, private communication). 
 The remaining velocities  were taken from the literature 
 as given in the notes to Tables \ref{t:bhb2} and \ref{t:rrl2}.

 The absolute proper motions given in this paper come primarily from 
 astrometric data assembled from the Second Guide Star Catalog (GSC-II; Lasker
 et al. 2008), and the Seventh Data Release of the Sloan Digital Sky Survey 
  (SDSS DR7; Abazajian et al. 2009; Yanny et al. 2009). Details are given in
 Appendix C. In the case of a few of the brighter stars ($V$ $<$ 12.3) we have
  chosen   the proper motions given in the NOMAD Catalog (Zacharias et al. 
  2004). In the case of th BHB star P 30-38, we chose the proper motion 
   given by the SDSS DR7 because the
  GSC-II---SDSS proper motion has unusually large errors. The SDSS DR7 
  proper motions have also been used for the stars (mostly BHB stars)
 for which GSC-II---SDSS proper motions were not available. 

  We have only used stars that have radial velocities 
  to compute U, V and W, and since only 4 of the BHB stars that
  only have SDSS DR7 proper motions also have radial velocities,
  possible systematic differences between the SDSS DR7 and the GSC-II---SDSS
 proper motions should have little effect on our overall results\footnote{ In
 our discussion of halo stars at the North Galactic Pole (Kinman et al. 2007b),
 the BHB and RR Lyrae stars were both closely grouped near the NGP and it 
 seemed reasonable to adopt a zero radial velocity for stars whose radial 
velocity was not known in computing their galactic velocity V. This assumption
 has not been made for the more widely spread stars in the Anticentre.}.
  A comparison of the SDSS DR7 and the 
  GSC-II---SDSS proper motions (Appendix C) shows good agreement at the
  1 mas y$^{-1}$ level; this corresponds to a tangential velocity of 40
  km s$^{-1}$ at a distance of 8.5 kpc.

% Figure 3 
\begin{figure*}
\includegraphics[width=18.0cm, bb=30 270 545 517, clip=true]{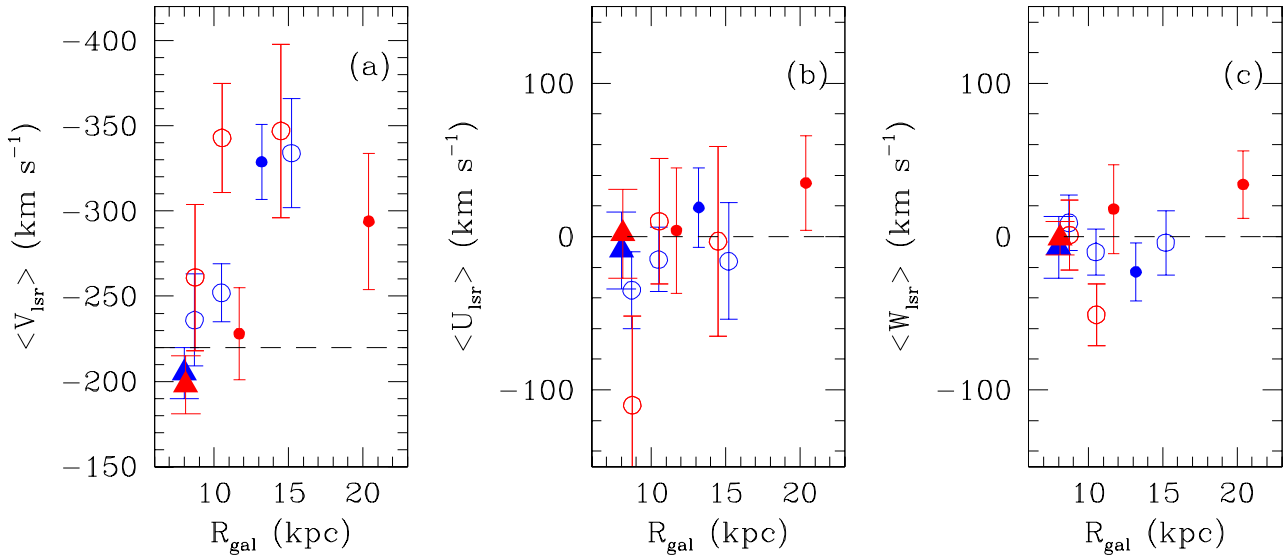}
\caption{In (a) the ordinate is $\langle$V$_{lsr}$$\rangle$ in km s$^{-1}$.  
In (b) the ordinate is $\langle$U$_{lsr}$$\rangle$ in km s$^{-1}$. 
In (c) the ordinate is $\langle$W$_{lsr}$$\rangle$ in km s$^{-1}$. The
abscissa is the Galactocentric distance in kpc. NGP BHB stars (blue open circles
); NGP RR Lyrae stars (red open circles); Anticentre BHB stars (blue full circles); 
Anticentre RR Lyrae stars (red full circles); Local BHB stars (blue full triangles); 
Local RR Lyrae (red full triangles). It is seen that 
$\langle$V$_{lsr}$$\rangle$ becomes more retrograde while           
$\langle$U$_{lsr}$$\rangle$ and $\langle$W$_{lsr}$$\rangle$ remain unchanged 
with increasing Galactocentric distance.
}
\label{f:3}
\end{figure*}

\begin{center}
\begin{table*}
\caption{Space Motions U, V \& W with respect to LSR and their dispersions 
$\sigma_{u}$,$\sigma_{v}$ \& $\sigma_{w}$ in km s$^{-1}$. \label{t:uvw}}
%\begin{flushleft}
\begin{tabular}{@{}lccccccccccc@{}}
\noalign{\smallskip} \hline
 Sample& N& U & $\sigma_{u}$ & V & $\sigma_{v}$ & W& $\sigma_{w}$ &$\langle$Z$
\rangle$ & $\langle$ D $\rangle$ & $\langle$ R$_{gal}$ $\rangle$& Notes \\
       &      &km s$^{-1}$&km s$^{-1}$&km s$^{-1}$ & km s$^{-1}$ & km s$^{-1}$ &   km s$^{-1}$ & kpc & kpc & kpc &    \\
\hline \hline
 bhb         & 8 &--73$\pm$41 &115$\pm$29 &--192$\pm$45 & 121$\pm$30& --55$\pm$48 &
 134$\pm$34 & 3.8 & 6.2 &13.4&  1  \\
 BHB (A)            & 28&  +19$\pm$26 & 131$\pm$18 &--329$\pm$21& 99$\pm$13 &  -14$\pm$20  &
 99$\pm$13  &  3.7 & 6.1  &13.2 & 2 \\
 RR  (A )           & 15& +3$\pm$40 & 152$\pm$28 &--228$\pm$26& 95$\pm$17 &  +18$\pm$28  &
 108$\pm$20  &  2.8 & 4.5  &11.7 & 3 \\
 BHB (B)    & 3&--177$\pm$69 &  89$\pm$36 &--181$\pm$95&107$\pm$44 & +86$\pm$98 &
 118$\pm$48  &  6.0 & 9.8  &16.9 & 4 \\
 RR  (B)            & 16& +31$\pm$29 & 97$\pm$17 &--263$\pm$40&121$\pm$21 & +6$\pm$28  &
  78$\pm$14  &  6.8 & 13.4 &20.7 & 5 \\
 LOCAL BHB  & 27 & +9$\pm$25 &129$\pm$18 &--205$\pm$15 & 79$\pm$ 11 &--7$\pm$20 &
 101$\pm$ 14 & ... & ... & 8.0 & 6  \\
 LOCAL RR   & 32 &--12$\pm$29 &160$\pm$20 &--204$\pm$19 & 107$\pm$13 &+11$\pm$20 &   113$\pm$14 & ... & ... & 8.0 & 7  \\
\hline
\end{tabular}

Notes to Table: \\
\bf{(1)} Stars classified as bhb (possible BHB stars)  ~~~~~~
\bf{(2)} Halo BHB stars  with D $<$ 8.5 kpc. ~~~~~~
\bf{(3)} Halo RR Lyrae stars  with D $<$ 8.5 kpc. ~~~~~~
\bf{(4)} Halo BHB stars  with D $>$ 8.5 kpc. ~~~~~~
\bf{(5)} Halo RR Lyrae stars with 8.5 $<$ D $<$ 17 kpc. ~~~~~~
\bf{(6)} Local sample of BHB stars (Kinman et al. (2007b). ~~~~~~~
\bf{(7)} Local sample of Halo RR Lyrae stars within 1 kpc from Maintz \&
        de Boer (2005). Velocity dispersions are upper limits. For further
details see text. \\
\end{table*}
\end{center}

\subsection{Discussion of the space motions U, V \& W}

  Table \ref{t:uvw} gives the space motions for various subgroups of our program stars;
  all velocities are with respect to the LSR. The possible 
  BHB stars (type bhb) have a V motion and velocity dispersions ($\sigma_{u}$,
  $\sigma_{v}$ and $\sigma_{w}$) that are similar to those of the local BHB
  stars. This suggests that the majority are BHB stars but, conservatively,
  we have not included them in any of the other samples. 

 We have used the angular momenta (L$_{\perp}$ and (L$_{z}$) to distinguish
 between {\it disc} and {\it halo} stars. These quantities are defined in the
 Appendix of Kepley et al. (2007) and are given for our program stars in 
 columns 15 and 16 of Tables \ref{t:bhb2} and \ref{t:rrl2}. 
 Following the discussion given in our
 Appendix D2, we identify the BHB star RR7 064 and the RR Lyrae stars
 TW Lyn and P 82 06 as probable {\it disc} stars and have excluded them 
 from further discussion. 

 We assume that $\langle$V$\rangle$ = --V$_{LSR}$ = --220 km s$^{-1}$ 
 for zero halo rotation although higher values are possible\footnote{e.g.
 $-$236$\pm$15 km s$^{-1}$(Reid \& Brunthaler 2004); 
  $-$246$\pm$7 km s$^{-1}$ (Brunthaler et al. 2011).}.
  We divide our program stars into samples according to distance: (A) those
 nearer than 8.5 kpc, and (B) those with distances between 8.5 and 17.0 kpc.
  At 17 kpc, an error of 1 mas in the proper motion will give an error of 80
  km s$^{-1}$ in the transverse velocity. With our relatively small samples,
  the inclusion of more distant stars would not add useful information.
 Our results are given in Table \ref{t:uvw}.
  The samples that contain an adequate number (N) of stars, namely  
 BHB(A), RR (A) and RR (B), have mean 
 U and W velocities that are essentially zero; this suggests that
 the systematic errors in our proper motions are not having a significant 
 effect on the results for these samples.
  The velocity dispersions in Table \ref{t:uvw} were corrected following 
  Jones \& Walker (1988); if the observed dispersion in U is Disp(U), and
  $\xi_{i}$ is the error in U of star $i$, then the corrected dispersion 
  $\sigma_{u}$ is given by: 

\begin{eqnarray}
\sigma_u^2 = (Disp(U))^2 - \frac{1}{n} \sum_{i=1}^{n} \xi_{i}^{2} \nonumber %\\
\end{eqnarray}

 The corrected dispersions of U, V \& W in our Anticentre samples are also
 comparable (within their errors) with those of the local samples.

 Fig.s 3(a), 3(b)And 3(c) are plots of the galactic velocity components 
  $\langle$V$_{lsr}$$\rangle$,
  $\langle$U$_{lsr}$$\rangle$ and
  $\langle$W$_{lsr}$$\rangle$ respectively against 
 galactocentric distance (R$_{gal}$) for both our Anticentre stars and
 those at the North Galactic Pole (Kinman et al. 2007b). Although there is
 considerable scatter between the different samples, there is clearly a trend
 in $\langle$V$_{lsr}$$\rangle$ from zero
  galactic rotation in the solar neighbourhood to a strong retrograde 
 rotation for R$_{gal}$ greater than 12.5 kpc. On the other hand, 
 both $\langle$U$_{lsr}$$\rangle$ and 
 $\langle$W$_{lsr}$$\rangle$ are essentially zero at all Galactocentric 
 distances. This supports our conclusion that the trend of
 $\langle$V$_{lsr}$$\rangle$ with galactocentric distance 
 is real and not produced by systematic errors in the proper motions.  
  If the outer halo has a significantly retrograde rotation, as originally found by
 Carollo et. al (2007, 2010), and confirmed by Beers et al. (2012), this suggests that
 the outer halo dominates beyond R$_{gal}$=12.5~kpc.
 
% Figure 4 
\begin{figure*}
\includegraphics[width=15.0cm]{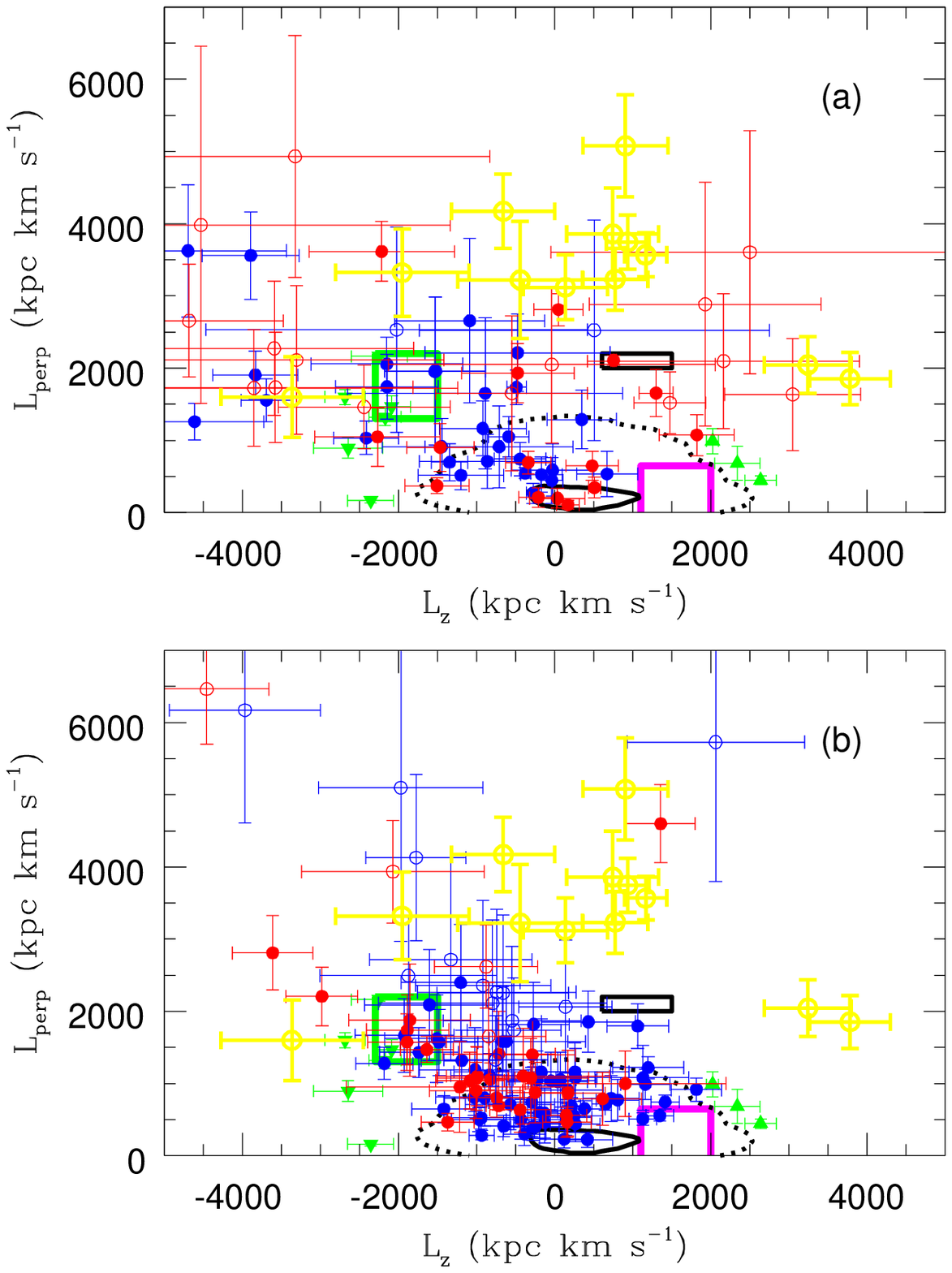} 
\caption{A plot of L$_{\perp}$ against L$_{z}$ for (a) stars in the Anticentre, 
and (b) stars at the North Galactic Pole. The black dotted curve is the outer
contour of the majority of stars studied by Morrison et al. (2009); the black
full curve is the flattened distribution that they discovered. The black 
and green rectangles are the locations of the groups discovered by Helmi et 
al. (1999) and Kepley et al. (2007) respectively. The magenta box shows the
location of the Thick disc. BHB and RR Lyrae stars are shown by blue and red
filled circles respectively. Selected outliers from Kepley et al. are shown
by green triangles and subdwarf outliers from Smith et al. (2009) by yellow
 open circles. 
}
\label{f:4}
\end{figure*}

% Figure 5 
\begin{figure*}
\includegraphics[width=15.0cm]{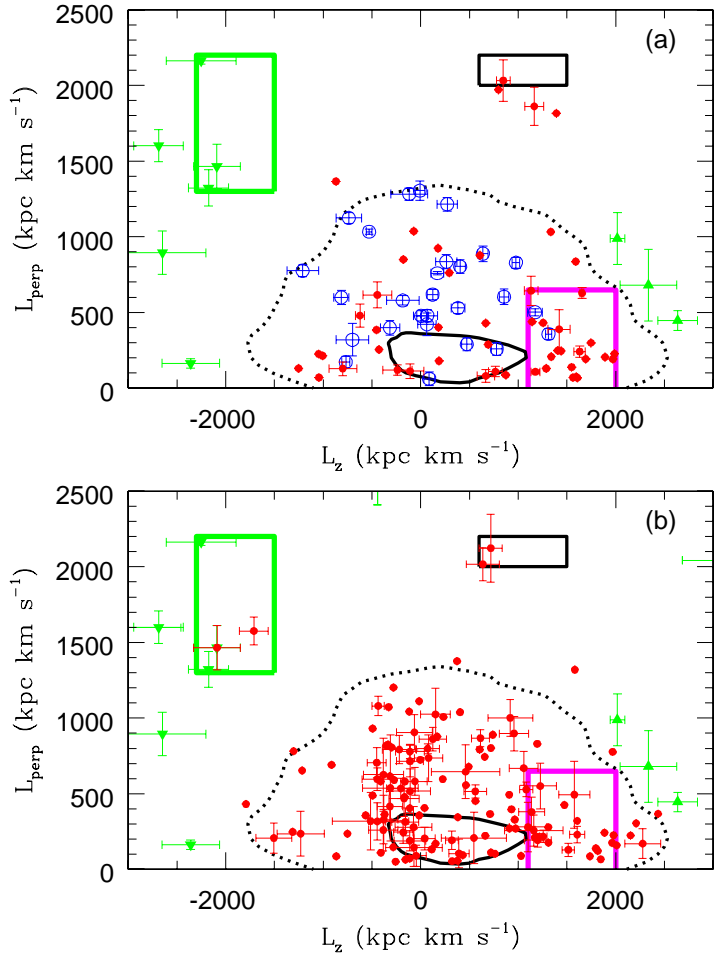} 
\caption{A plot of L$_{\perp}$ against L$_{z}$ for (a) stars within 1 kpc and 
(b) those with distances from between one and two kpc.  The red filled circles 
 are RR Lyrae stars taken from the catalogue of Maintz \& de Boer (2005). Local
BHB stars are shown by blue open circles and the outliers from Kepley et al.
(2007) are shown by green filled triangles. The black contours and black, green and
magenta boxes are described in Fig. \ref{f:4}.
}
\label{f:5}
\end{figure*}

% Figure 6 
\begin{figure*}
\includegraphics[width=13.0cm, bb=95 270 490 500, clip=true]{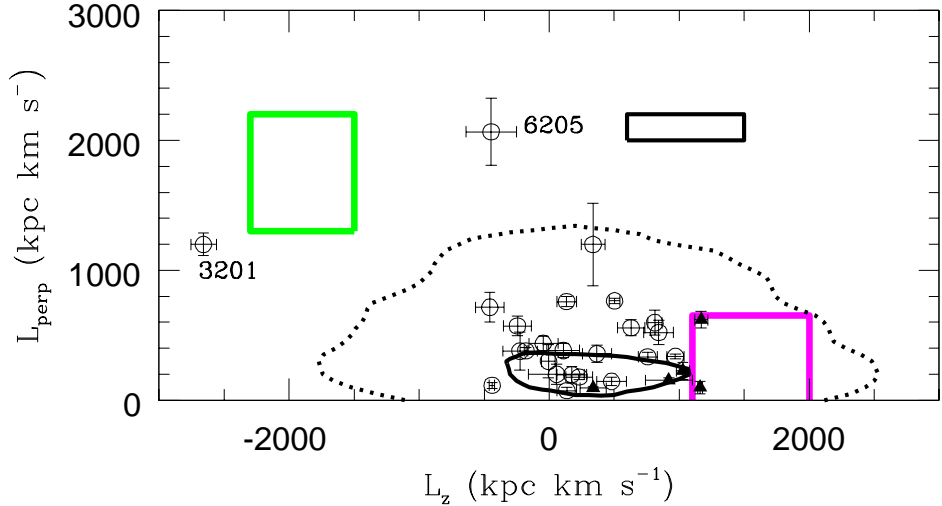}
\caption{A plot of L$_{\perp}$ against L$_{z}$ for globular clusters that are 
 within 10 kpc. The two outlying clusters NGC 3201 and NGC 6205 (M13) are 
indicated by their NGC numbers. Globular clusters with [Fe/H] $>$ -1.0 are 
shown by black filled triangles and those with [Fe/H] $<$ -1.0 by black open 
circles. The black curves and black, green and 
magenta boxes are described in Fig. \ref{f:4}.
}
\label{f:6}
\end{figure*}

% Figure 7 
\begin{figure*}
\includegraphics[width=15.0cm, bb=50 260 545 490, clip=true]{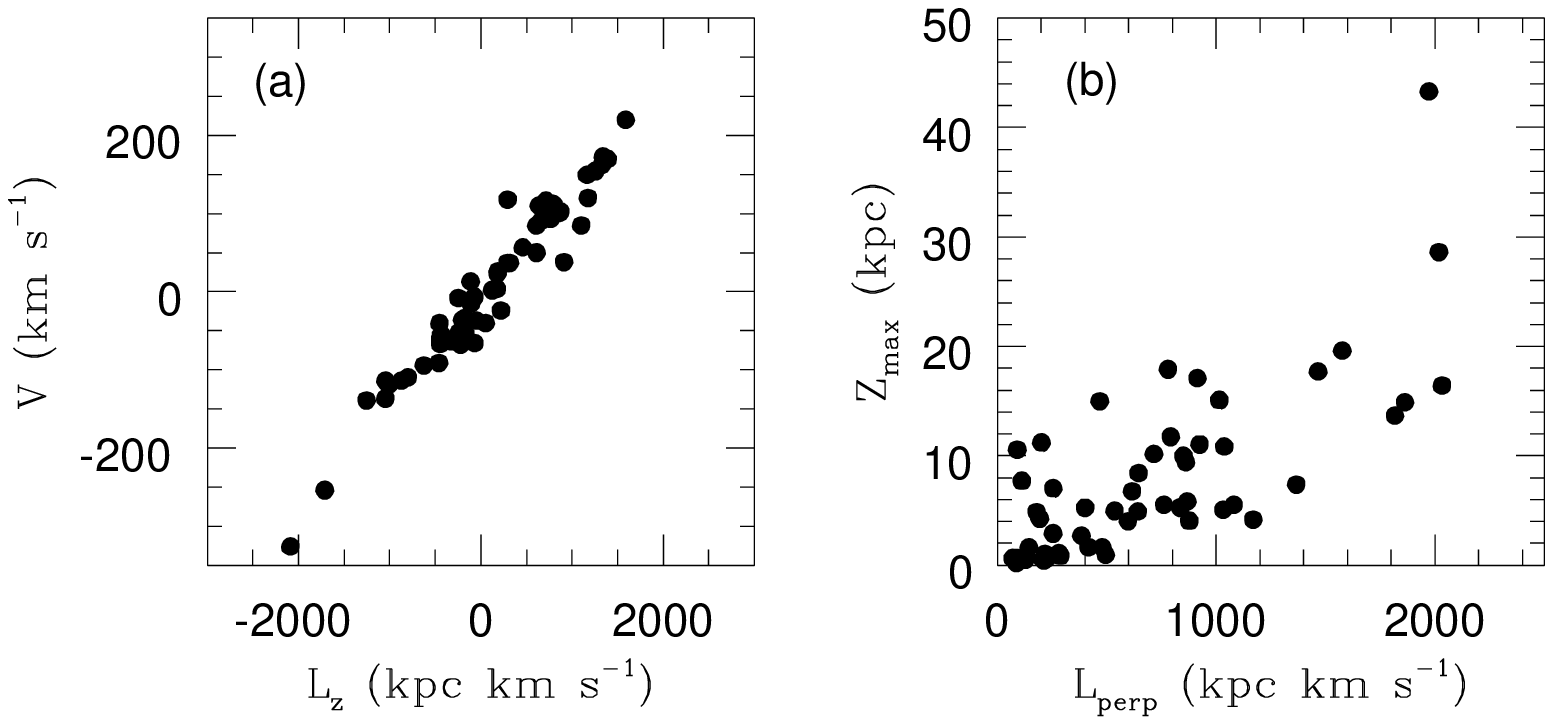}
\caption{ Plots showing the correlation between (a) the galactic rotation V 
 and  L$_{z}$, and (b) the maximum height of the orbit above the plane
 (z$_{max}$) and L$_{\perp}$. 
The V and z$_{max}$ are from Maintz \& de Boer (2005) 
for 56 halo RR Lyrae stars.
}
\label{f:7}
\end{figure*}

\section{Structure in the motions of our halo stars.}

Plots of the angular momenta L$_{\perp}$ and L$_{z}$ can be  used to 
demonstrate kinematic structure among halo stars (e.g. Helmi et al. 1999).
We give a L$_{\perp}$ {\it vs.} L$_{z}$ plot for our Anticentre BHB and 
RR Lyrae stars in Fig \ref{f:4}(a) and in Fig \ref{f:4}(b) for our North 
Galactic Pole sample of these stars (Kinman et  al. 2007b). 
Similar plots are shown in Fig \ref{f:5} for local RR Lyrae and BHB stars 
and in Fig  \ref{f:6} for the globular clusters within 10 kpc.
Definitions of L$_{\perp}$ and L$_{z}$ are given by Kepley et al. (2007). 
We calculated these quantities and their errors with a program 
that was kindly made available by Heather Morrison and modified for our
  use by Carla Cacciari. The values of L$_{\perp}$ and L$_{z}$ for the
North Galactic Pole RR Lyrae stars and BHB stars, the local BHB stars and the
globular clusters within 10 kpc are tabulated in Appendix E, where we also 
compare our L$_{\perp}$ and L$_{z}$ with those calculated by Re Fiorentin et al.
(2005) and Morrison et al. (2009) for a small sample of halo 
stars\footnote{The data for the local RR Lyraes in
Fig. \ref{f:5} were either taken from Morrison et al. (2009) or calculated 
from the data given by Maintz \& de Boer (2005); in this latter case no errors 
 are given since  Maintz \& de Boer do not give errors for their data.
 The L$_{\perp}$ and L$_{z}$ for the globular clusters were calculated 
 from data given in Table 3 of Vande Putte and Cropper (2009).}. 

L$_{z}$ correlates with galactic rotation: in the left-handed system of 
coordinates, objects with positive L$_{z}$ are prograde (the Sun has
L$_{z}$$\sim$1760 kpc km s$^{-1}$) and those with negative L$_{z}$ are retrograde. 
L$_{\perp}$ correlates with the maximum height of the orbit above the plane 
(Fig. \ref{f:7}).

Morrison et al. (2009) investigated the L$_{\perp}$ {\it vs.} L$_{z}$ plot for
246 local metal-poor stars. The majority ($\sim$90\%) of their sample, 
 which we will call the {\it main concentration}, are in the location
bounded by the black dotted line in Figs. \ref{f:4}, \ref{f:5} \& \ref{f:6}. 
This is taken from
the outer contour of their Fig. 3. They also discovered a flattened component 
whose location is shown by the full black contour in our Figs. \ref{f:4}, 
\ref{f:5} \& \ref{f:6}. The remaining 10\% of their sample lie outside the 
black dotted contour and, following Kepley et al. (2007) and Smith et al. 
(2009), we call them {\it outliers}. 
About a third of these {\it outliers} in the Morrison et al.
 (2009) sample belong to the prograde group (H99) discovered by Helmi et al.
 (1999) and further investigated by Re Fiorentin et al. (2005); its location 
 of the majority of stars in this group is
 shown by the black rectangle in Figs \ref{f:4}, \ref{f:5} \& \ref{f:6}. 
 The green rectangle shows
 the location of another group suggested by Kepley et al. (2007). The 
 magenta box in Figs. \ref{f:4}, \ref{f:5} \& \ref{f:6} shows the location 
 of stars in the {\it main
  concentration} that have a high probability belonging to the Thick Disc;
 this is discussed in Appendix D.

The review by Klement (2010) lists sixteen halo ``streams"  that have been 
 identifed among stars in the solar neighbourhood. 
 All except the H99 and Kapteyn Group lie within the
   {\it main concentration} in the L$_{\perp}$ {\it vs.} L$_{z}$ plot. An 
example of structure within the {\it main concentration} is shown by the RR
Lyrae stars at distances between 1 and 2 kpc (Fig \ref{f:5}b) which are less evenly
distributed than those at distances less than 1 kpc (Fig. \ref{f:5}a). In general, the
identification of structure in this {\it main concentration} is only possible
for stars with relatively large proper motions and well determined distances.
In discussing our program stars, we shall therefore largely confine ourselves to
  discussing the {\it outliers} and the ratio of the number of 
  {\it outliers} to the the number in the {\it main concentration}.

% Figure 8 
\begin{figure*}
\includegraphics[width=15.0cm, bb=30 214 550 600, clip=true]{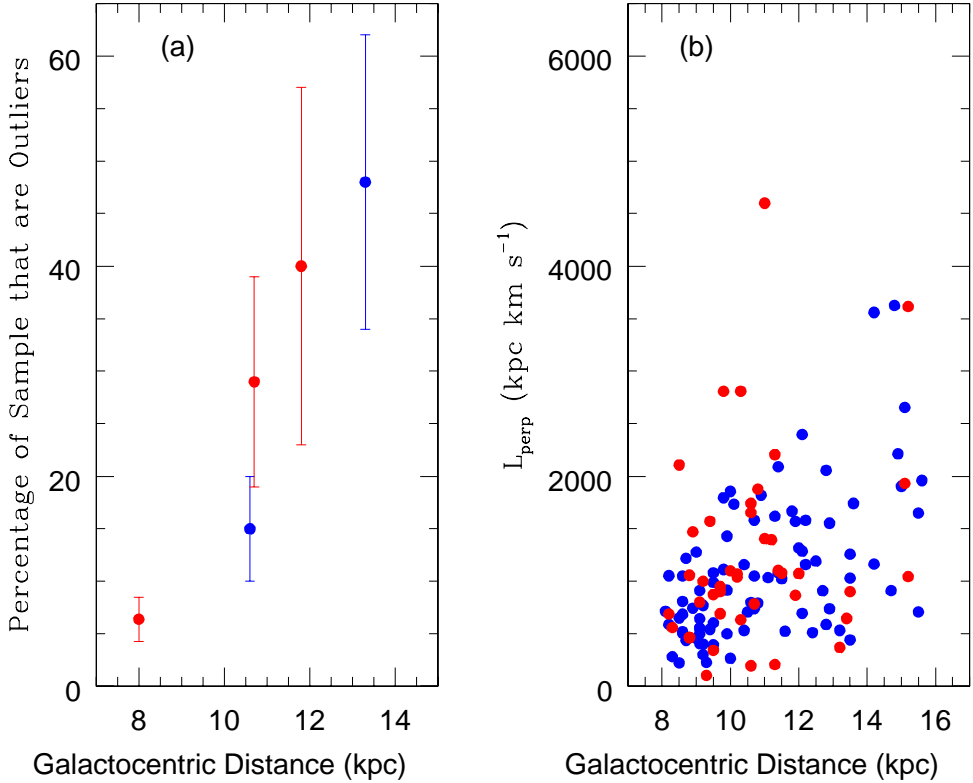}
\caption{ (a) The percentage of outliers in each sample of BHB stars (blue filled 
 circles) and RR Lyrae stars (red filled circles) as a function of 
Galactocentric distance in kpc. The samples are described in Table \ref{t:out-mc}. 
 (b) The angular momentum L$_{\perp}$ (in kpc km s$^{-1}$) for the BHB stars
 (blue filled circles) and RR Lyrae stars (red filled circles) for the NGP and
 Anticentre samples within 8.5 kpc as a function of Galactocentric distance.
{\it outliers} have a more spherical distribution and more retrograde orbits 
than those in the {\it main concentration} and they consitute a larger fraction 
of the halo with increasing galactocentric distance.
}
\label{f:8}
\end{figure*}

\begin{center}
\begin{table*}
\caption{Numbers of stars that are $outliers$ and in the $Main~Concentration$ 
 in the Anticentre and North Galactic Pole Fields. \label{t:out-mc}
 }
%\begin{flushleft}
\begin{tabular}{@{}lccccc@{}}
\noalign{\smallskip} \hline
  Field$^{\dagger}$  &  No. in & No. of & Per cent. &$\langle$R$_{gal}$$\rangle$ & $\langle$
 R$_{gal}$$\rangle$  \\
      & Main & Outliers& of    & $M. C.$ & $outliers$ \\
      & Conc.&         & Outliers &(kpc)  & (kpc)     \\
\hline \hline
 ANTICENTRE BHB & 15 & 14&48\%&12.5$\pm$0.4 & 14.1$\pm$0.4   \\
 ANTICENTRE RR  &  9 &  6&40\%&11.2$\pm$0.6 & 12.5$\pm$0.7   \\
 ANTICENTRE BHB + RR  & 24 & 20&45\%&12.0$\pm$0.4 & 13.6$\pm$0.5   \\
 NGP        BHB & 51 &  9&15\%& 9.7$\pm$0.2 & 11.4$\pm$0.4   \\
 NGP        RR  & 22 &  9&29\%&10.2$\pm$0.3 & 11.2$\pm$0.6   \\
 NGP        BHB + RR  & 73 & 18&20\%&9.8$\pm$0.2 & 11.3$\pm$0.3   \\
\hline
\end{tabular}

Notes to Table: \\
$^{\dagger}$~~~ The Anticentre fields are those described in this paper and the
Fields at the NGP are those described in Kinman et al. (2007b). \\
\end{table*}
\end{center}

\begin{center}
\begin{table*}
\caption{Candidates for membership of the K07 group. \label{t:k07}} 
%\begin{flushleft}
\begin{tabular}{@{}lccccccc@{}}
\noalign{\smallskip} \hline
 Star& Type$^{\dagger}$& L$_{\perp}$ & L$_{z}$ &$\langle$$V$$\rangle$ & D & [Fe/H] & Period \\
    &  &kpc km s$^{-1}$&kpc km s$^{-1}$ & mag. & kpc &     & days             \\
\hline \hline
 HD 214925  & RG$^{a}$ & 1322$\pm$119 &--2177$\pm$205& 9.30 & 2.15 &--2.15  &  ...    \\
 AT VIR     & RR$^{a}$ & 1576$\pm$93  &--1712$\pm$149& 11.34& 1.30 &--1.60  & 0.5257  \\  
 RV CAP     & RR$^{a}$ & 1464$\pm$146 &--2090$\pm$239& 11.04& 1.06 &--1.72  & 0.4477  \\
            &     &      &      &      &      &        &         \\  
 RR7-066    & BHB$^{b}$ & 1960$\pm$1028 &--1533$\pm$645& 15.31& 8.47 & ...    & ...     \\  
 CHSS 608   & BHB$^{b}$ & 1742$\pm$452  &--2152$\pm$657& 14.81& 6.76 & ...    & ...     \\
            &     &      &      &      &      &        &         \\  
 AF-115     & BHB$^{c}$ & 1571$\pm$457 &--1484$\pm$494 & 15.42& 8.13 & ...    & ...     \\ 
 SA57-032   & BHB$^{c}$ & 1621$\pm$604 &--1514$\pm$670 & 15.13& 7.81 & ...    & ...     \\  
 AF-041     & BHB$^{c}$ & 1668$\pm$511 &--1928$\pm$635 & 15.02& 7.46 & ...    & ...     \\
 AF-053     & BHB$^{c}$ & 2091$\pm$768 &--1609$\pm$828 & 15.19& 6.94 & ...    & ...     \\  
 SA57-001   & BHB$^{c}$ & 1430$\pm$342 &--1742$\pm$503 & 14.43& 5.68 & ...    & ...     \\ 
 AF-108     & BHB$^{c}$ & 1279$\pm$224 &--2185$\pm$370 & 13.86& 3.82 & ...    & ...     \\
            &     &      &      &      &      &        &         \\  
 IP COM     & RR$^{c}$  & 1878$\pm$780 &--1860$\pm$780 & 14.85& 7.25 &--1.48  & 0.6406  \\ 
 EO COM     & RR$^{c}$  & 1743$\pm$483 &--1889$\pm$528 & 14.74& 6.94 &--1.67  & 0.6320  \\  
 MQ COM     & RR$^{c}$  & 1572$\pm$393 &--1895$\pm$501 & 14.23& 5.40 & ...    & 0.6224  \\ 
 IS COM     & RR$^{c}$  & 1472$\pm$177 &--1642$\pm$338 & 13.80& 4.44 & ...    & 0.3146  \\  
\hline
\end{tabular}

Notes to Table: \\
$^{\dagger}$~~~ RG = red giant; BHB = blue horizontal branch star; RR = RR
 Lyrae star. The superscripts a, b \& c indicate that the star belongs to 
 the solar neighbourhood, the Anticentre fields of the present paper or 
 the NGP fields of Kinman et al. (2007b) respectively.   \\
\end{table*}
\end{center}

\subsection{Ratio of the number of $outliers$ to 
 the number in  the {\it main concentration}.}

The ratio of the number of $outliers$ to the number in the {\it main 
 concentration} is a simple measure of the spread of halo stars in the
 L$_{\perp}$ $vs$ L$_{z}$ plot. Our sample of 
 globular clusters within 10 kpc contains 5 (with [Fe/H] $>$ --1.0) 
that belong to the disc or bulge. Among the remainder, 24 belong to the main
concentration and 2 (or 8\%) are {\it outliers}. The two {\it outliers}  
are NGC 3201 and NGC 6205 (M13). Although these two clusters are listed as
 ``young"  by Mar\'{i}n-Franch et al. (2009), Dotter et al. (2010) 
 give ages of 12.00$\pm$0.75 and 13.0$\pm$0.50 Gyr for NGC 3201 and
NGC 6205 respectively, which does not support this description.
 NGC 6205 (together with NGC 5466, NGC 6934 and NGC 7089) is one of a group
 of four globular clusters with similar L$_{\perp}$ and L$_{z}$ that are
 discussed by Smith et al. (2009) in connection with an overdensity in the
 subdwarfs that they studied. Smith et al. list the properties of 12 
 {\it outlier} subdwarfs that lie at heliocentric distances up to 5 kpc.
 They are shown by yellow open circles in Figs. \ref{f:4}(a) and (b).    
 They show some tendency to occur in groups among themselves in their 
 L$_{\perp}$ {\it vs} L$_{z}$ plot
 (as discussed by Smith et al.) but their locations in our plot 
 (Figs. \ref{f:4}(a) and (b)) show 
 little in common with those of our RR Lyrae and BHB {\it outliers}. 

The local BHB stars within 1 kpc all belong to the {\it main concentration}
 and have no {\it outliers}. The RR Lyrae stars within 1 kpc have 4 
{\it outliers} that belong to the H99 group (RZ CEP, XZ CYG, CS ERI and TT 
 LYN); MT TEL is a possible retrograde {\it outlier} that lies just outside
 the {\it main concentration}. The RR Lyrae stars at distances between 1 and 2
  kpc have 2 {\it outliers} that belong to the H99 group (TT CNC and AR SER), 
 2 that belong to the Kepley retrograde group (AT VIR and RV CAP),  and one 
prograde {\it outlier} (U CAE) besides a number that are on the edge of the 
{\it main concentration}. Kepley et (2007) found that XZ CYG belongs to the  
 H99 group and CS ERI is also likely to be a member of this group. 
 The H99 and K07 {\it outlier} groups are discussed further in Appendix F.

  Of the 188 RR Lyrae stars within 2 kpc 
  for which we have data, 41 are likely to belong to the thick disc.
 Of the remaining 147 halo stars, there are 5 in Fig. 5a and 10 in Fig.
 5b that formally lie outside the {\it main concentration} and so would 
 formally be considered {\it outliers}. Five of those in Fig.  5b,
 however, lie so close the boundary of the {\it main concentration} that
 (with reasonable assumptions as to their error bars) it seems likely 
 that most belong to the {\it main concentration}. The 10 certain 
 {\it outliers} comprise 7$\pm$2\% of the total. If we include the 5 that
 lie close to the {\it main concentration} boundary, there are 15 
 {\it outliers} or 10$\pm$3\% of the total. These percentages are 
 comparable with those found (10\%) by both Helmi et al. (1999) and
 Morrison et al. (2009) among their local samples of metal-poor halo stars. 

 Table \ref{t:out-mc} gives the number of stars in the {\it main concentration} and the 
 number of {\it outliers} for the BHB and RR Lyrae stars within 8.5 kpc in 
 both the Anticenter and NGP (Kinman et al. 2007) fields. The percentage of
 {\it outliers} and the L$_{\perp}$ of the stars in these
 fields is shown plotted against galactocentric distance in Figs. \ref{f:8}(a)
 and  \ref{f:8}(b) respectively. It can be seen 
 from Fig. \ref{f:4} and Fig \ref{f:8}(b) that the majority of  {\it outliers} have 
greater L$_{\perp}$ and more negative L$_{z}$ than the stars in the {\it main 
concentration}. This shows, according to the correlations shown in Fig. 7, 
that the orbits of the {\it outliers} tend to be more retrograde and reach 
larger $\mid$z$_{max}$$\mid$ than the stars in the {\it main concentration}.
 The increase in the  percentage of {\it outliers} with galactocentric 
 distance shown in Fig. \ref{f:8}(a) therefore implies that 
 {\it as the galactocentric distance increases, the halo has 
 an increasing contribution from stars that have more retrograde orbits and 
 a more spherical distribution than the stars in  the {\it main concentration}
that predominate in the solar neighbourhood.} This result is in general 
agreement with the observational results of Carollo et al. (2007,2010) and Beers et
al. (2012) and the simulations of Oser et al. (2010), Font et al. (2011) and 
McCarthy et al. (2012). Our observational support for the duality of the halo is
important because (as the simulations have shown) dual halos are a general property
 of the stellar spheroids of disk galaxies whose masses are comparable with that
 of the Milky Way. We are grateful to the referee for asking us to emphasize 
 this point.

We note that a simulation of a ``smooth halo" with a Gaussian distribution of
velocities (e.g. the right-handed L$_{\perp}$ $vs.$ L$_{z}$ plot of Fig. 5 in
Smith et al. 2009) gives a {\it main concentration} that is similar in shape
but smoother than that shown by the observations. In this connection we note 
that Hattori \& Yoshii (2011) conclude that violent relaxation has been effective
for stars within a scale radius of 10 kpc from the Galactic centre. 
{\it We suggest that the stars of the {\it main concentration} are those 
where this relaxation has been most effective}.

\section{Summary and Conclusions}

   Fifty one BHB stars and 12 possible BHB stars are identified in the 
   Anticentre. Our selection criteria for these stars give results 
   that agree with those used by Smith et al. (2010) and  
   Ruhland et al. (2011). Fifty eight RR Lyrae stars are identified in the
   Anticentre; 7 of these are new and their light curves are given in
   Appendix B. Photometric data for the BHB and RR Lyrae stars are given in
   Tables \ref{t:bhb} and \ref{t:rrl} respectively. Five methods are used to get distances for the
   BHB stars and three methods for the RR Lyrae stars; these are compared and 
   combined to give distances on a uniform scale.  Absolute proper motions
   (largely derived from the GSCII and SDSS DR7 databases) are given for 
   all these stars and also radial velocities for 31 of the BHB and 37 of
   the RR Lyrae stars (Tables \ref{t:bhb2} and \ref{t:rrl2}). 
   Our conclusions are itemized below:

  \begin{enumerate}
   \item
   All but 4 of the 58 RR Lyrae stars in the Anticentre fields are of 
   Oosterhoff type I; this agrees with the Oo II stars being more centrally
   concentrated in the Galaxy than those of Oo type I (Miceli et al. 2008).  
   Oo I globular clusters tend to have retrograde orbits 
  (Lee \& Carney, 1999; van den Bergh, 1993); the field RR Lyrae stars in the
   Anticentre tend to have retrograde orbits. 

   \item
   We combined the kinematic data of our Anticentre stars with 
   those of the stars in the North Galactic Pole fields (Kinman et al. 2007b).
   In the combined data, the Galactic V motion (Fig. \ref{f:3}) is significantly
  retrograde for {\it both} BHB and RR Lyrae stars with R$_{gal}$ $>$ 10 kpc.
  This agrees with the findings of Carollo et al. (2007), Carollo et al. (2010) 
 and Beers et al. (2012) that the {\it outer halo}  
 shows retrograde rotation compared with the rotation of the stars in the
 solar neighbourhood where the {\it inner halo} predominates. The lack of any
 similar trend in the Galactic U motion makes it unlikely that the trend in
 the V motion is caused by a systematic error in the proper motions.

   \item
Angular momenta plots (L$_{\perp}$ $vs.$ L$_{z}$) for the BHB and RR Lyrae stars
  in the Anticentre fields and the North Galactic Pole fields are compared with
  similar plots for these stars in the solar neighbourhood and for the globular
  clusters nearer than 10 kpc. We suggest that halo stars belong to either of 
  two groups --- either the
  {\it main concentration} or the {\it outliers} ---  according to whether they lie
 inside or outside a contour in this plot which encloses the majority of 
 metal-poor stars in the solar neighbourhood (as defined by Morrison et al.  
 2009). We suggest that the stars in the {\it main concentration} are those for
 which violent relaxation has been most effective (Hattori \& Yoshii, 2011).
 The ratio of {\it outliers} to {\it main concentration} stars increases with
 galactocentric distance (Fig. \ref{f:8}). The {\it outliers} primarily have retrograde
 orbits. Since L$_{\perp}$ correlates with z$_{max}$ (the orbit's maximum height
 above the galactic plane), this also implies that the halo becomes more
 spherical with increasing galactocentric distance
 (c.f. Schmidt, 1956; Kinman et al., 1966, Miceli et al., 2008 Table 2).
 It also agrees with the simulations (McCarthy et al., 2012) that predict
 that the inner halo should be more flattened than the outer halo.
  
   \item
 A review of the RR Lyrae stars in the H99 group  of {\it outliers} (Helmi 
 et al. 1999) shows that there are six RR Lyrae stars that are likely
 members  (all probably of Oo type I) and that their mean [Fe/H] is --1.59.
 Their mean {\it rms} scatter in [Fe/H] is 0.16 which is comparable with the 
 likely errors in these metallicities.  These  RR Lyrae stars therefore form a 
 more homogeneous set than the later-type stars in H99 (Roederer et al. 2010) 
  and they could have originated from a single globular cluster.
 Another grouping with similar L$_{\perp}$ and L$_{z}$) (which we
 call K07) contains 15 BHB and RR Lyrae stars at distances in the range 1.1  
 to 8.5 kpc. K07 contains two pairs of RR Lyrae stars (AT VIR \& RV CAP and 
 IP COM \& EO COM); the stars in each pair have similar properties. Better data
 are needed to verify  membership of the other stars in K07.

\end{enumerate} 

\section*{Acknowledgments}

We thank D.R. Soderblom for kindly making  available the 
program to calculate the UVW space motions and Heather Morrison for allowing us
to use her program for computing L$_{\perp}$ and L$_{z}$  and the referee 
for comments which helped improve the paper.
This research has made use of 2MASS data provided by the NASA/IPAC 
Infrared Science Archive, which is operated by the Jet Propulsion Laboratory,
California Institute of Technology, under contract with the National 
Aeronautics and Space Administration.\\
The GSCII is a joint project of the Space Telescope Science
Institute (STScI) and the INAF-Osservatorio Astronomico di Torino
(INAF-OATo). \\
This work is partly based on observations made with the Italian Telescopio 
 Nazionale Galileo (TNG) operated on the island of La Palma by the Fundacion 
Galileo Galilei of the INAF (Istituto Nazionale di Astrofisica) at the Spanish
Observatorio del Roque de los Muchachos of the Instituto de Astrofisica de
Canarias.\\
This work has been partly supported by the MIUR (Mi\-ni\-ste\-ro 
dell'Istruzione, dell'Universit\`a e della Ricerca) under 
PRIN-2001-1028897 and PRIN-2005-1060802.

%\newpage

%\smallskip
% Table 1 
\begin{center}
\begin{landscape} 
\topmargin 45mm 
\thispagestyle{empty}
\begin{table*} 
\caption{Positions and Photometry for the BHB candidate stars. The equatorial 
 coordinates are for J2000. The magnitudes and colours $V$,$B$,$(u-B)_{K}$,
 $NUV$ and $K$ are defined in the text.
 W is a weight and its relation to the Type (column 14) is given in 
 Appendix A. \label{t:bhb}
}
\label{t:bhb1}
\begin{tabular}{@{}clccccccccccccc@{}} 
\hline 
 No &ID& RA&DEC&l & b&$V$&$B-V$&$(u-B)_{K}$&E$(B-V)$& $NUV$&$K$   & W & Type& Note\\ 
(1) &(2)&(3) &(4) &(5) &(6) &(7) &(8) &(9) &(10) &(11) &(12)&(13)&(14)&(15)   \\
\hline 
  1&BS 17438-0126&08 04 07.0 &+38 10 30 &182.5&+30.00&13.55$\pm$0.01&
 $+$0.215$\pm$0.006&1.985$\pm$0.010&0.046&16.146$\pm$0.004&12.800$\pm$0.024&
       $-$6 & A &   \\
  2&P 54-32.5    &08 05 30.9 &+41 08 02 &179.2&+30.9&15.38$\pm$0.01&
     $+$0.137$\pm$0.013&2.026$\pm$0.013&0.064&17.904$\pm$0.033&14.813$\pm$0.094
        &$+$4 & bhb &   \\
  3&AF 186       &08 06 11.1 &+40 15 01 &180.3&+30.8&15.66$\pm$0.01&
     $+$0.155$\pm$0.013&2.121$\pm$0.013&0.055&18.238$\pm$0.014&15.281$\pm$0.120
        &$+$8 & BHB &  \\
  4&AF 189       &08 09 21.8 &+38 18 00 &182.6&+31.0&15.20$\pm$0.01&
     $+$0.085$\pm$0.016&2.137$\pm$0.043&0.048&17.616$\pm$0.029&15.108$\pm$0.133
        &$+$8 & BHB &  \\
  5&P 54-111     &08 12 06.5 &+38 50 53 &182.1&+31.7&14.58$\pm$0.01&
     $+$0.136$\pm$0.009&2.119$\pm$0.013&0.039&17.232$\pm$0.008&14.053$\pm$0.053
        &$+$8 & BHB & \\
  6&P 54-122     &08 16 00.3 &+40 10 00 &180.7&+32.6&15.17$\pm$0.01&
     $+$0.036$\pm$0.008&1.965$\pm$0.013&0.045&  .....         &15.152$\pm$0.140
        &$+$4 &bhb &  \\
  7&P 54-120     &08 16 02.9 &+39 25 11 &181.6&+32.5&12.49$\pm$0.01&
     $+$0.216$\pm$0.002&1.984$\pm$0.005&0.039&15.400$\pm$0.010&11.940$\pm$0.022
        &$-$3 & A  &  \\
  8&P 54-119     &08 17 41.6 &+39 04 29 &182.1&+32.8&14.26$\pm$0.01&
     $+$0.214$\pm$0.007&2.087$\pm$0.011&0.038&17.254$\pm$0.023&13.532$\pm$0.039
        &$+$6 &BHB &  \\
  9&BS 17444-0025&08 21 33.5 &+42 31 36 &178.2&+34.0&10.11$\pm$0.01&
     $+$0.136$\pm$0.005&2.138$\pm$0.005&0.052&   .....        &09.684$\pm$0.013
        &$+$4 &bhb & 1\\
 10&AF 209       &08 21 48.5 &+42 27 36 &178.2&+34.1&16.41$\pm$0.02&
     $+$0.056$\pm$0.022&2.042$\pm$0.032&0.051&18.889$\pm$0.031&$>$15.43
        &$+$6 &BHB &  \\
 11&AF 210       &08 21 59.7 &+42 18 55 &178.4&+34.1&15.42$\pm$0.01&
     $+$0.189$\pm$0.012&  .....        &0.051&18.090$\pm$0.020&15.339$\pm$0.189
        &$+$4 &bhb &  \\
 12&AF 214       &08 22 51.5 &+36 18 04 &185.6&+33.3&15.66$\pm$0.01&
     $+$0.226$\pm$0.014&2.088$\pm$0.020&0.058&18.751$\pm$0.049&15.126$\pm$0.126
        &$+$8 &BHB &  \\
 13&RR7 002      &08 22 00.6 &+37 09 40 &184.5&+33.3&15.13$\pm$0.01&
     $+$0.201$\pm$0.010&2.167$\pm$0.020&0.058&18.119$\pm$0.031&14.412$\pm$0.086
        &$+$8 &BHB &2 \\
 14&P 81-42      &08 23 47.9 &+44 32 44 &175.8&+34.7&14.43$\pm$0.01&
     $+$0.050$\pm$0.004&2.039$\pm$0.011&0.045&16.622$\pm$0.010&14.142$\pm$0.059
        &$+$8 &BHB &  \\
 15&RR7 008      &08 24 09.6 &+41 43 47 &179.2&+34.4&15.14$\pm$0.01&
     $+$0.118$\pm$0.010&2.120$\pm$0.020&0.043&17.601$\pm$0.010&14.510$\pm$0.081
        &$+$8 &BHB &3 \\
 16&RR7 015      &08 26 30.4 &+38 10 16 &183.5&+34.3&11.75$\pm$0.01&
     $+$0.208$\pm$0.010&2.081$\pm$0.020&0.042&   .....        &11.007$\pm$0.020
        &$+$10&BHB &4 \\
 17&P 81-39      &08 27 16.7 &+45 18 10 &174.9&+35.3&15.69$\pm$0.01&
     $+$0.035$\pm$0.008&2.002$\pm$0.011&0.031&17.902$\pm$0.023&15.200$\pm$0.176
        &$+$8 &BHB &  \\
 18&RR7 021      &08 28 02.3 &+40 21 57 &181.0&+34.9&15.33$\pm$0.01&
     $+$0.104$\pm$0.010&2.118$\pm$0.020&0.042&18.088$\pm$0.032&14.885$\pm$0.131
        &$+$1 &bhb &  \\
 19&P 81-72      &08 28 14.1 &+44 40 46 &175.7&+35.5&11.79$\pm$0.01&
     $+$0.202$\pm$0.001&2.024$\pm$0.003&0.028&    .....       &11.281$\pm$0.019
        &$+$1 &bhb & 5\\
 20&P 81-79      &08 28 19.9 &+45 26 09 &174.8&+35.5&13.37$\pm$0.01&
     $+$0.243$\pm$0.002&1.999$\pm$0.005&0.026&16.445$\pm$0.004&12.732$\pm$0.025
        &$-$6 & A  &  \\
 21&RR7 023      &08 28 29.6 &+40 27 47 &180.9&+35.0&12.64$\pm$0.01&
     $+$0.059$\pm$0.010&2.111$\pm$0.020&0.044&14.926$\pm$0.007&12.347$\pm$0.022
        &$+$9 &BHB & 6\\
 22&RR7 036      &08 32 26.5 &+39 27 25 &182.2&+35.7&15.25$\pm$0.01&
     $+$0.165$\pm$0.010&2.137$\pm$0.020&0.044&18.000$\pm$0.046&14.765$\pm$0.102
        &$+$12&BHB & 7\\
 23&P 81-101     &08 34 24.5 &+46 00 23 &174.2&+36.6&15.43$\pm$0.01&
     $+$0.154$\pm$0.015&2.116$\pm$0.020&0.029&18.146$\pm$0.029&14.620$\pm$0.070
        &$+$5 &bhb &  \\
 24&P 81-121     &08 35 25.7 &+44 00 41 &176.7&+36.7&15.65$\pm$0.01&
     $+$0.045$\pm$0.012&1.975$\pm$0.014&0.030&17.676$\pm$0.011&15.316$\pm$0.149
        &$+$8 &BHB &  \\
 25&RR7 043      &08 35 29.7 &+42 01 19 &179.1&+36.5&16.62$\pm$0.01&
     $+$0.036$\pm$0.015&1.986$\pm$0.020&0.027&18.721$\pm$0.034&15.853$\pm$0.238
        &$+$8 &BHB & 8\\
 26&P 28-045     &08 37 42.5 &+36 07 28 &186.5&+36.2&14.315$\pm$0.01&
     $+$0.135$\pm$0.008&2.133$\pm$0.021&0.035&     .....      &13.809$\pm$0.042
        &$+$4 &bhb &  \\
 27&RR7 053      &08 38 52.8 &+40 54 32 &180.6&+37.0&15.05$\pm$0.01&
     $+$0.160$\pm$0.010&2.175$\pm$0.020&0.041&17.782$\pm$0.033&14.4:           
        &$+$8 &BHB & 9 \\
 28&P 81-162     &08 39 57.3 &+46 26 45 &173.7&+37.6&16.05$\pm$0.01&
     $+$0.071$\pm$0.021&2.047$\pm$0.015&0.027&18.341$\pm$0.010&15.819$\pm$0.228
        &$+$8 &BHB &10  \\
 29&RR7 058      &08 40 47.5 &+38 13 49 &184.0&+37.1&15.43$\pm$0.01&
     $+$0.064$\pm$0.010&2.020$\pm$0.020&0.041&17.641$\pm$0.026&15.055$\pm$0.124
        &$+$8 &BHB &11  \\
 30&RR7 060      &08 42 16.8 &+36 45 35 &185.9&+37.2&14.49$\pm$0.01&
     $+$0.160$\pm$0.010&2.102$\pm$0.020&0.035&17.164$\pm$0.023&13.836$\pm$0.055
        &$+$11&BHB & 12 \\
 31&P 82-04      &08 42 41.3 &+42 47 07 &178.3&+37.9&15.84$\pm$0.02&
     $+$0.061$\pm$0.014&2.069$\pm$0.029&0.033&18.123$\pm$0.027&15.544$\pm$0.208
        &$+$8 &BHB &13  \\
 32&RR7 064      &08 43 04.9 &+41 39 05 &179.7&+37.9&11.22$\pm$0.01&
     $+$0.115$\pm$0.010&2.100$\pm$0.010&0.032&    .....       &10.820$\pm$0.018
        &$+$8 &BHB & 14 \\
 33&RR7 066      &08 43 40.9 &+39 49 20 &182.1&+37.8&15.31$\pm$0.01&
     $+$0.130$\pm$0.010&2.087$\pm$0.020&0.035&17.901$\pm$0.030&14.771$\pm$0.108
        &$+$8 & BHB& 15 \\
 34&P 81-167     &08 45 20.2 &+46 41 34 &173.4&+38.5&14.44$\pm$0.01&
     $+$0.109$\pm$0.008&2.119$\pm$0.014&0.035&16.957$\pm$0.027&13.959$\pm$0.049
        &$+$8 &BHB &  \\
 35&P 11419-01   &08 46 48.8 &+29 49 04 &194.6&+36.8&12.79$\pm$0.01&
     $+$0.123$\pm$0.006&2.079$\pm$0.024&0.043&15.376$\pm$0.010&12.348$\pm$0.022
        &$+$9 &BHB &16  \\
 36&RR7 082      &08 47 09.9 &+42 16 04 &179.0&+38.7&10.69$\pm$0.01&
     $+$0.069$\pm$0.002&1.972$\pm$0.003&0.028&     .....      &10.475$\pm$0.018
        &$-$3 & A  &17  \\
 37&AF 293       &08 47 41.0 &+36 10 50 &186.8&+38.2&16.24$\pm$0.01&
     $+$0.193$\pm$0.013&2.055$\pm$0.036&0.031&18.904$\pm$0.015&15.435$\pm$0.189
        &$+$8 &BHB &  \\
 38&P 11419-04   &08 47 59.8 &+31 31 06 &192.6&+37.4&14.88$\pm$0.01&
     $+$0.044$\pm$0.004&2.017$\pm$0.017&0.038&17.028$\pm$0.022&14.739$\pm$0.087
        &$+$8 &BHB &  \\
 39&RR7 084      &08 48 15.0 &+40 28 45 &181.3&+38.8&15.76$\pm$0.01&
     $+$0.141$\pm$0.010&2.097$\pm$0.020&0.028&18.476$\pm$0.024&15.148$\pm$0.143
        &$+$8 &BHB &18  \\
 40&RR7 091      &08 49 10.7 &+39 40 06 &182.4&+38.9&14.16$\pm$0.01&
     $+$0.042$\pm$0.010&2.046$\pm$0.020&0.026&16.343$\pm$0.010&13.906$\pm$0.043
        &$+$12&BHB &19  \\	
\hline
\end{tabular}
\end{table*}
\end{landscape}
\end{center}

\begin{center}
\begin{landscape} 
\topmargin 50mm 
\addtocounter{table}{-1}
\thispagestyle{empty}
\begin{table*} 
\caption{Continued.
}
\label{t:bhb1}
\begin{tabular}{@{}clccccccccccccc@{}} 
\hline 
 No &ID& RA&DEC&l & b&$V$&$B-V$&$(u-B)_{K}$&E$(B-V)$& $NUV$&$K$   & W & Type& Note\\ 
(1) &(2)&(3) &(4) &(5) &(6) &(7) &(8) &(9) &(10) &(11) &(12)&(13)&(14)&(15)   \\
\hline 

 41&RR7 090      &08 49 23.3 &+40 23 40 &181.5&+39.0&15.65$\pm$0.01&
     $+$0.089$\pm$0.010&2.090$\pm$0.020&0.024&18.028$\pm$0.019&15.285$\pm$0.125
        &$+$8 &BHB &20  \\
 42&P 82-49      &08 53 25.8 &+44 26 21 &176.3&+39.9&15.48$\pm$0.01&
     $+$0.108$\pm$0.008&2.099$\pm$0.017&0.029&17.905$\pm$0.032&14.977$\pm$0.103
        &$+$8 &BHB & 21 \\
 43&BS 16473-0090&08 57 34.5 &+43 28 21 &177.6&+40.7&10.91$\pm$0.01&
     $+$0.215$\pm$0.010&1.991$\pm$0.010&0.023&     .....      &10.414$\pm$0.020
        &$ $0 & A  &  \\
 44&BS 16473-0102&08 58 27.1 &+47 04 08 &172.8&+40.8&13.93$\pm$0.01&
     $+$0.103$\pm$0.005&2.092$\pm$0.017&0.021&16.338$\pm$0.014&13.504$\pm$0.033
        &$+$8 &BHB &  \\
 45&BS 17139-0069&09 06 14.4 &+30 58 33 &194.3&+41.2&14.45$\pm$0.01&
     $+$0.146$\pm$0.004&2.073$\pm$0.015&0.025&17.052$\pm$0.044&13.879$\pm$0.046
        &$+$8 &BHB &  \\
 46&TON 384      &09 06 56.8 &+30 04 20 &195.5&+41.1&15.25$\pm$0.01&
     $+$0.056$\pm$0.011&2.126$\pm$0.061&0.028&17.721$\pm$0.035&14.759$\pm$0.078
        &$+$4 &bhb & 22 \\
 47&BS 16468-0026&09 07 13.9 &+40 23 48 &181.7&+42.4&14.72$\pm$0.01&
     $+$0.077$\pm$0.005&2.089$\pm$0.054&0.020&17.101$\pm$0.018&14.186$\pm$0.045
        &$+$8 &BHB &  \\
 48&AF 379       &09 10 37.1 &+38 55 10 &183.8&+43.0&15.26$\pm$0.02&
     $+$0.064$\pm$0.010&2.018$\pm$0.023&0.019&17.547$\pm$0.026&15.149$\pm$0.139
        &$+$8 &BHB &  \\
 49&AF 386       &09 13 30.0 &+36 49 22 &186.7&+43.4&15.01$\pm$0.01&
     $+$0.088$\pm$0.005&2.005$\pm$0.032&0.020&17.476$\pm$0.370&14.679$\pm$0.089
        &$+$6 &BHB &  \\
 50&AF 390       &09 15 35.7 &+38 36 12 &184.3&+43.9&15.37$\pm$0.01&
     $+$0.042$\pm$0.025&2.091$\pm$0.020&0.019&17.497$\pm$0.022&15.117$\pm$0.149
        &$+$8 &BHB &  \\
 51&BS 16468-0078&09 16 19.4 &+40 16 02 &181.9&+44.1&11.67$\pm$0.01&
     $+$0.024$\pm$0.001&1.998$\pm$0.010&0.014&       .....    &11.446$\pm$0.017
        &$+$7 &BHB &23\\
 52&P 30-16      &09 16 53.1 &+35 52 17 &188.1&+44.0&14.40$\pm$0.01&
     $+$0.123$\pm$0.006&2.058$\pm$0.030&0.019&       .....    &13.879$\pm$0.055
        &$+$4 &bhb &  \\
 53&BS 16468-0080&09 17 13.4 &+41 20 43 &180.5&+44.2&14.09$\pm$0.01&
     $+$0.023$\pm$0.009&1.900$\pm$0.017&0.021&15.915$\pm$0.012&14.009$\pm$0.057
        &$+$4 &bhb &  \\
 54&P 30-28      &09 17 44.4 &+33 20 52 &191.6&+43.9&10.28$\pm$0.01&
     $+$0.119$\pm$0.004&2.083$\pm$0.003&0.018&     .....      &09.757$\pm$0.016
        &$+$7 &BHB & 24 \\
 55&BS 16468-0090&09 18 25.8 &+39 29 59 &183.0&+44.5&14.10$\pm$0.01&
     $+$0.132$\pm$0.011&2.075$\pm$0.004&0.016&16.674$\pm$0.017&13.676$\pm$0.045
        &$+$8 &BHB &  \\
 56&CHSS 608     &09 18 59.0 &+29 40 46 &196.7&+43.6&14.81$\pm$0.01&
     $+$0.089$\pm$0.011&2.045$\pm$0.030&0.021&17.276$\pm$0.023&14.353$\pm$0.071
        &$+$11&BHB &  \\
 57&P 11424-28   &09 20 23.3 &+31 17 11 &194.5&+44.2&14.50$\pm$0.01&
     $+$0.104$\pm$0.007&2.049$\pm$0.024&0.023&17.052$\pm$0.022&14.008$\pm$0.053
        &$+$8 &BHB &  \\
 58&P 30-38      &09 21 27.6 &+35 24 14 &188.8&+44.9&14.39$\pm$0.01&
     $+$0.142$\pm$0.015&2.046$\pm$0.050&0.018&17.029$\pm$0.016&13.929$\pm$0.057
        &$+$8 &BHB &  \\
 59& 57-121      &09 25 48.5 &+39 04 30 &183.7&+45.9&15.25$\pm$0.02&
     $+$0.005$\pm$0.020&2.040$\pm$0.033&0.014&17.310$\pm$0.025&15.396$\pm$0.209
        &$+$6 &BHB &  \\
 60& AF 419      &09 25 56.3 &+28 50 07 &198.2&+45.0&15.19$\pm$0.02&
     $+$0.113$\pm$0.016&2.100$\pm$0.043&0.019&17.636$\pm$0.029&14.611$\pm$0.093
        &$+$9 &BHB & 25 \\
 61&P 11424-70   &09 30 02.7 &+31 54 14 &194.1&+46.3&14.57$\pm$0.01&
     $+$0.072$\pm$0.019&2.072$\pm$0.032&0.020&16.939$\pm$0.019&14.236$\pm$0.073
        &$+$8 &BHB &  \\
 62&BS 16927-22  &09 32 43.6 &+39 28 31 &183.1&+47.3&11.18$\pm$0.01&
     $+$0.103$\pm$0.005&2.084$\pm$0.011&0.017&      .....     &10.710$\pm$0.019
        &$+$4 &bhb &  \\
 63&CHSS 663     &09 33 48.8 &+29 07 14 &198.2&+46.7&15.18$\pm$0.01&
     $+$0.056$\pm$0.011&2.077$\pm$0.035&0.018&17.282$\pm$0.024&14.959$\pm$0.112
        &$+$12&BHB &  \\
 64&P 11424-82   &09 34 23.5 &+29 48 10 &197.3&+46.9&14.99$\pm$0.01&
     $+$0.136$\pm$0.017&2.115$\pm$0.048&0.019&17.554$\pm$0.026&14.379$\pm$0.068
        &$+$8 &BHB &  \\
 65&BS 16940-45  &09 37 07.2 &+36 09 48 &188.0&+48.1&13.55$\pm$0.01&
     $+$0.021$\pm$0.011&2.009$\pm$0.008&0.014&15.602$\pm$0.015&13.336$\pm$0.030
        &$+$8 &BHB &  \\
 66&BS 16927-55  &09 40 31.5 &+41 48 32 &179.5&+48.6&14.53$\pm$0.01&
     $+$0.003$\pm$0.005&1.993$\pm$0.046&0.012&16.592$\pm$0.023&14.276$\pm$0.064
        &$+$6 &BHB &  \\
 67&BS 16940-0070&09 42 36.7 &+34 38 18 &190.4&+49.1&14.98$\pm$0.01&
     $+$0.016$\pm$0.008&2.002$\pm$0.083&0.011&17.201$\pm$0.031&14.815$\pm$0.077
        &$+$4 &bhb &  \\
 68&BS 16940-0072&09 43 10.3 &+33 57 10 &191.4&+49.2&13.99$\pm$0.01&
     $+$0.107$\pm$0.008&2.032$\pm$0.004&0.014&16.453$\pm$0.015&13.524$\pm$0.027
        &$+$8 &BHB &  \\
\hline
\end{tabular}
\noindent  Notes to table: \\
\bf{(1)} BD +42 1850;~                            
\bf{(2)} AF 211;~            
\bf{(3)} AF 217;~            
\bf{(4)} Str\"{o}mgren $\beta$ = 2.758; CHSS (class 3);~
\bf{(5)} Str\"{o}mgren $\beta$ = 2.855;~  \\
\bf{(6)} Str\"{o}mgren $\beta$ = 2.879; CHSS (class 3);~
\bf{(7)} CHSS (class 4);~    
\bf{(8)} AF 241;~            
\bf{(9)} AF 256;~            
\bf{(10)} US 1430;~  \\       
\bf{(11)} AF 262.~            
\bf{(12)} CHSS (class 3);~   
\bf{(13)} US 1513;~          
\bf{(14)} Str\"{o}mgren $\beta$ = 2.856; CHSS (class 3);~
\bf{(15)} AF 271;~    \\     
\bf{(16)} Str\"{o}mgren $\beta$ = 2.801;~
\bf{(17)} Str\"{o}mgren $\beta$ = 2.892;~
\bf{(18)} AF 297;~            
\bf{(19)} CHSS (class 4);~  
\bf{(20)} AF 307;~ \\         
\bf{(21)} BD +42 1926; US 1862;~           
\bf{(22)} AF 368;~           
\bf{(23)} Str\"{o}mgren $\beta$ = 2.858; \\
\bf{(24)} Str\"{o}mgren $\beta$ = 2.827; BD +33 1834;~
\bf{(25)} CHSS 632 (class 1).  \\  
\noindent References to Notes:  \\
 AF nnn (Pesch \& Sanduleak, 1989); CHSS (Brown et al., 2003); \\
 US nnnn (Usher \& Mitchell, 1982). \\
\end{table*}
\end{landscape}
\end{center}

\begin{center}
\topmargin 45mm 
\thispagestyle{empty}
\begin{table*} 
\caption{Positions, Photometry and Abundances for the RR Lyrae stars. The 
 equatorial coordinates are for J2000. The magnitudes and colours $V$
  and $K$ are defined in the text. Sources are given in the Notes. \label{t:rrl}
}
\label{t:rr1}
\begin{tabular}{@{}clcccccccccccc@{}} 
\hline 
 No &ID& RA&DEC&l & b&Type&$\log$P& [Fe/H] & $\langle V \rangle$ &$V_{amp}$ &
 $\langle K \rangle$ & E$(B-V)$ & Notes \\
(1) &(2)&(3) &(4) &(5) &(6) &(7) &(8) &(9) &(10) &(11) &(12)&(13)&(14)   \\
\hline 

  1 & V385 Aur &07 25 56.0&+38 12 59&180.3&+22.8&$ab$&$-$0.266 &...    &
  (17.41)  &(0.59) & ...           &0.053&             \\
  2 & V386 Aur &07 26 13.2&+40 52 50&177.6&+23.6&$ c$&$-$0.516 &$-$1.75&
  (16.75)  &(0.50) & ...           &0.063 &  5    \\
  3 & V387 Aur &07 27 01.0&+36 38 46&182.0&+22.5&$ab$&$-$0.308 &$-$1.32&
  (16.92)  &(1.05) & ...           &0.056 &  5    \\
  4 & V389 Aur &07 30 10.8&+38 21 54&180.4&+23.6&$ab$&$-$0.249 &...    &
  (17.53)  &(0.95) & ...           &0.057 &       \\
  5 & VX Lyn   &07 31 51.9&+39 07 47&179.7&+24.1&$ab$&$-$0.257 &$-$1.58&
    17.01  &(0.82) & ...           &0.057 &  4   \\
  6 & VY Lyn   &07 32 26.0&+38 50.05&180.1&+24.2&$c$&$-$0.451 &$-$1.57&
    15.75  &(0.37) & 14.61$\pm$0.12&0.062&   4  \\
  7 & VZ Lyn   &07 32 40.8&+41 37 38&177.1&+25.0&$c$&$-$0.487 &$-$1.48&
    16.20  &(0.42) & 15.43$\pm$0.19&0.054 &  4  \\
  8 & WX Lyn   &07 35 38.5&+39 15 27&179.8&+24.9&$ab$&$-$0.257 &$-$1.72&
    16.84  &(0.69) & ...           &0.049 &  4  \\
  9 & AS Lyn   &07 40 32.9&+41 11 37&178.0&+26.3&$ab$&$-$0.298 &$-$1.2:&
  (18.35)  &1.05) & ...           &0.049  & 1         \\
 10 & WZ Lyn   &07 40 45.7&+39 18 51&180.1&+25.8&$ab$&$-$0.207 &$-$1.89&
  (14.25)  &(0.95) & 13.11$\pm$0.03&0.049  & 2   \\
 11 & XZ Lyn   &07 44 48.4&+40 12 44&179.3&+26.8&$c$&$-$0.549 &...    &
  (16.32)  &(0.50) & ...           &0.050  &      \\
 12 & TW Lyn   &07 45 06.3&+43 06 42&176.1&+27.5&$ab$&$-$0.317 &$-$0.43&
    11.99  &1.00   & 10.78$\pm$0.02&0.046  &  3,8   \\
 13 & YY Lyn   &07 45 30.1&+37 22 59&182.4&+26.2&$c$&$-$0.476 &$-$1.87&
    14.98  &(0.46) & 14.08$\pm$0.11&0.065  &  4  \\
 14 & YZ Lyn   &07 45 40.9&+40 22 32&179.2&+27.0&$ab$&$-$0.304 &$-$0.6:&
  (17.47)  &(0.80) & ...           &0.052  &  1       \\
 15 & AU Lyn   &07 49 35.3&+41 42 57&177.9&+28.0&$ab$&$-$0.197 &$-$1.8:&
  (17.82)  &(0.76) & ...           &0.048  &  1      \\
 16 & ZZ Lyn   &07 50 21.8&+37 42 00&182.3&+27.3&$ab$&$-$0.313 &$-$1.42&
    15.80  &(1.03) & 15.08$\pm$0.14&0.048  &  4      \\
 17 & RW Lyn   &07 50 39.2&+38 27 15&181.5&+27.5&$ab$&$-$0.302 &$-$1.53&
    12.90  &(1.20) & 11.66$\pm$0.02&0.040  &  4,9  \\
 18 & AV Lyn   &07 54 09.6&+42 49 04&176.9&+29.1&$ab$&$-$0.233 &$-$1.7:&
  (16.62)  &(0.76) & ...           &0.048  &  1      \\
 19 & AC Lyn   &07 54 42.1&+38 54 20&181.2&+28.4&$ab$&$-$0.256 &$-$1.50&
    16.38  &(0.85) & ...           &0.047  &  4 \\
 20 & AD Lyn   &07 56 23.0&+39 22 58&180.8&+28.8&$c$&$-$0.450 &$-$1.46&
   15.85   &(0.49) & 15.13$\pm$0.15&0.061  &  4    \\
 21 & AW Lyn   &07 57 24.5&+43 12 29&176.5&+29.8&$ab$&$-$0.333 &$-$1.6:&
  (16.13)  &(0.92) & 15.00$\pm$0.11&0.036  &  1  \\
 22 & AX Lyn   &07 59 46.4&+39 16 30&181.1&+29.4&$ab$&$-$0.331 &...    &
  (18.54)  &(0.76) & ...           &0.048  &           \\
 23 & AY Lyn   &08 00 29.9&+40 39 24&179.6&+29.8&$c$&$-$0.503 &...    &
  (16.88)  &(0.47) & ...           &0.045  &         \\
 24 &P 54-13   &08 01 56.2&+41 01 18&179.2&+30.2&$ab$&$-$0.226 &...    &
    15.20  &0.90   & 13.52$\pm$0.04&0.058  &  6      \\
 25 & AZ Lyn   &08 03 39.8&+42 30 45&177.6&+30.7&$ab$&$-$0.324 &$-$2.24&
    16.47  &(0.84) & ...           &0.046  &  5      \\
 26 & BB Lyn   &08 04 36.2&+42 29 01&177.6&+30.9&$ab$&$-$0.253 &$-$1.36&
  (16.86)  &(0.92) & ...           &0.048  &  5      \\
 27 & BC Lyn   &08 09 37.4&+42 33 31&177.7&+31.9&$ab$&$-$0.281 &$-$1.6:&
  (17.00)  &(1.07) & ...           &0.048  &  1       \\
 28 & AF 194   &08 12 00.6&+40 39 20&180.0&+32.0&$ab$&$-$0.075 &...    &
    15.84  &0.45   & 14.37$\pm$0.08&0.048  &  7       \\
 29 & AF 197   &08 13 46.4&+38 03 02&183.1&+31.8&$c$&$-$0.411 &...    &
    15.50  &0.40   & 14.35$\pm$0.07&0.038  &  7      \\
 30 & DQ Lyn   &08 23 41.0&+37 28 11&184.2&+33.6&$c$&$-$0.306 &...    &
    11.41  &0.37   & 10.44$\pm$0.02&0.044  &  6   \\
 31 & RR7 032  &08 30 41.8&+40 24 24&181.0&+35.4&$ab$&$-$0.201 &...    &
    14.58  &0.65   & 13.22$\pm$0.03&0.047  &  6      \\
 32 & RR7 034  &08 31 52.2&+38 32 14&183.3&+35.4&$c$&$-$0.539 &...    &
    15.32  &0.29   & 14.64$\pm$0.09&0.039  &  6      \\
 33 &P 81 129  &08 32 49.6&+43 16 02&177.5&+36.2&$c$&$-$0.510 &...    &
    14.46  &0.52   & 13.67$\pm$0.04&0.022  &  6      \\
 34 &AF Lyn    &08 35 57.4&+41 01 11&180.4&+36.5&$ab$&$-$0.237 &$-$1.56 &
    16.12  &(0.76) & 14.87$\pm$0.09&0.039  &  4      \\
 35 &P 82 06   &08 43 56.7&+43 22 13&177.6&+38.2&$c$&$-$0.548 &...    &
    14.15  &0.30   & 13.45$\pm$0.03&0.024  &  6      \\
 36 & AI Lyn   &08 44 02.6&+38 54 48&183.2&+37.8&$ab$&$-$0.250 &...    &
  (17.10)  &(0.92) & ...           &0.032  &         \\
 37 & AK Lyn   &08 45 55.1&+39 14 55&182.8&+38.2&$ab$&$-$0.329 &$-$1.56&
    16.00  &(1.07) & 14.86$\pm$0.11&0.030  &  4        \\
 38 & EN Lyn   &08 46 07.0&+38 02 53&184.4&+38.1&$ab$&$-$0.204 &...    &
    13.53  &0.52   & 12.18$\pm$0.02&0.035  &  6       \\
 39 & RR7-086  &08 48 26.2&+36 20 08&186.6&+38.4&$c$&$-$0.451 &...    &
    16.14  &0.63   & 15.51$\pm$0.15&0.029  &  7        \\
 40 & AL Lyn   &08 49 13.1&+38 49 31&183.5&+38.8&$ab $&$-$0.293 &$-$1.90 &
    16.52  &(0.99) & 15.18$\pm$0.11&0.035  &  4        \\
 41 & AM Lyn   &08 49 50.2&+36 56 00&185.9&+38.7&$ab$&$-$0.294 &...    &
  (17.30)&(1.31)              & ...            & 0.033  &    \\
 42 &P 82-32   &08 50 39.5&+43 40 03&177.3&+39.4&$ab$&$-$0.304 &...    &
    15.07&1.16      & 14.24$\pm$0.06 & 0.031  & 6 \\
 43 & AF 316   &08 50.46.3&+41 18 54&180.3&+39.3&$c$&$-$0.462 &...    &
    16.13 &0.46     & 15.09$\pm$0.12 & 0.027  & 7 \\
 44 & RR7-101  &08 51 40.2&+40 17 11&181.6&+39.4&$c$&$-$0.482 &...    &
    16.15 &0.60     & 15.14$\pm$0.11 & 0.022 &  7\\
 45 &  TT Lyn  &09 03 07.8&+44 35 08&176.1&+41.7&$ab$&$-$0.224 &$-$1.35&
    09.85& 0.70     & 08.61$\pm$0.02 & 0.018  & 10,11  \\
 46 &AF 400    &09 18 17.0&+31 58 49&193.5&+43.9&$c$&$-$0.403 &...    &
    14.10& 0.40     & 13.37$\pm$0.03 & 0.018  & 7 \\
 47 &AF 430    &09 30 23.3&+33 53 11&191.2&+46.6&$c$&$-$0.515 &...    &
    14.90& 0.40     & 14.21$\pm$0.04 & 0.016 &  7  \\
48&BS 16927-123&09 44 36.4&+41 08 39&180.4&+49.4&$c$&$-$0.445 &...    &
    13.18  &0.46    & 12.36$\pm$0.02 & 0.017 &  6   \\
 49 & X LMi    &10 06 06.7&+39 21 28&182.5&+53.7&$ab$&$-$0.165 &$-$1.41&
    12.35  &1.02    & 11.06$\pm$0.01 & 0.018  & 12,13   \\
 50 &AG UMa    &10 48 56.3&+42 40 14&172.9&+60.7&$ab$&$-$0.335 &...    &
  (15.42)  &1.71    & 14.53$\pm$0.10 & 0.012 &  14  \\
\hline
\end{tabular}
\end{table*}
\end{center}

\begin{center}
\topmargin 50mm 
\addtocounter{table}{-1}
\thispagestyle{empty}
\begin{table*} 
\caption{Continued.
}
\label{t:bhb1}
\begin{tabular}{@{}clcccccccccccc@{}} 
\hline 
 No &ID& RA&DEC&l & b&Type&$\log$P& [Fe/H] & $\langle V \rangle$ & $V_{amp}$ &
 $\langle K \rangle$ & E$(B-V)$ & Notes \\
(1) &(2)&(3) &(4) &(5) &(6) &(7) &(8) &(9) &(10) &(11) &(12)&(13)&(14)   \\
\hline 
 51 &BK UMa    &10 50 18.9&+42 34 08&172.9&+61.0&$ab$&$-$0.197 &$-$1.29&
    12.91& 0.54     & 11.50$\pm$0.02 & 0.012& 12,15     \\
 52 &AK UMa    &10 53 13.2&+41 19 02&174.9&+61.9&$c$&$-$0.309 &...    &
  (16.08)& (0.46)   & 15.00$\pm$0.12 & 0.012  & 16  \\
 53 &AO UMa    &11 07 39.8&+40 33 58&174.1&+64.7&$ab$&$-$0.251 &...    &
  (15.54)& (1.22)   & 14.62$\pm$0.10 & 0.015  &     \\
 54 &BN UMa    &11 16 22.9&+41 14 02&170.9&+65.9&$d$&$-$0.398 &...    &
    13.50 &  0.50   & 12.58$\pm$0.03 & 0.014  & 6,17    \\
 55 &CK UMa    &12 01 36.4&+31 54 12&186.2&+78.2&$ab$&$-$0.214 &...    &
    14.08 & 0.53     & 12.72$\pm$0.03 & 0.024 & 6   \\
\hline
\end{tabular}
 Notes to table: \\
\bf{(1)} [Fe/H] from Saha \& Oke (1984).~~
\bf{(2)} [Fe/H] from private communication from Suntzeff (1990).~~
\bf{(3)} [Fe/H] from Jurcsik et al. (2006).~~
\bf{(4)} [Fe/H] and $\langle V \rangle$ from Pier, Saha \& Kinman (2003).~~
\bf{(5)} [Fe/H] and $\langle V \rangle$ from Kinman, Saha \& Pier (2004).~~
\bf{(6)} $\langle V \rangle$ from Kinman \& Brown (2010).~~
\bf{(7)} $\langle V \rangle$ from this paper (appendix).~~
\bf{(8)} $\langle V \rangle$ from Schmidt, Chab \& Reiswig (1995).~~
\bf{(9)} $\langle V \rangle$ from Schmidt \& Seth (1996).~~
\bf{(10)} $\langle V \rangle$ from Liu \& Janes (1990).~~
\bf{(11} [Fe/H] from Sodor, Jurcsik \& Szeidl (2009).~~
\bf{(12)} $\langle V \rangle$ from Schmidt (2002).~~
\bf{(13} [Fe/H] from Jurcsik \& Kovacs (1996).~~
\bf{(14)} $\langle V \rangle$ from Kinemuchi et al. (2006).~~
\bf{(15)} [Fe/H] from Kemper (1982).~~
\bf{(16)} Bailey type and period uncertain.~~
\bf{(17)} McClusky (2008) showed that this star is an RRd. The period given
          is that of the first overtone.   \\
\end{table*}
\end{center}

\begin{center}
\begin{landscape} 
\topmargin 45mm 
\thispagestyle{empty}
\begin{table*} 
\caption{Parallaxes, Proper Motions, Radial Velocities, Galactic Distances and 
 Galactic Space Velocities for the BHB stars. \label{t:bhb2}
}
\begin{tabular}{@{}clcccccccccccccc@{}} 
\hline 
 No &ID&$\Pi$ &$\mu_{\alpha}$ & $\mu_{\delta}$&S$_{\mu}$&RV&S$_{RV}$&D &Z &R$_{gal}$ &U &V  & W &L$_{\perp}$ & L$_{z}$ \\ 
    &    & (mas) & (mas y$^{-1}$)  & (mas y$^{-1}$)&  &(km s$^{-1}$)&  &(kpc)&(kpc)&(kpc)&(km s$^{-1}$)&(km s$^{-1}$)&(km s$^{-1}$)&(kpc km s$^{-1})$ &(kpc km s$^{-1}$)  \\
(1) &(2)&(3) &(4) &(5) &(6) &(7) &(8) &(9) &(10) &(11) &(12)&(13)&(14)&(15)&(16)   \\
\hline 
   2&P 54-32.5     & 0.119$\pm$ 0.002&--1.3$\pm$ 0.6&--0.2$\pm$ 1.6&1&--017.4$\pm$ 4& 1& 8.4& 4.3 &15.8&    +002$\pm$015 &  +007$\pm$061      &--047$\pm$024 &  + 1299$\pm$264 &+3445$\pm$929 \\
   3&AF 186        & 0.104$\pm$0.002&  0.0$\pm$3.0&--10.0$\pm$3.0&3& ...  &..& 9.6& 4.9 &17.0&        ...    &    ...          &    ...    &   ...  &   ... \\
   4&AF 189        & 0.135$\pm$0.004& +4.0$\pm$3.0&--5.0$\pm$3.0&3& ... &..& 7.4& 3.8 &14.8&   ...         &     ...          &   ... & ... & ... \\
   5&P 54-111      & 0.165$\pm$0.003& +4.1$\pm$1.4&--19.6$\pm$2.4&1&--054.5$\pm$ 4& 1& 6.1& 3.2 &13.5&  --081$\pm$022&--570$\pm$067      &--016$\pm$036        &+1258$\pm$255&--4615$\pm$884 \\
   6&P 54-122      & 0.144$\pm$0.004& +3.9$\pm$0.5&--7.1$\pm$0.8&1&--128.7$\pm$ 4& 1& 6.9& 3.7 &14.3& --168$\pm$009&--244$\pm$027     &+011$\pm$014            &+0808$\pm$225&--0348$\pm$374\\ 
   8&P 54-119      & 0.185$\pm$0.003& +1.7$\pm$1.2&--20.5$\pm$0.9&1&--194.2$\pm$ 4& 1& 5.4& 2.9 &12.9& --166$\pm$017&--513$\pm$025     &--139$\pm$026          &+1554$\pm$297 &--3697$\pm$320\\  
   9&BS 17444-0025 & 1.348$\pm$0.031& +5.9$\pm$0.6&--27.0$\pm$ 0.6&2&  ...&..& 0.7& 0.4 & 8.6&   ...        &     ...          &   ...       & ...  &  ...  \\
  10&AF 209        & 0.075$\pm$0.002& +2.5$\pm$3.0&--4.0$\pm$3.0&3& ... & ..&13.3& 7.5 &20.5&   ...        &     ...          &    ...   & ... & ...   \\
  11&AF 210        & 0.111$\pm$0.002& +9.0$\pm$3.8&  0.8$\pm$1.3&1&--084.9$\pm$ 4& 1& 9.0& 5.0 &16.3& --295$\pm$091& --015$\pm$060      &+277$\pm$135         & +5938$\pm$2401&+3204$\pm$925 \\
  12&AF 214        & 0.100$\pm$0.002&--1.0$\pm$2.0&--1.0$\pm$ 2.0&3& ... &..&10.0& 5.5 &17.2&    ...       &     ...          &    ...  &  ...  & ...   \\
  13&RR7 002       & 0.128$\pm$0.002&--5.6$\pm$2.2&--8.9$\pm$1.7&1&+249.0$\pm$  4& 1& 7.8& 4.3 &15.1&  +322$\pm$047&--308$\pm$066    &--081$\pm$069         &   +2655$\pm$1139&--1091$\pm$957\\ 
  14&P 81-42       & 0.191$\pm$0.003&--3.0$\pm$3.0&--10.0$\pm$3.0&3& ... &..& 5.2& 3.0 &12.7&  ...         & ...              & ...    & ... &   \\
  15&RR7 008       & 0.129$\pm$0.002&--1.2$\pm$0.7&--13.6$\pm$ 1.0&1&--005.8$\pm$ 4& 1& 7.8& 4.4 &15.0&  +054$\pm$016&--487$\pm$038      &--085$\pm$023        &+1905$\pm$329&--3838$\pm$548\\ 
  16&RR7 015       & 0.592$\pm$0.009&--20.2$\pm$1.0&--35.0$\pm$0.7&2&+238.4$\pm$  4& 1& 1.7& 1.0 & 9.4&  +287$\pm$006&--263$\pm$006      &--028$\pm$007        &+0544$\pm$075 &--0380$\pm$066\\
  17&P 81-39       & 0.106$\pm$0.002&--2.0$\pm$3.0&--2.0$\pm$3.0&3& ...&  ..& 9.4& 5.5 &16.6&     ...      &     ...          &     ...  & ...  & ... \\
  18&RR7 021       & 0.120$\pm$0.002&--1.9$\pm$0.9&--8.3$\pm$1.2&1&+092.9$\pm$  4& 1& 8.3& 4.8 &15.6&  +129$\pm$019 &--310$\pm$046      &--037$\pm$027       & +1287$\pm$454&--1318$\pm$686\\ 
  19&P 81-72       & 0.575$\pm$0.016&--2.5$\pm$0.9&--2.8$\pm$0.6&2& ...&  ..& 1.7& 1.0 & 9.5&   ...        &      ...         &   ...   & ... & ...     \\
  21&RR7 023       & 0.436$\pm$0.007& +7.7$\pm$1.6&--22.6$\pm$0.7&2&--059.3$\pm$ 4&1& 2.3& 1.3 &10.0& --091$\pm$010 &--248$\pm$009      &  +014$\pm$014      & +0267$\pm$140&--0281$\pm$087\\  
  22&RR7 036       & 0.121$\pm$0.002&--1.5$\pm$3.0&--7.5$\pm$3.0&3&+160.0$\pm$ 40& 3& 8.3& 4.8 &15.5&  +171$\pm$074 &--284$\pm$120    &  +017$\pm$099        &  +1650$\pm$1048&--894$\pm$1765\\
  23&P 81-101      & 0.107$\pm$0.002& +1.0$\pm$3.0&--2.0$\pm$3.0&3& ...&  ..& 9.3& 5.6 &16.5&    ...       &     ...          &    ...   &  ...  &  ...    \\
  24&P 81-121      & 0.100$\pm$0.006&  0.0$\pm$3.0&--8.0$\pm$3.0&3& ...& ..&10.0& 6.0 &17.1&      ...     &      ...         &     ...  & ... &  ...      \\ 
  25&RR7 043       & 0.067$\pm$0.002& +1.0$\pm$3.0&--3.0$\pm$3.0&3& ...&  ..&14.9& 8.9 &21.9&     ...      &    ...           &     ...  & ... & ...     \\
  26&P 28-045      & 0.186$\pm$0.003&--5.0$\pm$1.6&--12.2$\pm$0.8&1&--356.6$\pm$ 4& 1& 5.4& 3.2 &12.7& --225$\pm$025 &--257$\pm$021    &  --345$\pm$034       & +3528$\pm$498&--0568$\pm$265\\
  27&RR7 053       & 0.132$\pm$0.002&--2.9$\pm$2.0&--16.0$\pm$2.5&1&--241.0$\pm$ 5& 2& 7.6& 4.6 &14.8& --111$\pm$43 &--555$\pm$090     & --266$\pm$057        & +3626$\pm$916&--4700$\pm$1258\\
  28&P 81-162      & 0.086$\pm$0.001& +1.0$\pm$3.0&--1.0$\pm$3.0&3& ... & ..&11.6& 7.1 &18.6&      ...     &     ...          &      ...   & ... &  ...   \\
  29&RR7 058       & 0.120$\pm$0.002& +1.0$\pm$0.9&--7.1$\pm$1.5&1&+030.0$\pm$  5& 2& 8.3& 5.0 &15.5&--006$\pm$023 &--279$\pm$060   &  +027$\pm$029         &   +0709$\pm$374&--0866$\pm$871\\
  30&RR7 060       & 0.169$\pm$0.003& +0.3$\pm$1.5&--6.1$\pm$1.1&1&+063.2$\pm$  4& 1& 5.9& 3.6 &13.2& +033$\pm$025 &--168$\pm$031     &  +033$\pm$032      &  +0534$\pm$314&+0668$\pm$391\\ 
  31&P 82-04       & 0.097$\pm$0.001& +2.0$\pm$3.0&--3.0$\pm$3.0&3& ... &..&10.3& 6.3 &17.3&     ...      &      ...         &    ...  &... & ...      \\
  32&RR7 064       & 0.784$\pm$0.016&--5.1$\pm$0.7&--5.4$\pm$0.7&2&+034.8$\pm$  4& 1& 1.3& 0.8 & 9.0&  +038$\pm$004 & --024$\pm$004      &+002$\pm$004        &+0159$\pm$007&+1758$\pm$038\\   
  33&RR7 066       & 0.118$\pm$0.002&--1.9$\pm$2.1&--8.5$\pm$1.1&1&--055.0$\pm$ 5& 2& 8.5& 5.2 &15.6&  +003$\pm$054&--325$\pm$044     & --114$\pm$069         & +1960$\pm$1028&--1533$\pm$645\\ 
  34&P 81-167      & 0.178$\pm$0.003&--10.0$\pm$3.0&--2.0$\pm$3.0&3& ...& ..& 5.6& 3.5 &12.9&    ...       &      ...         &    ...  & ...  &  ...     \\  
  35&P 11419-01    & 0.386$\pm$0.007& +6.1$\pm$2.2&--18.3$\pm$0.9&1&+287.1$\pm$  4& 1& 2.6& 1.6 &10.1&  +141$\pm$017&--277$\pm$012     &  +196$\pm$021     &    +1734$\pm$238& --0494$\pm$124\\
  37&AF 293        & 0.074$\pm$0.001&--5.0$\pm$3.0&--6.0$\pm$3.0&3& ...  &..&13.5& 8.4 &20.4&     ...      &      ...         &     ...  & ... & ...     \\
  38&P 11419-04    & 0.157$\pm$0.003&--0.8$\pm$1.2&--7.1$\pm$0.3&1&+170.7$\pm$  4& 1& 6.4& 3.9 &13.5&  +116$\pm$023&--233$\pm$010     & +059$\pm$028       &    +0444$\pm$319&--0040$\pm$139\\  
  39&RR7 084       & 0.094$\pm$0.001&  0.0$\pm$3.0&--7.0$\pm$3.0&3&--065.0$\pm$40& 3&10.6& 6.7 &17.6&--057$\pm$099&--344$\pm$151     & --050$\pm$120     &       +2533$\pm$1425&--2029$\pm$2446 \\
  40&RR7 091       & 0.211$\pm$0.003&--0.7$\pm$1.4&--15.3$\pm$.9&1&--041.0$\pm$ 4& 1& 4.7& 3.0 &12.1&--027$\pm$020&--335$\pm$020     & --054$\pm$024      &      +0696$\pm$256&--1349$\pm$235\\  
\hline
\end{tabular}
\end{table*}
\end{landscape}
\end{center}

%\pagebreak

\begin{center}
\begin{landscape} 
\topmargin 50mm 
\addtocounter{table}{-1}
\thispagestyle{empty}
\begin{table*} 
\caption{Continued.\label{t:bhb2}
}
\begin{tabular}{@{}clcccccccccccccc@{}} 
\hline 
 No &ID&$\Pi$ &$\mu_{\alpha}$ & $\mu_{\delta}$&S$_{\mu}$&RV&S$_{RV}$&D &Z &R$_{gal}$ &U &V  & W &L$_{\perp}$ & L$_{z}$ \\ 
    &    & (mas) & (mas y$^{-1}$)  & (mas y$^{-1}$)&  &(km s$^{-1}$)&  &(kpc)&(kpc)&(kpc)&(km s$^{-1}$)&(km s$^{-1}$)&(km s$^{-1}$)&(kpc km s$^{-1}$)&(kpc km s$^{-1}$)  \\
(1) &(2)&(3) &(4) &(5) &(6) &(7) &(8) &(9) &(10) &(11) &(12)&(13)&(14)&(15)&(16)   \\
\hline 
  41&RR7 090       & 0.101$\pm$  0.002& +3.0$\pm$ 3.0&--4.0$\pm$ 3.0&3&--112.0$\pm$40& 3& 9.9& 6.2 &16.9&--180$\pm$095&--185$\pm$143  & +032$\pm$111         &  +2525$\pm$1525& +0502$\pm$2242 \\
  42&P 82-49       & 0.111$\pm$0.002&--2.0$\pm$3.0&--4.0$\pm$  3.0&3& ...&  ..& 9.0 & 5.8 &16.0&     ...    &    ...      &    ...   & ...  &  ...    \\
  44&BS 16473-0102 & 0.220$\pm$0.004& +8.0$\pm$3.0&--12.0$\pm$ 3.0&3& ...&  ..& 4.5& 3.0 &11.8&       ...    &      ...         &      ...   & ... & ...     \\
  45&BS 17139-69   & 0.170$\pm$0.003&--3.3$\pm$0.7&--9.2$\pm$  1.0&1&+105.4$\pm$ 4& 1& 5.9& 3.9 &12.9&  +096$\pm$014&--265$\pm$027     & --027$\pm$015       & +0740$\pm$231& --0446$\pm$331\\ 
  46&TON 384       & 0.124$\pm$0.002&--0.7$\pm$1.3&--3.6$\pm$  1.1&1&--174.7$\pm$ 4& 1& 8.1& 5.3 &14.9& --140$\pm$035&--094$\pm$044     &  --146$\pm$039        &+1415$\pm$582&+1510$\pm$610\\   
  47&BS 16468-0026 & 0.154$\pm$0.003&--1.8$\pm$1.2&--13.4$\pm$ 1.0&1&+211.2$\pm$  4& 1& 6.5& 4.4 &13.5&  +181$\pm$024&--412$\pm$032      & +098$\pm$027        &+1031$\pm$231&--2418$\pm$418 \\
  48&AF 379        & 0.123$\pm$0.003& +3.5$\pm$3.0&--9.5$\pm$  3.0&3& ...& ..& 8.1& 5.5 &15.0&      ...     &     ...          &      ...   &  ...   & ...   \\
  49&AF 386        & 0.136$\pm$0.002& +2.2$\pm$0.9&--8.4$\pm$  1.2&1&+009.6$\pm$  4& 1& 7.4& 5.1 &14.2&--075$\pm$022&--286$\pm$042     &  +053$\pm$023       & +1164$\pm$379&--0925$\pm$557   \\
  50&AF 390        & 0.118$\pm$0.002&--2.0$\pm$3.0&--5.0$\pm$  3.0&3& ...&  ..& 8.5& 5.9 &15.3&      ...    &       ...         &      ...  &  ...   &   ...  \\
  51&BS 16468-0078 & 0.654$\pm$0.010&--19.8$\pm$0.6&--29.1$\pm$ 0.7&2& ...& ..& 1.5& 1.1 & 9.2&       ...    &     ...          &      ...  & ... & ...   \\
  52&P 30-16       & 0.173$\pm$0.003& +1.0$\pm$3.0&--4.7$\pm$  3.0&3& ...&  ..& 5.8& 4.0 &12.8&       ...    &      ...         &      ...   & ...  & ...  \\
  53&BS 16468-0080 & 0.224$\pm$0.007& +1.4$\pm$1.2&--17.9$\pm$ 1.6&1&--001.7$\pm$ 4& 1& 4.5& 3.1 &11.6&--037$\pm$017&--372$\pm$036    &   +030$\pm$018       &  +0691$\pm$184&--1700$\pm$405 \\ 
  54&P 30-28       & 1.161$\pm$0.017& +9.5$\pm$0.7&--20.6$\pm$ 0.7&2& ...&  ..& 0.9& 0.6 & 8.6&      ...     &     ...          &      ...    & ... & ... \\
  55&BS 16468-0090 & 0.200$\pm$0.004&--0.4$\pm$1.6&--8.0$\pm$  1.6&1&+221.6$\pm$  4&..& 5.0& 3.5 &12.1&  +145$\pm$027&--193$\pm$039     &   +155$\pm$027     &  +1287$\pm$404& +0345$\pm$446 \\
  56&CHSS 608      & 0.148$\pm$0.002&--3.4$\pm$1.2&--12.5$\pm$ 1.6&1&+029.0$\pm$  4&..& 6.8& 4.7 &13.6& +023$\pm$029&--393$\pm$051    & --104$\pm$029        &  +1742$\pm$452&--2152$\pm$657\\ 
  57&P 11424-28    & 0.169$\pm$0.003& +1.3$\pm$1.4&--7.9$\pm$  1.6&1&+011.6$\pm$  4& 1& 5.9& 4.1 &12.8&--061$\pm$030&--217$\pm$043   & +017$\pm$030          & +0589$\pm$363&--0030$\pm$527   \\
  58&P 30-38       & 0.175$\pm$0.003& +2.0$\pm$3.0&--11.0$\pm$ 3.0&3&--179.3$\pm$ 4& 1& 5.7& 4.0 &12.7& --204$\pm$059&--269$\pm$079      &--096$\pm$059       &  +0911$\pm$568&--0716$\pm$946\\ 
  59& 57-121       & 0.125$\pm$0.005&--1.8$\pm$1.2&--8.6$\pm$  1.4&1&+066.1$\pm$  5& 2& 8.0& 5.7 &14.7&  +066$\pm$032&--329$\pm$056      & +005$\pm$031      &  +0912$\pm$391&--1451$\pm$760\\  
  60& AF 419       & 0.121$\pm$0.003& +3.3$\pm$1.0&--6.2$\pm$  2.2&1&+072.7$\pm$  4& 1& 8.3& 5.8 &14.9& --098$\pm$033&--242$\pm$086   & +116$\pm$030         &  +2213$\pm$539& --0479$\pm$1186 \\
  61&P 11424-70    & 0.168$\pm$0.003& +3.3$\pm$1.2&--13.8$\pm$ 2.9&1&+152.7$\pm$  4& 1& 6.0& 4.3 &12.8&--035$\pm$028&--397$\pm$080 & +148$\pm$024          &     +2057$\pm$372&--2154$\pm$970\\
  62&BS 16927-22   & 0.773$\pm$0.012& +6.8$\pm$0.8&--29.8$\pm$ 0.6&2&+082.2$\pm$  4& 1& 1.3& 1.0 & 8.9&  +003$\pm$004&--177$\pm$005  &  +098$\pm$004      &    +0872$\pm$039&+0378$\pm$041     \\
  63&CHSS 663      & 0.130$\pm$0.003&--9.0$\pm$2.0&--11.0$\pm$ 2.0&3& ...&  ..& 7.7& 5.6 &14.3&     ...      &     ...          &    ...  & ... &  ...      \\
  64&P 11424-82    & 0.132$\pm$0.002&--6.6$\pm$1.4&--14.7$\pm$ 1.3&1&+030.7$\pm$  4& 1& 7.6& 5.5 &14.2&  +087$\pm$038&--533$\pm$047  &--187$\pm$035        &    +3559$\pm$605&--3896$\pm$620  \\
  65&BS 16940-45   & 0.276$\pm$0.004&--3.7$\pm$1.8&--16.7$\pm$ 1.5&1&--101.0$\pm$  4& 1& 3.6& 2.7 &10.7&--0063$\pm$023&--275$\pm$025   &--116$\pm$021       &    +1050$\pm$278 &--0594$\pm$258\\ 
  66&BS 16927-55   & 0.181$\pm$0.006&--2.5$\pm$0.9&--12.3$\pm$ 1.8&1&+043.1$\pm$  4& 1& 5.5& 4.1 &12.4&  +050$\pm$018&--323$\pm$047  &  +015$\pm$016        &   +0513$\pm$193&--1204$\pm$548  \\
  67&BS 16940-0070 & 0.144$\pm$0.002& +2.6$\pm$1.2&--3.5$\pm$  1.3&1&--082.3$\pm$  4& 1& 6.9& 5.3 &13.6& --146$\pm$030& --091$\pm$043 & --001$\pm$025       &   +1080$\pm$374&+1473$\pm$538    \\
  68&BS 16940-0072 & 0.210$\pm$0.003& +5.2$\pm$1.2&--11.5$\pm$ 1.0&1&--135.1$\pm$  4& 1& 4.8& 3.6 &11.6& --225$\pm$020&--223$\pm$024 & --029$\pm$018        &   +0525$\pm$246 &--0175$\pm$265     \\
\hline
\end{tabular}
 Notes to table: \\
\bf{(1)} Sources of proper motions (S$_{\mu}$) (1) GSCII-SDSS (2) Nomad 
  (3) SDSS (DR 7) \\
\bf{(2)} Sources of Radial Velocities (S$_{RV}$): (1) Bologna (2) Kitt Peak
  4-m (3) Kinman et al. (1994) \\
\bf{(3)} Distances: (D) Heliocentric distance; (Z) Height above Plane; R$_{gal}$ 
 Galactocentric distance assuming Solar Galactocenttric Distance = 8.0 kpc \\
\end{table*}
\end{landscape}
\end{center}

\begin{center}
\begin{landscape} 
\topmargin 45mm 
\thispagestyle{empty}
\begin{table*} 
\caption{Parallaxes, Proper Motions, Radial Velocities, Galactic Distances and 
 Galactic Space Velocities for the RR Lyrae stars. \label{t:rrl2}
}
\begin{tabular}{@{}clcccccccccccccc@{}} 
\hline 
 No &ID&$\Pi$ &$\mu_{\alpha}$ & $\mu_{\delta}$&S$_{\mu}$&RV&S$_{RV}$&D &Z &R$_{gal}$ &U &V  & W &L$_{\perp}$ & L$_{z}$ \\ 
    &    & (mas) & (mas y$^{-1}$)  & (mas y$^{-1}$)&  &(km s$^{-1}$)&  &(kpc)&(kpc)&(kpc)&(km s$^{-1}$)&(km s$^{-1}$)&(km s$^{-1}$)&(kpc km s$^{-1}$)&(kpc km s$^{-1}$ )  \\
(1) &(2)&(3) &(4) &(5) &(6) &(7) &(8) &(9) &(10) &(11) &(12)&(13)&(14)&(15)&16 \\
\hline 
   1 & V385 Aur  & 0.044$\pm$0.002&--3.4$\pm$3.3& +0.6$\pm$3.2&  1 & ...            & ..&22.7 & 8.8& 30.3 &       ...    &   ...        & ... & ...& ...   \\  
   2 & V386 Aur  & 0.060$\pm$0.003& +2.1$\pm$2.1&--5.0$\pm$1.7&  1 & +116.0$\pm$30.0& 3 &16.7 & 6.7& 24.2 & +091$\pm$071 &--412$\pm$135 & +106$\pm$152       &  +3982$\pm$2474&--4534$\pm$3191\\
   3 & V387 Aur  & 0.057$\pm$0.003& +1.4$\pm$1.2&--0.6$\pm$0.9&  1 &--003.0$\pm$30.0& 3 &17.5 & 6.7& 25.1 &--050$\pm$046 &--074$\pm$080  & +092$\pm$087    &    +3223$\pm$1794&+3486$\pm$1954 \\
   4 & V389 Aur  & 0.042$\pm$0.002&--1.2$\pm$3.0&  3.3$\pm$0.8&  1 &  ...           & ..&23.8 & 9.5& 31.3 &   ...    &  ...     & ... & ... &  ... \\
   5 & VX Lyn    & 0.053$\pm$0.003&--1.8$\pm$1.2&--4.1$\pm$1.9&  1 & +001.3$\pm$15.0& 1 &18.9 & 7.7& 26.4 & +099$\pm$049  &--296$\pm$168 &--228$\pm$103     &    +6730$\pm$2927&--1919$\pm$4259\\ 
   6 & VY Lyn    & 0.098$\pm$0.003& +2.2$\pm$1.0&--1.6$\pm$2.1&  1 & +114.5$\pm$15.0& 1 &10.2 & 4.2& 17.8 & +061$\pm$026   &--095$\pm$101   &129$\pm$050     &   +2097$\pm$931&+2160$\pm$1743  \\ 
   7 & VZ Lyn    & 0.078$\pm$0.004&--1.9$\pm$1.4&--2.2$\pm$2.2&  1 &--182.1$\pm$15.0& 1 &12.8 & 5.4 &20.3 &--109$\pm$044  &--095$\pm$129  &--200$\pm$086     &   +3604$\pm$1684&+2498$\pm$2548  \\  
   8 & WX Lyn    & 0.056$\pm$0.003&--2.1$\pm$1.1&--3.3$\pm$1.0&  1 & +026.3$\pm$15.0& 1 &17.9 & 7.5 &25.3 & +121$\pm$042  &--214$\pm$087 &--211$\pm$086     &   +6093$\pm$2401&+148$\pm$2131     \\  
   9 & AS Lyn    & 0.030$\pm$0.002& +1.1$\pm$0.7&--3.1$\pm$2.0&  1 &--179.0$\pm$46.0& 3 &33.3 &14.8 &40.7 &--175$\pm$078  &--502$\pm$309 &--026$\pm$119      &   +7604$\pm$3882 &--10531$\pm$11848\\ 
  10 & WZ Lyn    & 0.180$\pm$0.004& +1.3$\pm$0.6&--13.0$\pm$1.2& 1 & +197.0$\pm$20.0& 4 & 5.6 & 2.4 &13.2 & +193$\pm$020  &--337$\pm$031  &+043$\pm$018     &   +371$\pm$107&--1511$\pm$411     \\
  11 & XZ Lyn    & 0.072$\pm$0.004& +2.1$\pm$1.6&--1.5$\pm$1.0&  1 &--032.0$\pm$20.0& 4 &13.9 & 6.3 &21.3 &--086$\pm$051    &--125$\pm$073 &+087$\pm$096     &   +2885$\pm$1689&+1926$\pm$1482     \\
  12 & TW Lyn    & 0.592$\pm$0.011& +0.6$\pm$3.8&  2.6$\pm$2.9&  1 &--039.0$\pm$05.0& 5 & 1.7 & 0.8 & 9.5 &--050$\pm$015   &+021$\pm$022 & --002$\pm$027     &    +307$\pm$121&+2294$\pm$213       \\
  13 & YY Lyn    & 0.133$\pm$0.002& +5.7$\pm$0.7&--5.9$\pm$1.4&  1 &--093.1$\pm$15.0& 1 & 7.5& 3.3  &15.1 &--165$\pm$018 & --249$\pm$049& +092$\pm$026      &   +1932$\pm$412&--478$\pm$723      \\
  14 & YZ Lyn    & 0.047$\pm$0.002&--0.5$\pm$0.7&--3.7$\pm$1.3&  1 & +033.0$\pm$40.0& 3 &21.3 & 9.7 &28.6 & +087$\pm$051 &--349$\pm$133 &--098$\pm$070      &    +4038$\pm$2027&--3482$\pm$3594  \\ 
  15 & AU Lyn    & 0.035$\pm$0.002&--0.8$\pm$1.7&--0.8$\pm$2.1&  1 & +096.0$\pm$45.0& 3 &28.6 &13.4 &35.8 & +139$\pm$120 & --056$\pm$283 &--060$\pm$209     &    +8978$\pm$5151&+5307$\pm$9477    \\ 
  16 & ZZ Lyn    & 0.089$\pm$0.001&--1.4$\pm$1.4&--1.4$\pm$0.9&  1 & +147.2$\pm$15.0& 1 &11.2 & 5.1 &18.7 & +160$\pm$034 &--054$\pm$048  & --006$\pm$061    &    +1636$\pm$775 &+3045$\pm$872 \\  
  17 & RW Lyn    & 0.358$\pm$0.016& +7.3$\pm$1.7&--15.7$\pm$2.5& 1 &--149.8$\pm$15.0& 1 & 2.8 & 1.3 &10.6 &--169$\pm$016  &--216$\pm$034 & --023 $\pm$022    &    +197$\pm$139&+32$\pm$354  \\  
  18 & AV Lyn    & 0.062$\pm$0.003& +0.5$\pm$0.8&--4.9$\pm$1.1&  1 &--126.0$\pm$30.0& 3 &16.1 & 7.8 &23.4 &--081$\pm$041   &--373$\pm$084 & --87$\pm$056     &   +2114$\pm$1030 &--3311$\pm$1887 \\
  19 & AC Lyn    & 0.071$\pm$0.004& +1.1$\pm$0.9&--5.9$\pm$1.2&  1 &--025.8$\pm$15.0& 1 &14.1 & 6.7 &21.5 &--030$\pm$031   &--395$\pm$085 & --024$\pm$054    &   +1733$\pm$766 &--3581$\pm$1761 \\
  20 & AD Lyn    & 0.088$\pm$0.006& +1.2$\pm$1.1&--8.0$\pm$2.6&  1 & +123.6$\pm$15.0& 1 &11.4 & 5.5 &18.8 &+108$\pm$033 &--435$\pm$143 & +037$\pm$058      &   +1724$\pm$805&--3855$\pm$2606\\
  21 & AW Lyn    & 0.080$\pm$0.003&--1.2$\pm$1.1&--4.2$\pm$1.9&  1 & +93.0$\pm$30.0 & 3 &12.5 & 6.2 &19.8 &+142$\pm$044  &--217$\pm$111& --047$\pm$060     &   +2048$\pm$1087&--41$\pm$2092  \\ 
  22 & AX Lyn    & 0.026$\pm$0.001& +3.4$\pm$1.4&--2.0$\pm$2.8&  1 & ...            & ..&38.5 &18.9 &45.6 &   ...       &   ...        &      ... & ... & ... \\
  23 & AY Lyn    & 0.055$\pm$0.003& +1.2$\pm$1.4&--6.0$\pm$1.2&  1 & ...            & ..&18.2 & 9.0 &25.4 &  ...       &   ...        &  ... & ... & ...      \\
  24 &P 54-13    & 0.130$\pm$0.011& +2.6$\pm$1.0&--10.1$\pm$1.1&  1 & +069.0$\pm$10.0&2  & 7.7 & 3.9 &15.2 &+044$\pm$021 &--374$\pm$051 &+061$\pm$032     &    +1046$\pm$404&--2273$\pm$801 \\ 
  25 & AZ Lyn    & 0.063$\pm$0.003& +3.3$\pm$1.0&--4.6$\pm$1.5&  1 & +087.0$\pm$33.0& 3 &15.9 & 8.1 &23.1 &--012$\pm$050  &--374$\pm$114 & 207$\pm$070      &    +4933$\pm$1673&--3328$\pm$2492 \\ 
  26 & BB Lyn    & 0.058$\pm$0.003&--0.6$\pm$0.7&--6.5$\pm$1.4&  1 & +058.0$\pm$33.0& 3 &17.2 & 8.9 &24.4 & +136$\pm$044 &--505$\pm$118 &--083$\pm$054     &     +4183$\pm$1429&--6557$\pm$2752 \\
  27 & BC Lyn    & 0.053$\pm$0.003& +3.3$\pm$1.2&--3.1$\pm$0.7&  1 & +239.0$\pm$45.0& 3 &18.9 &10.0 &26.0 & +075$\pm$068 &--308$\pm$067 &+345$\pm$091      &     +7628$\pm$2718 &--2159$\pm$1627  \\
  28 & AF 194    & 0.088$\pm$0.002& +0.5$\pm$0.8&--4.7$\pm$1.3&  1 & ...            & ..&11.4 & 6.0 &18.6 &   ...        &  ...         &   ... & ... & ...   \\
  29 & AF 197    & 0.105$\pm$0.003&--1.0$\pm$3.9&--8.2$\pm$1.8&  1 & ...            & ..& 9.5 & 5.0 &16.9 &    ...       &  ...         &   ... & ...& ...  \\
  30 & DQ Lyn    & 0.649$\pm$0.043&--1.9$\pm$0.8&--28.7$\pm$1.0& 2 & +053.0$\pm$10.0& 2 & 1.5 & 0.9 & 9.3 & +047$\pm$009  &--202$\pm$016 &--006$\pm$008    &     +106$\pm$61&+167$\pm$144  \\ 
  31 & RR7 032   & 0.163$\pm$0.004&--1.4$\pm$1.1&--11.9$\pm$1.1& 1 & +42.0$\pm$10.0 & 2 & 6.1 & 3.6 &13.5 & +069$\pm$020   &--334$\pm$033  & --039$\pm$026    & +902$\pm$330 &--1470$\pm$432  \\
  32 & RR7 034   & 0.112$\pm$0.002& +0.1$\pm$0.7&--2.9$\pm$0.7&  1 & +316.0$\pm$10.0& 2 & 8.9 & 5.2 &16.1 & +249$\pm$019 &--130$\pm$030 &+178$\pm$025        &   +1519$\pm$427&+1474$\pm$459 \\
  33 &P 81 129   & 0.164$\pm$0.003&--0.5$\pm$2.0&--7.3$\pm$0.9&  1 & +003.0$\pm$10.0& 2 & 6.1 & 3.6 &13.4 & +022$\pm$035 &--204$\pm$026 & --019$\pm$047     &   +648$\pm$477 &+198$\pm$336           \\
  34 &AF Lyn     & 0.079$\pm$0.002& +1.2$\pm$0.9&--5.9$\pm$1.0&  1 &--121.7$\pm$15.0& 1 &12.7 & 7.5 &19.7 &--132$\pm$034  &--354$\pm$061 & --036$\pm$044    &    +1462$\pm$557&--2444$\pm$1102  \\
  35 &P 82 06    & 0.190$\pm$0.003& +4.0$\pm$3.0&--3.0$\pm$3.0&  3 &--211.0$\pm$10.0& 2 & 5.3 & 3.3 &12.6 &--234$\pm$047  &--84$\pm$075  & --045$\pm$059     &    +919$\pm$470&+1684$\pm$903 \\ 
  36 & AI Lyn    & 0.049$\pm$0.003&--1.5$\pm$1.2&--0.2$\pm$2.0&  1 &  ...           &.. &20.4 &12.5 &27.2 &    ...       &   ...        &   ... & ... & ...  \\
  37 & AK Lyn    & 0.084$\pm$0.004& +0.8$\pm$1.2&--8.7$\pm$1.1&  1 & +237.6$\pm$15.0& 1 &11.9 & 7.4 &18.8 & +157$\pm$044 &--495$\pm$067 & 148$\pm$054  &         +2656$\pm$783&--4686$\pm$1203 \\
  38 & RR7-079   & 0.260$\pm$0.007& +1.0$\pm$1.1&--13.5$\pm$1.2& 1 &--032.0$\pm$10.0& 2 & 3.8 & 2.4 &11.3 &--047$\pm$015   &--238$\pm$023  & --021$\pm$017    &   +209$\pm$137 &--212$\pm$253 \\
  39 & RR7-086   & 0.073$\pm$0.005&--0.7$\pm$1.1&--4.0$\pm$0.7&  1 &  ...           & ..&13.7 & 8.5 &20.6 &  ...          &  ...         &  ...               &  ...           &   ...        \\
  40 & AL Lyn    & 0.066$\pm$0.005& +1.0$\pm$0.8&--5.6$\pm$1.2&  1 &--065.8$\pm$15.0& 1 &15.2 & 9.5 &22.0 &--105$\pm$038  &--398$\pm$087 & --011$\pm$046      &  +2273$\pm$932&--3595$\pm$1789 \\
\hline
\end{tabular}
\end{table*}
\end{landscape}
\end{center}

\begin{center}
\begin{landscape} 
\topmargin 50mm 
\addtocounter{table}{-1}
\thispagestyle{empty}
\begin{table*} 
\caption{Continued. \label{t:rrl2}
}

\begin{tabular}{@{}clcccccccccccccc@{}} 
\hline 
 No &ID&$\Pi$ &$\mu_{\alpha}$ & $\mu_{\delta}$&S$_{\mu}$&RV&S$_{RV}$&D &Z &R$_{gal}$ &U &V  & W &L$_{\perp}$ & L$_{z}$ \\ 
    &    & (mas) & (mas y$^{-1}$)  & (mas y$^{-1}$)&  &(km s$^{-1}$)&  &(kpc)&(kpc)&(kpc)&(km s$^{-1}$)&(km s$^{-1}$)&(km s$^{-1}$)&(kpc km s$^{-1}$) &(kpc km s$^{-1}$ ) \\
(1) &(2)&(3) &(4) &(5) &(6) &(7) &(8) &(9) &(10) &(11) &(12)&(13)&(14)&(15)&16 \\
\hline 
  41 & AM Lyn    & 0.045$\pm$0.002& +2.7$\pm$1.0&--3.7$\pm$1.3&  1 &  ...           & ..&22.2 &13.9 &28.9 &    ...    &   ...     &   ...& ... & ...    \\
  42 &P 82-32    & 0.122$\pm$0.004& +5.9$\pm$0.7&--9.5$\pm$1.6&  1 & +060.0$\pm$10.0& 2 & 8.2 & 5.2 &15.2 &--084$\pm$019  &--377$\pm$065 & +213$\pm$022   &     +3615$\pm$414 &--2219$\pm$931\\
  43 & AF 316    & 0.078$\pm$0.002& +1.9$\pm$1.1&--4.8$\pm$1.5&  1 &  ...           & ..&12.8 & 8.1 &19.7 &    ...    &   ...    &   ... & ... & ...     \\
  44 & RR7-101   & 0.077$\pm$0.003& +0.5$\pm$1.0&--4.8$\pm$1.2&  1 &  ...           & ..&13.0 & 8.2 &19.8 &    ...   &   ...    &   ... & ... &  ...  \\
  45 &  TT Lyn   & 1.408$\pm$0.029&--81.9$\pm$1.5&--41.8$\pm$0.9&2 &--065.0$\pm$05.0& 5 & 0.7 & 0.5 & 8.5 &+132$\pm$006  &--131$\pm$004  &--240$\pm$007    &    +2108$\pm$62&+751$\pm$36 \\
  46 &AF 400     & 0.184$\pm$0.011& +0.2$\pm$1.3&--8.5$\pm$1.4&  1 &  ...           & ..& 5.4 & 3.8 &12.4 &     ...  &    ...  &   ... &  ... & ... \\
  47 &AF 430     & 0.132$\pm$0.003&--1.4$\pm$1.1&--8.2$\pm$0.8&  1 &  ...           & ..& 7.6 & 5.5 &14.3 &     ...  &  ...  &   ... & ... & ...   \\
 48&BS 16927-123 & 0.289$\pm$0.007&--15.6$\pm$2.4&--4.6$\pm$1.3& 1 & +070.0$\pm$10.0& 2 & 3.5 & 2.6 &10.6 &+223$\pm$029 &--094$\pm$022  &--100$\pm$026  &        +1654$\pm$327&+1295$\pm$220\\ 
  49 & X LMi     & 0.437$\pm$0.007& +7.8$\pm$1.3&--17.3$\pm$0.7& 2 &--082$\pm$20.0  & 6 & 2.3 & 1.8 & 9.5 &--152$\pm$016  &--165$\pm$008  &+004$\pm$018    &      +345$\pm$144 &+507$\pm$78    \\  
  50 &AG UMa     & 0.104$\pm$0.002&--1.6$\pm$1.0& --8.4$\pm$1.9&1 &  ...           & ..& 9.6 & 8.4 &15.2 &   ...   & ...   &  ... & ... & ...      \\
  51 &BK UMa     & 0.348$\pm$0.013&--13.0$\pm$1.5&--16.9$\pm$2.2&1 & +171.4$\pm$05.0& 7 & 2.9 & 2.5 & 9.7 &+174$\pm$019 &--253$\pm$031 & +120$\pm$012     &     +692$\pm$159&--341$\pm$292\\
  52 &AK UMa     & 0.075$\pm$0.002&--3.7$\pm$6.1&--3.3$\pm$6.3  &1 &  ...           & ..&13.3 &11.8 &18.5 &  ...   & ...   &  ... & ... &  ...       \\
  53 &AO UMa     & 0.096$\pm$0.003&--2.8$\pm$1.3&--6.3$\pm$1.3  &1 &  ...           & ..&10.4 & 9.4 &15.6 &  ...   &  ...   &  ...  & ... & ...   \\ 
  54 &BN UMa     & 0.248$\pm$0.006&+12.5$\pm$1.3&--17.3$\pm$1.7 &1 & +019.0$\pm$10.0& 2 & 4.0 & 3.7 &10.3 &--306$\pm$024  &--223$\pm$033 & +175$\pm$014      &   +2810$\pm$223&+45$\pm$316 \\ 
  55& CK UMa     & 0.201$\pm$0.003&--3.8$\pm$1.3& +0.8$\pm$2.4  &1 & +016.0$\pm$10.0& 2 & 5.0 & 4.9 &10.2 & +080$\pm$037 & --018$\pm$053 &+004$\pm$013     &    +1073$\pm$285&+1817$\pm$479\\
\hline
\end{tabular}
 Notes to table: \\
\bf{(1)} Sources of proper motions (S$_{\mu}$) (1) GSCII-SDSS (2) Nomad 
  (3) SDSS (DR 7) \\
\bf{(2)} Sources of Radial Velocities (S$_{RV}$):  
 (1) Pier, Saha, Kinman (2003); (2) Kinman, Brown (2010);
 (3) Saha, Oke (1984); (4) Pier (unpublished); (5) Fernley \& Barnes (1997); 
 (6) Layden (1994); (7) Jeffery et al. (2007). \\
\bf{(3)} Distances: (D) Heliocentric distance; (Z) Height above Plane; R$_{gal}$ 
 Galactocentric distance assuming Solar Galactocenttric Distance = 8.0 kpc \\
\end{table*}
\end{landscape}
\end{center}

\clearpage

\appendix

\section{The BHB stars.}

\subsection{Our Selection Methods.}
   Kinman et al. (1994) used both photometric and spectroscopic criteria 
   to identify  15 BHB stars in the Anticentre field RR VII 
  (l = 183$^{\circ}$, b = +37$^{\circ}$) which Kinman et al. (1982) had 
   previously searched for RR Lyrae stars. Brown et al. (2003) independently
   confirmed these classifications for nine stars in this field (6 BHB and 
   3 non-BHB stars). The classification of these 15 stars (RR 7-02, -08,
   -15, -21, -23, -36, -43, -53, -58, -60, -64, -66, -84, -90, -91)\footnote{
  The identifications  [KSK94]RR 7 nnn, Case A-F nnn and BPS BS nnnnn-nnnn used
  by $SIMBAD$ are abbreviated in this paper to RR7-nnn, AF-nnn and BS nnnnn-nnn
  respectively.} is  therefore considered to be secure.

% Figure A1
\begin{figure}
\includegraphics[width=8.5cm, bb=95 150 495 610, clip=true]{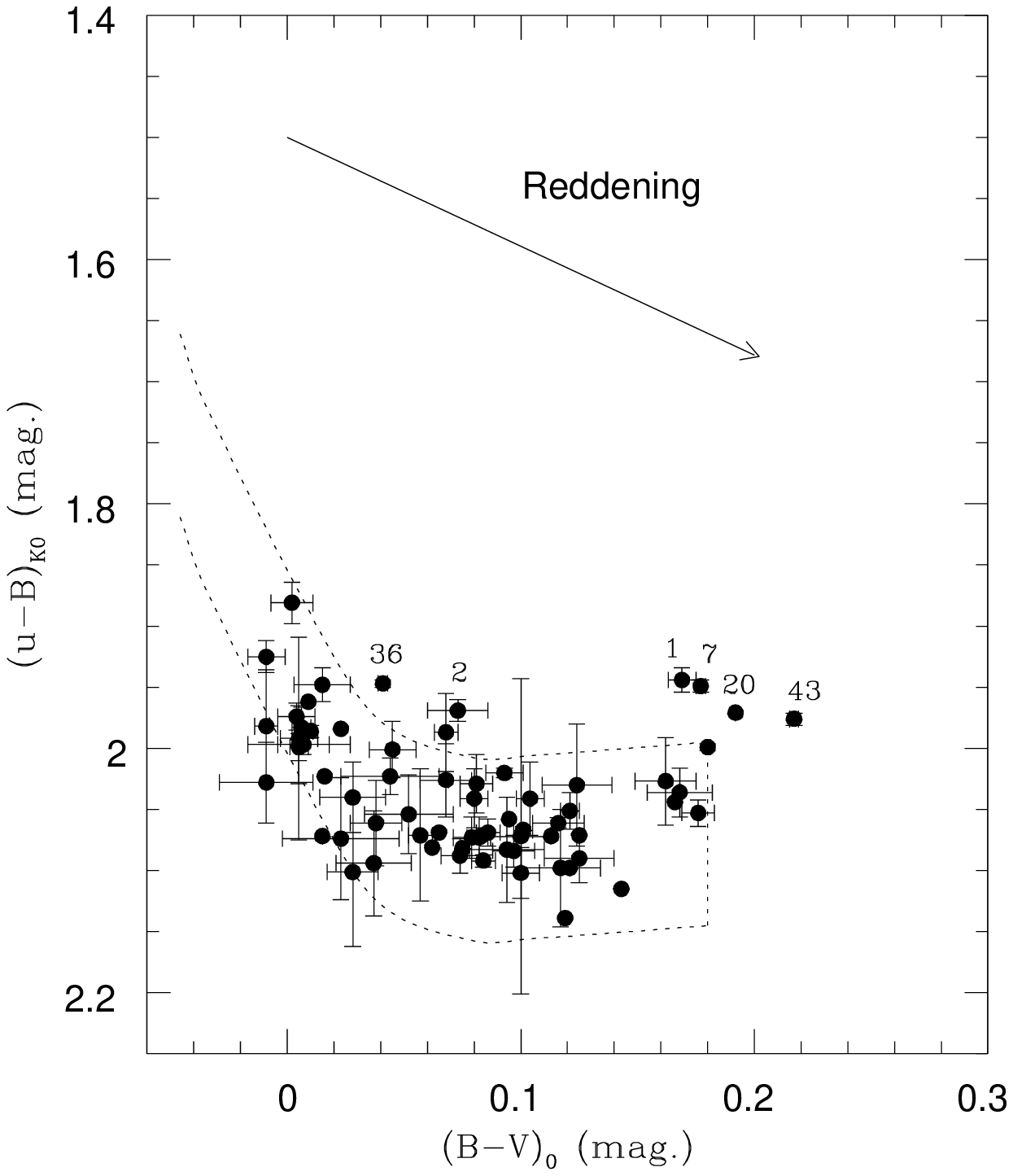}
\caption{The ordinate ($u-B)_{K0}$ is defined in Kinman et al. (1994). The
 abscissa is Johnson ($B-V)_{0}$. The running numbers of stars in Table 1 are
 shown next to stars whose error bars lie outside the zone outlined by the
 dotted lines in which the BHB stars are located.
}
\label{f:a1}
\end{figure}

% Figure A2
\begin{figure}
\includegraphics[width=8.5cm, bb=95 150 495 610, clip=true]{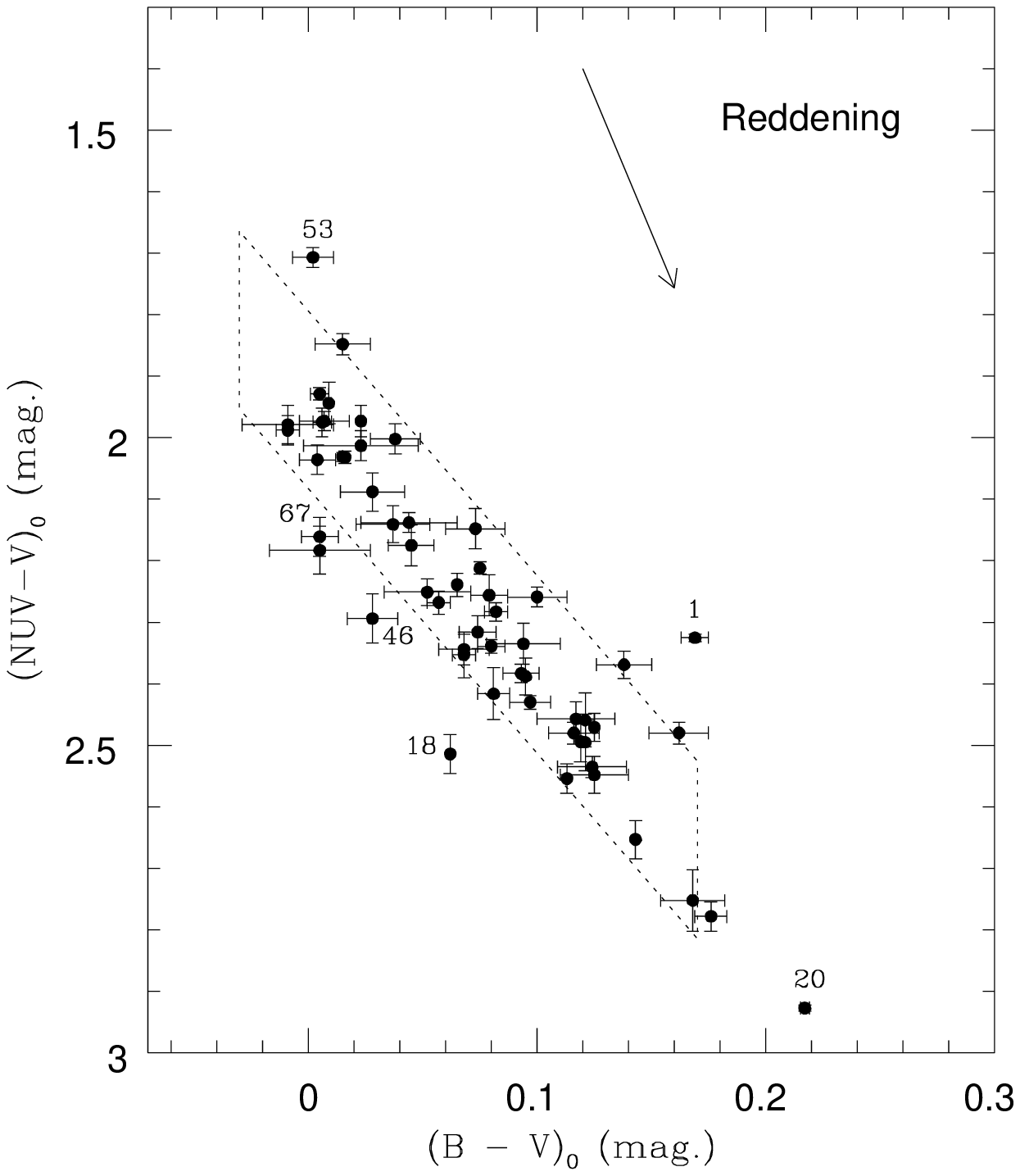}
\caption{The ordinate is the de-reddened difference between the $GALEX$ $NUV$
 magnitude (effective wavelength 2267 \AA) and the Johnson $V$ magnitude. 
The abscissa is the Johnson ($B-V)_{0}$ colour. The dotted parallelogram is the
expected location of BHB stars according to Kinman et al. (2007). The running
numbers of stars in Table 1 whose error bars lie outside this parallelogram are 
shown next to these stars.
}
\label{f:A2}
\end{figure}

% Figure A3
\begin{figure}
\includegraphics[width=8.5cm, bb=80 360 545 680, clip=true]{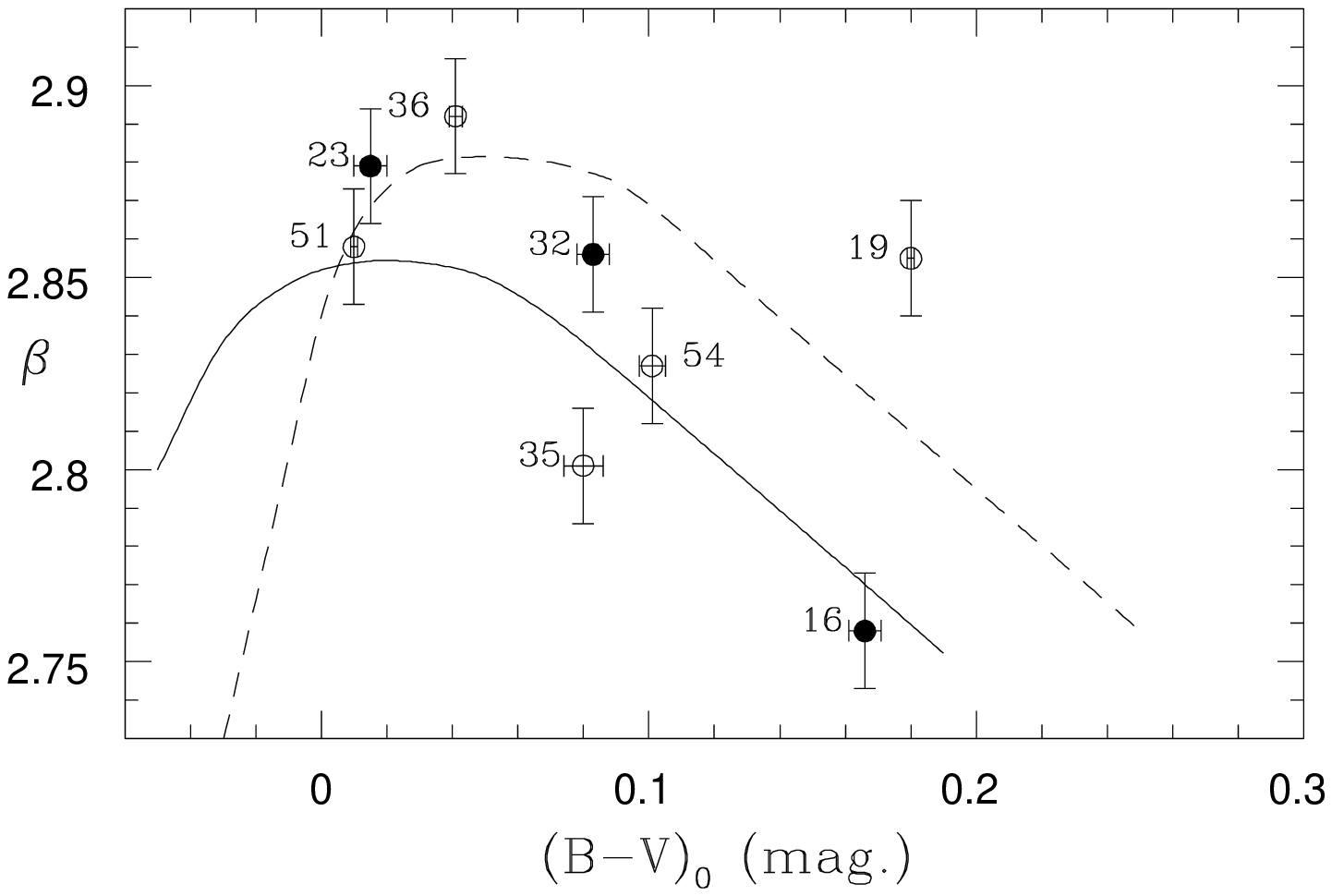}
\caption{The ordinate is Str\"{o}mgren $\beta$ and the abscissa is Johnson
 ($B-V)_{0}$. The solid curve shows the location of BHB stars; the dashed 
 curve shows the lower limit of $\beta$ for non-BHB stars. Further details are
 given in Kinman \& Brown (2011).
}
\label{f:A3}
\end{figure}

% Figure A4
\begin{figure}
\includegraphics[width=8.5cm, bb=100 145 500 695, clip=true]{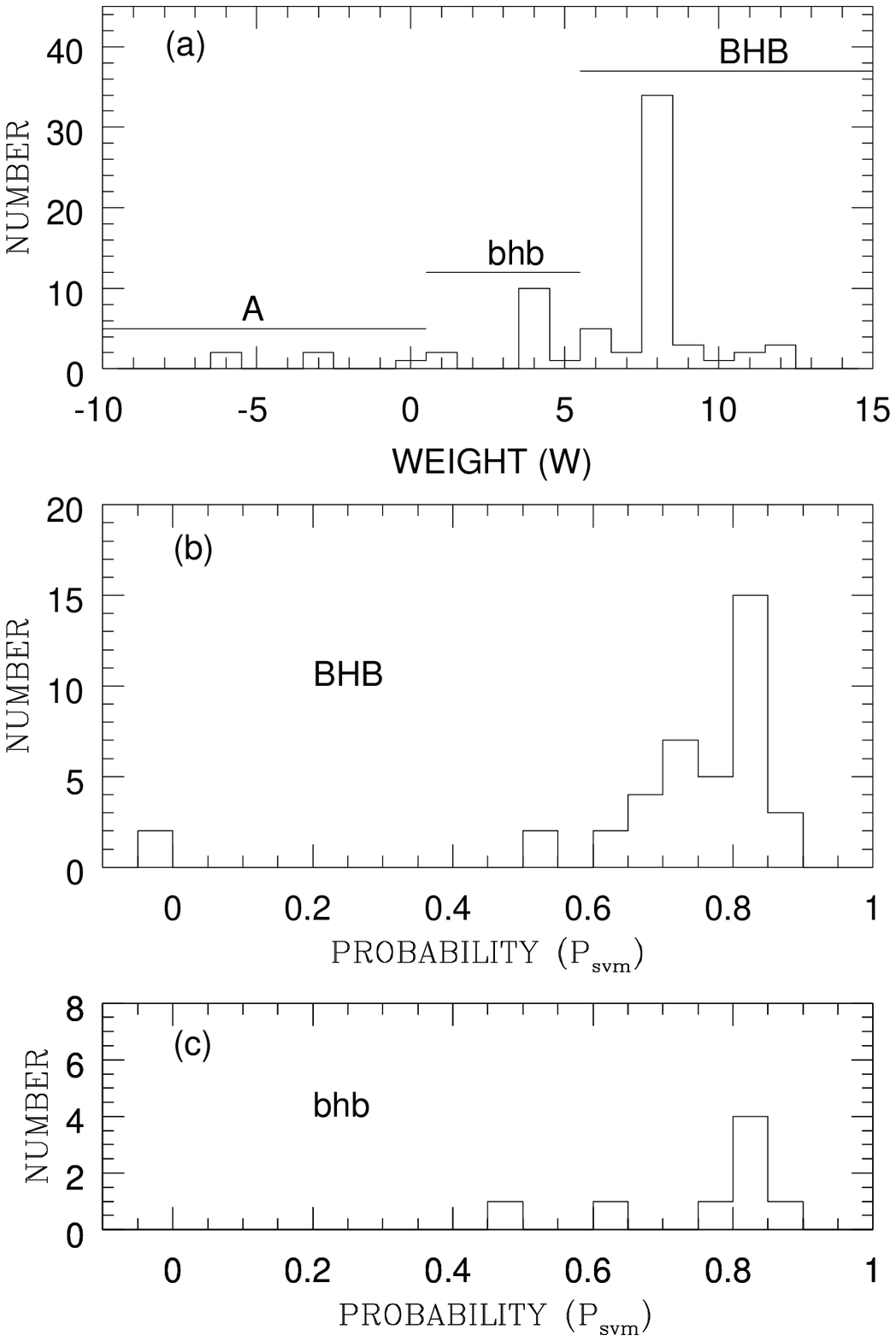}
\caption{ (a) Frequency distribution of the weights given to the program stars:
   the more positive the weight, the greater the probability that the star is
   a BHB star. Stars with weights greater or equal to 6 are classified as 
   BHB stars. Star with weights of zero or less are classified as A stars.
   Stars of intermediate weight are given the intermediate classification of
   bhb.
   (b) Frequency distribution of the probabilty (P$_{svm}$) that a star 
   classified as BHB is a BHB star.  
   (c) Frequency distribution of the probabilty (P$_{svm}$) that a star 
   classified as bhb is a BHB star. P$_{svm}$ is described in the text.
}
\label{f:A4}
\end{figure}

% Figure A5
\begin{figure}
\includegraphics[width=8.5cm, bb=100 145 500 615, clip=true]{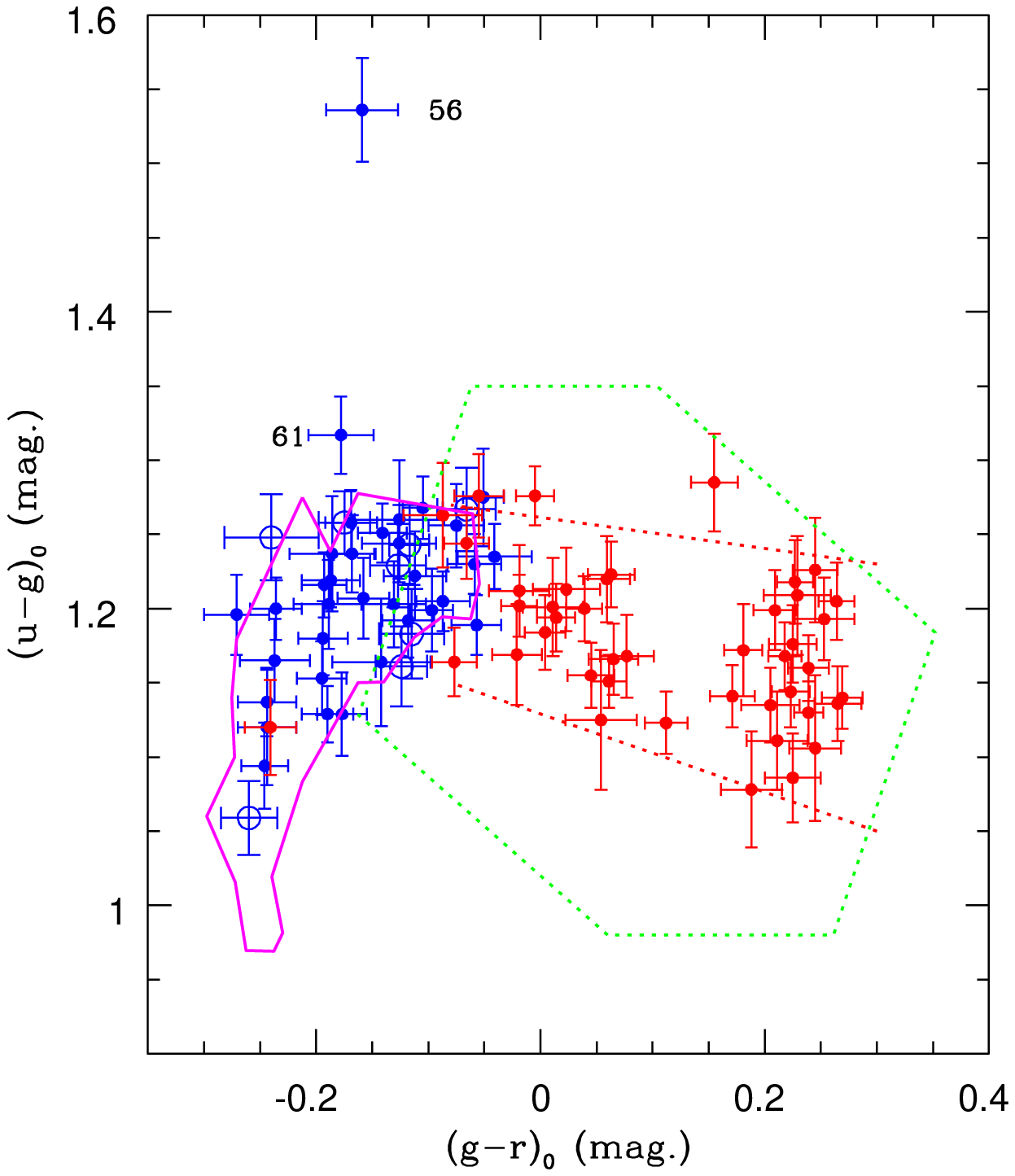}
\caption{ The ordinate is SDSS $(u-g)_{0}$ and the abscissa is $(g-r)_{0}$.
   Stars classified as BHB are shown as blue filled circles. Stars classified
   as bhb are shown as blue open circles. RR Lyrae stars are shown as red
   filled circles. The area enclosed by the magenta line is the locus of the
   BHB stars according to Ruhland et al. (2011). The green dotted hexagon is
   the locus of the RR Lyrae stars according to Watkins et al. (2009). The red
   dotted lines show our colour limits for the RR Lyrae stars. 
}
\label{f:A5}
\end{figure}

 Additional BHB star {\it candidates} with (170$^{\circ}$$<$l$<$207$^{\circ}$)
  were taken from BHB candidates discovered in the objective prism surveys of
  Pesch \& Sanduleak (1989) (AF-nnn) and Beers et al. (1996) 
  (BS nnnnn-nnn). We also included unpublished BHB candidates from the Case 
  Survey that were kindly made available to us by Dr Peter Pesch; we call 
  these P nn.nn stars. All these stars are included in Table 4 (main section
  of this paper) with a running number which is also used as a means of
  identification in the figures. 
 
  Following Kinman \& Brown (2011), BHB stars were selected from these 
  {\it candidates} by 3 methods:

(a)  The ($u-B)_{K0}$ {\it vs.} ($B-V)_{0}$ plot where $u$ is a Str\"{o}mgren
 magnitude and $B$ and $V$ are Johnson magnitudes. The plot is shown in Fig. A1 
     where the the dotted lines enclose an area in which ($B-V)_{0}$ $\leq$
     0.18 and ($u-B)_{K0}$ is within $\pm$0.075 mag of a curve defined by 
     nearby BHB stars whose classification rests on high-resolution 
     spectroscopy (Kinman et al. 2000; Behr 2003). 

(b) The ($NUV-V)_{0}$ {\it vs.} ($B-V)_{0}$ plot where $NUV$ is the near-$UV$
   $GALEX$ magnitude (effective wavelength 2267\AA) taken from the $MAST$
   (Multimission Archive at STSci, http://archive.stsci.edu/). The plot
    is shown in Fig. A2 where the dotted parallelogram (taken from Kinman et al.
    2007a) shows the expected location of BHB stars. This method can only be 
    used for stars fainter than about $V$ = 12 because the $GALEX$ magnitudes
    of brighter stars are affected by saturation.

(c) The Str\"{o}mgren $\beta$ {\it vs.} ($B-V)_{0}$ plot is shown in Fig. A3 
    which is taken from Fig 9(a) of Kinman \& Brown (2011). The full curve 
    shows the expected location of BHB stars and dashed curve shows the lower
    limit of $\beta$ for non-BHB stars. This method was only used for 8 of the
    brighter stars. 

     In these figures, a running number (from Table 1) is given against each
     star in Fig. A3 and for those stars whose error bars lie outside the 
     defining boxes in Figs. A1 \& A2. 
     For stars in (a) Fig. A1 and (b) Fig. A2, those whose colours fall
     within the defining box were given weight +4; those whose error bars 
     intersected the defining box were given weight +2 and the rest were given
     weights of 0 or $-$3 according to their distance from the defining box.
     For the stars in (c) Fig. A3, those whose error bars intersect the full
     curve are given weight 3, those whose error bars lie above this line but
     below the dashed line are given weight 1 and those with larger $\beta$ 
     are given weight $-$3. The weights from the three methods were added to
     give a total weight (W) that is given in column 14 of Table 1.  
      Stars with W$\geq$ 6 are taken to have a high probability of being BHB
      stars; those with zero or negative weights are taken to have a high 
     probability of not being a BHB star (class A) while those with intermediate
     W are considered to be an intermediate class which we call ``bhb". The 
     distribution of these three classes as a function of the weight (W) is 
     shown in Fig. A4 (a).

\begin{figure*}
\includegraphics[width=13.0cm, bb=40 200 570 620, clip=true]{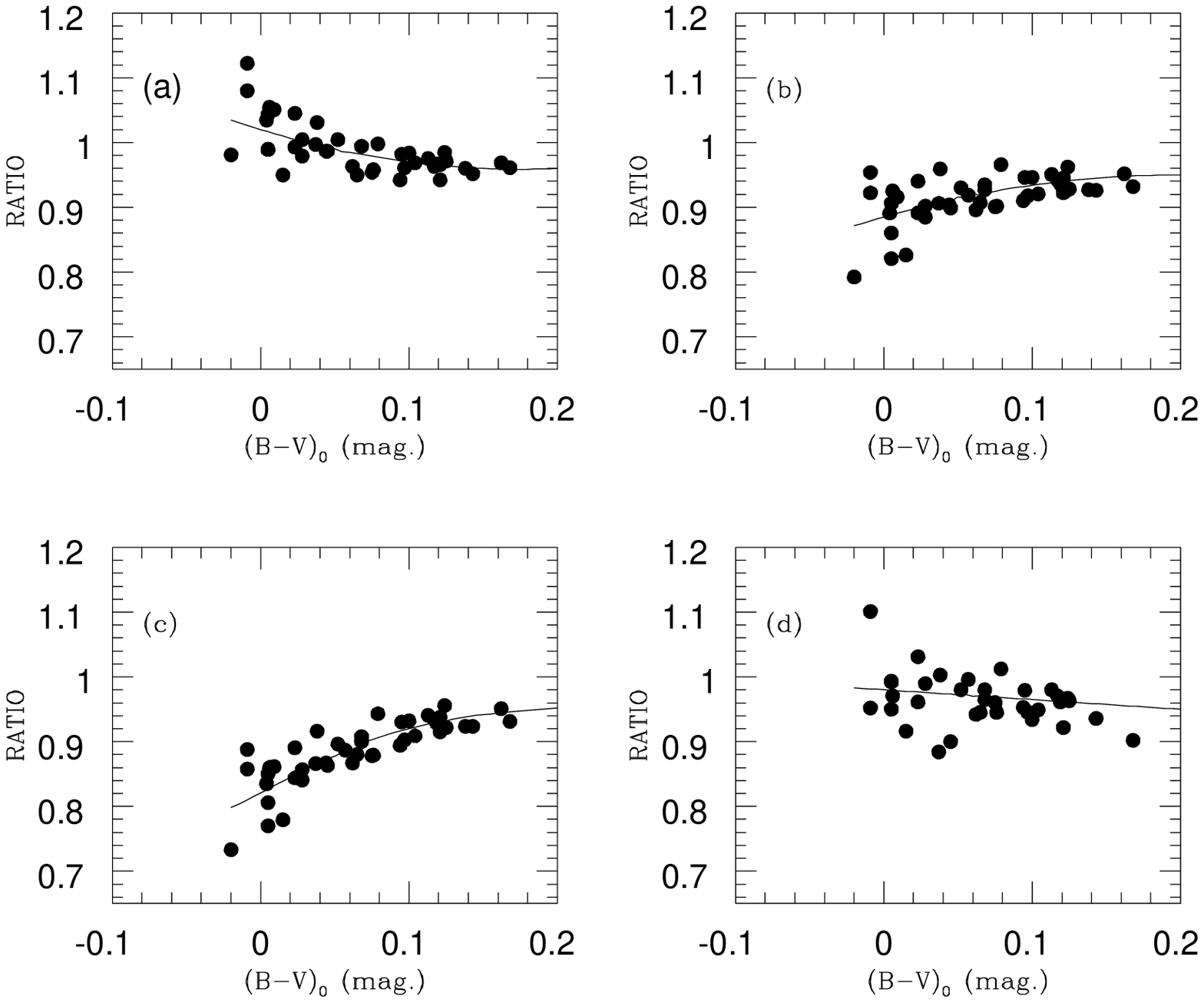}
\caption{ The ordinate is the ratio of the distance of a star by a given method
   divided by the distance (D2) for the same star as a function of $(B-V)_{0}$.
  (a) Distances D1 given by Smith et al. (2010). (b) Distances (D3) derived from
   the cubic in $(B-V)_{0}$ given by Preston et al. (1991). (c) Distances D4 are
   derived from the cubic 
   fit in $(B-V)_{0}$ for M3 and M13.  (d) Distances (D5) derived from a cubic 
   fit in $(V-K)_{0}$. For further details see text.
}
\label{f:A6}
\end{figure*}

\subsection{Comparison with other Selection Methods.}

    Smith et al. (2010) have used machine-learning methods to estimate the
    probability that a star is a BHB star from its SDSS photometry. 
    Their preferred probability (P$_{svm}$) is derived 
    from the {\it support vector machine} method and is available for those 
    of our program stars whose SDSS magnitudes are unsaturated 
    (roughly those with $V$ $>$ 14.5). The distribution of P$_{svm}$ for the
  stars that we classify  as BHB and bhb are shown in Fig. A4(b) and Fig. A4(c) 
    respectively. Most of the stars that we classify as BHB and bhb have a 
    high probability of being BHB stars on the Smith et al. criterion. The only 
    exception is our program star 47 (BS 16468-00260) which is missing from
    the Smith et al. (2010) catalogue although its $(u-g)_{0}$ and 
    $(g-r)_{0}$ are similar to those of stars that are given a high value
    of P$_{svm}$. The stars that we classify as bhb all have P$_{svm}$ 
    greater than 0.6; this suggests that most are likely to be BHB stars.

    Ruhland et al. (2011) have shown that the BHB stars lie in a relatively 
    well-defined locus in the $(u-g)_{0}$ $vs.$ $(g-r)_{0}$ diagram; this is 
  enclosed by the full magenta line in Fig. A5.  All the stars with unsaturated 
    SDSS magnitudes that we classify as BHB or bhb (shown by filled blue
    circles and blue open circles respectively) lie within this line except
    for stars 56 and 61 for which we assigned weights +11 and +8 respectively.
    There are two discrepant $u$ magnitudes for the object in the position
    of star 56 and this presumably explains the location of this star in Fig.
    A5. Star 61 has V = 14.57 which is close to the saturation limit of SDSS
    magnitudes. For these reasons we think that the anomalous locations of stars
    56 and 61 are caused by errors in the SDSS magnitudes and that the high
    weights that we have assigned to these stars are trustworthy. 
    Overall, there is satisfactory agreement between our selection criteria 
    and the most recent selection criteria based on SDSS magnitudes. 

\subsubsection{Reddenings}

The reddenings E$(B-V)$ in this paper were taken from the total reddenings of 
 Schlegel et al. (1998). All but 5 of our BHB and RR Lyrae stars are more than 
 1 kpc above the galactic plane and so the reddenings to these stars will be
 close to the total reddenings in these sight-lines.
The reddening corrections for other colours are derived from the 
 relations E$(V-K$ )=2.75E$(B-V)$, 
A$_V$=3.1E$(B-V)$ and A$_K$=0.35E$(B-V)$ (Cardelli et al. 1989).

\subsection{Distances of the BHB stars.}

   We have used five methods to determine absolute magnitudes (and hence
   distances) for our BHB stars. 
  
   (1) Sirko et al. (2004) gave absolute SDSS $g$ magnitudes based on models by
   Dorman et al. (1993) for various BHB properties including SDSS colours. 
   Smith et al. (2010) used this data to derive distances for their sample of
   BHB stars which includes 40 of our BHB and bhb stars; these
   are given as D1 in Table A1.

  (2) Deason et al. (2011) derived the absolute magnitudes of BHB stars from the
   SDSS photometry of 11 globular clusters by An et al. (2008).
   They expressed the absolute magnitude ($M_{g}$) in their equation (7)
   as a quartic in $(g-r)_{0}$.
   We used this expression to derive distances (D2 in Table A1) for the 43 stars
  for which unsaturated SDSS colours were available from SDSS DR7 \footnote{This
   expression gives ($M_{g}$ $\sim$0.44 at the blue edge of the instability gap;
   this corresponds to $M_{V}$$\sim$ 0.58.}.

 (3) Preston, Schectman and Beers (1991) used the Johnson photometry of 
   15 globular clusters 
   to derive the absolute $V$ magnitude of BHB stars in terms of a cubic
   in $(B-V)_{0}$. A slightly adjusted version of this expression is given as 
   their equation (5) by Kinman et al. (2007b). We used this expression to 
   derive the distances (D3) in Table A1.

 (4) Kinman et al. (2007b) attempted to improve on the cubic equation in (c) by
 deriving another cubic based on the photometry of the intermediate-metallicity 
   globular clusters M3 (Ferraro et al. 1997) and M13 (Paltrinieri et al. 1998).
   This cubic is given as their equation (6) in Kinman et al. (2007b). The 
     corresponding distances are given as (D4) in Table A1.

 (5) Kinman et al. (2007b) gave a cubic expression for the infrared absolute 
     magnitude $M_{K}$ in terms of the $(V-K)_{0}$ colours. The calibration was
     derived from the colour magnitude diagrams of the globular clusters M3 and
     M13 given by Valenti et al. (2004). This cubic is given as their equation
     (3) by Kinman et al. (2007b). The corresponding distances are given as (D5)
      in Table A1.

  Methods (1) and (2) are only available for the stars with unsaturated SDSS 
  magnitudes (roughly $V$ $>$ 14.5). Methods (3) and (4) are possible for all 
  stars in the sample while method (e) is only used for the brighter stars whose
  2MASS $K$-magnitudes have errors less than 0.15 mag.. We assume that the 
  distances D2 are the most reliable because they are based on the recent 
  homogeneous photometry of 11 globular clusters that have well established 
  moduli. For each star, and for each method, we have computed a ratio F 
  that equals the distance for that method 
  divided by its distance D2. These ratios are plotted for each method as a 
  function of $(B-V)_{0}$ in Fig. A6.  The values  of F for 
  distances D4 (Fig 6(c)) are always less than those for D3 (Fig 6(b)). This
  shows that method (4) is inferior to method (3) in determining $M_{V}$
  as a function of $(B-V)_{0}$; the distances D4 are therefore not considered
  further. Smoothed curves were fitted to the plots in Fig A6 (a), (b) and (c) 
  and these have the following analytic expressions as a function of $(B-V)_{0}$:

\begin{eqnarray}
 F1 =  1.020 - 0.700(B-V)_{0} + 2.000(B-V)_{0}^2~~~~(D1) \nonumber
\end{eqnarray}

\begin{eqnarray}
 F2 = 1.000~~~~~~~~~~~~~~~~(D2)  \nonumber 
\end{eqnarray}

\begin{eqnarray}
 F3 = 0.885 + 0.655(B-V)_{0} - 1.650(B-V)_{0}^2~~~~(D3) \nonumber
\end{eqnarray}

\begin{eqnarray}
 F5 = 0.980 - 0.150(B-V)_{0}~~~~~~~~~(D5) \nonumber
\end{eqnarray}

 Here, for example, the expression F1 is the quantity by which the distance D1 must 
 be divided to get it onto the scale of the distances called D2. 
 There is no F4 because we do not use the distance called D4. The distances D1. 
 D2, D3 and D5 were then divided respectively by F1, F2, F3 and F5 to give the 
 corrected distances d1, d2, d3 and d5. The adopted distance (D) is the 
  unweighted mean of these corrected distances. All these distances  
 are given in Table \ref{t:A1}. 

 We assume that the random error in these distances is given by the rms scatter 
 among the distances d1, d2, d3 and d5. This is given (in kpc) in col. 15 in 
 Table \ref{t:A1}. We have used these errors in computing the error bars of the 
velocities and angular momenta (L$_{\perp}$, L$_{z}$) that are distant-dependent.
 It remains to consider the effect of possible systematic errors. The point at
 issue is whether these systematic errors depend on distance. If they do,
 there would be systematic errors in our velocities and angular momenta that
 would produce errors in these quantities as a function of galactocentric
 distance. The most likely source of a distance-dependent error is in the 
 absolute magnitudes and proper motions since neither the radial velocity, 
 apparent magnitude or
 extinction are likely to have distance-dependent errors. Such an error could arise 
 if the mean colours of our BHB stars varied with distance. 
 Using only the stars that we
 used in our velocity and angular momentum analyses, we find a mean $(B-V)_{0}$
 of +0.61$\pm$0.018, +0.075$\pm$0.012 and +0.081$\pm$0.012 mag at mean 
 galactocentric distances of 10.6, 13.4 and 15.9 kpc respectively. These data
 would be compatible with a difference of say 0.02 magnitudes in the mean
 $(B-V)_{0}$ between galactocentric distances of of 10.6 and 13.4 kpc. This
 corresponds to a difference in 0.05 mag  in the absolute magnitude or 2.3\% in the
 distance. This error is far too small to account for the change in galactic 
 rotation (V) of about 100 km s$^{-1}$ between these two distances. Likewise, it 
 would require a {\it systematic difference} of --1.6 mas y$^{-1}$ between the 
 proper motions of the stars at $V$ = 13.0  and those at $V$ = 15.0  to produce the 
 differences in galactic rotation (V) that we observe (Sec 5.2). This seems 
 unlikely since the galactic U and W velocities show no dependence on distance.

\begin{center}
\topmargin 25mm 
\thispagestyle{empty}
\begin{table*} 
\caption{Distances in kpc for  BHB and bhb stars. The distances D1, D2, D3, D4 \& D5 
 and the corrected distances d1, d2, d3 \& d5 are defined in the text. The
 adopted distance D and its rms error $\sigma$ are given in columns (14) and (15).
\label{t:A1}
}
\begin{tabular}{@{}clccccccccccccc@{}} 
\hline 
 No &ID& Class&D1 & D2&D3&D4&D5&$(B-V)_{0}$&d1& d2 &d3 & d5& D & $\sigma$\\ 
(1) &(2)&(3) &(4) &(5) &(6) &(7) &(8) &(9) &(10) &(11) &(12)&(13)&(14)&(15)   \\
\hline 
 02&  4-32.5 &bhb  &  8.210 & 8.567 & 7.728 & 7.534 & 8.096 & 0.076  & 8.392& 8.567 & 8.352 & 8.358 & 8.417 &0.058\\            
 03& AF-186  &BHB  &  9.450 & 9.602 & 9.081 & 8.948 & 8.972 & 0.100  & 9.742& 9.602 & 9.723 & 9.279 & 9.591 &0.119\\                
 04& AF-189  &BHB  &  7.580 & 7.606 & 6.893 & 6.583 & 6.727 & 0.037  & 7.604& 7.606 & 7.600 & 6.903 & 7.428 &0.202\\                  
 05& 54-111  &BHB  &  5.900 & 6.137 & 5.633 & 5.544 & 5.798 & 0.097  & 6.077& 6.137 & 6.037 & 6.005 & 6.064 &0.033\\                  
 06& 54-122  &bhb  &  7.360 & 6.813 & 6.280 & 5.846 & 6.486 &$-$0.009& 7.170& 6.813 & 7.145 & 6.609 & 6.934 &0.156\\                    
 08& 54-119  &BHB  &  ...    &  ...  & 5.166 & 5.158 & 5.134 & 0.176  & ...  & ...   & 5.443 & 5.384 & 5.413 &0.042\\                  
 09& 17444-25&bhb  &  ...    &  ...  & 0.697 & 0.682 & 0.708 & 0.084  & ...  & ...   & 0.751 & 0.732 & 0.742 &0.013\\                  
 10& AF-209  &BHB  &  13.670 &13.822 &11.343 &10.640 & ...   & 0.005  &13.447&13.822 &12.770 & ...   &13.346 &0.377\\                  
 11& AF-210  &bhb  &  8.730 & 9.093 &8.427 & 8.399 & ...   & 0.138  & 9.080& 9.093 & 8.927 & ...   & 9.033 &0.065\\                  
 12& AF-214  &BHB  &  9.810 &10.206 &9.514 & 9.504 & 9.209 & 0.168  &10.231&10.206 &10.031 & 9.645 &10.028 &0.156\\                 
 13& RR7-02  &BHB  &  7.530 & 7.912 &7.326 & 7.307 & 7.402 & 0.143  & 7.837& 7.912 & 7.753 & 7.722 & 7.806 &0.050\\                 
 14& 81-42   &BHB  &  5.290 & 5.349 &4.598 & 4.313 & 5.080 & 0.005  & 5.204& 5.349 & 5.177 & 5.188 & 5.230 &0.046\\                 
 15& RR7-08  &BHB  &  7.510 & 7.871 &7.093 & 6.911 & 7.559 & 0.075  & 7.673& 7.871 & 7.669 & 7.803 & 7.754 &0.058\\                 
 16& RR7-15  &BHB  &  ...    & ...   &1.606 & 1.605 & 1.608 & 0.166  & ...  &  ...  & 1.694 & 1.684 & 1.689 &0.007\\                 
 17& 81-39   &BHB  &  9.710 & 9.383 &8.361 & 7.839 & ...   & 0.004  & 9.546& 9.383 & 9.420 & ...   & 9.450 &0.060\\                 
 18& RR7-21  &bhb  &  8.210 & 8.525 &7.637 & 7.393 & 8.027 & 0.062  & 8.341& 8.525 & 8.308 & 8.269 & 8.361 &0.065\\                 
 19& 81-72   &bhb  &  ...    & ...   &1.678 & 1.676 & 1.633 & 0.174  & ...  & ...   & 1.768 & 1.712 & 1.740 &0.040\\                 
 21& RR7-23  &BHB  &  ...    & ...   &2.057 & 1.941 & 2.235 & 0.015  & ...  & ...   & 2.300 & 2.286 & 2.293 &0.010\\                 
 22& RR7-36  &BHB  &  7.940 & 8.426 &7.772 & 7.714 & 7.766 & 0.121  & 8.232& 8.426 & 8.267 & 8.074 & 8.250 &0.083\\                 
 23& 81-101  &bhb  &  9.050 & 9.323 &8.652 & 8.597 & 8.980 & 0.125  & 9.390& 9.323 & 9.194 & 9.342 & 9.312 &0.048\\                 
 24& 81-121  &BHB  &  9.650 &11.366 &8.393 & 7.916 & 9.304 & 0.015  & 9.555&11.366 & 9.383 & 9.516 & 9.955 &0.545\\                 
 25& RR7-043 &BHB  &  14.950 &14.222 &13.028 &12.247 & ...   & 0.009  &14.746&14.222 &14.626 &15.880 &14.868 &0.410\\                 
 26&  28-45  &bhb  &  ...    & ...   &5.041 & 4.967 & 5.171 & 0.100  & ...  & ...   & 5.397 & 5.359 & 5.378 &0.027\\                 
 27& RR7-053 &BHB  &  7.340 & 7.593 &7.107 & 7.050 & 7.294 & 0.119  & 7.606& 7.593 & 7.564 & 7.581 & 7.586 &0.010\\                 
 28& 81-162  &BHB  &  11.580 &11.735 &10.614 &10.176 & ...   & 0.044  &11.661&11.735 &11.656 & ...   &11.684 &0.031\\                  
 29& RR7-058 &BHB  &  8.410 & 8.051 &7.571 & 7.174 & 8.298 & 0.023  & 8.369& 8.051 & 8.420 & 8.497 & 8.334 &0.113\\                 
 30& RR7-60  &BHB  &  ...    & ...   &5.565 & 5.530 & 5.698 & 0.125  & ...  & ...   & 5.913 & 5.928 & 5.920 &0.011\\                 
 31& 82-04   &BHB  &  10.390 &10.345 &9.327 & 8.863 & ...   & 0.028  &10.370&10.345 &10.340 & ...   &10.352 &0.016\\                 
 32& RR7-64  &BHB  &  ...    & ...   &1.195 & 1.169 & 1.223 & 0.083  & ...  & ...   & 1.288 & 1.264 & 1.276 &0.017\\                          
 33& RR7-66  &BHB  &  8.220 & 8.372 &7.917 & 7.785 & 8.193 & 0.095  &8.461& 8.372 & 8.492 & 8.484 & 8.452 &0.032\\                 
 34& 81-167  &BHB  &  ...    & ...   &5.192 & 5.057 & 5.438 & 0.074  & ...  & ...   & 5.616 & 5.613 & 5.614 &0.002\\                 
 35& 11419-01&BHB  &  ...    & ...   &2.416 & 2.360 & 2.486 & 0.080  & ...  & ...   & 2.607 & 2.568 & 2.588 &0.028\\                 
 37& AF-293  &BHB  &  13.090 &13.513 &12.864 &12.852 & ...   & 0.162  &13.648&13.513 &13.572 & ...   &13.578 &0.048\\                 
 38& 11419-04&BHB  &  6.580 & 6.242 &5.772 & 5.371 & 6.061 & 0.006  & 6.477& 6.242 & 6.494 & 6.190 & 6.351 &0.091\\                 
 39& RR7-84  &BHB  &  10.250 &10.509 &9.990 & 9.891 &10.300 & 0.113  &10.606&10.509 &10.651 &10.695 &10.615 &0.046\\                 
 40& RR7-91  &BHB  &  ...    & ...   &4.257 & 4.017 & 4.629 & 0.016  & ...  & ...   & 4.756 & 4.735 & 4.746 &0.015\\                 
 41& RR7-90  &BHB  &  9.540 &10.042 &9.113 & 8.836 & 9.486 & 0.065  & 9.705&10.042 & 9.899 & 9.777 & 9.856 &0.085\\                 
 42& 82-49   &BHB  &  8.780 & 8.798 &8.497 & 8.296 & 8.907 & 0.079  & 8.985& 8.798 & 9.172 & 9.200 & 9.039 &0.108\\                 
 44&16473-102&BHB  &  ...    & ...   &4.223 & 4.129 & 4.373 & 0.082  & ...  & ...   & 4.553 & 4.519 & 4.536 &0.024\\                 
 45& 17139-69&BHB  &  5.650 & 5.847 &5.523 & 5.482 & 5.635 & 0.121  & 5.857& 5.847 & 5.875 & 5.859 & 5.867 &0.011\\                 
 46& TON 384 &bhb  &  7.920 & 8.089 &7.157 & 6.801 & 8.011 & 0.028  & 7.904& 8.089 & 7.934 & 8.210 & 8.034 &0.082\\                 
 47& 16468-26&BHB  &  ...    & 6.436 &5.913 & 5.709 & 6.410 & 0.057  & ...  & 6.436 & 6.448 & 6.598 & 6.494 &0.064\\                 
 48& AF-379  &BHB  &  8.200 & 8.310 &7.473 & 7.168 & 7.476 & 0.045  & 8.262& 8.310 & 8.202 & 7.681 & 8.114 &0.168\\                 
 49& AF-386  &BHB  &  7.280 & 7.324 &6.848 & 6.650 & 7.068 & 0.068  & 7.416& 7.324 & 7.428 & 7.288 & 7.364 &0.040\\                 
 50& AF-390  &BHB  &  8.470 & 8.533 &7.599 & 7.200 & 8.204 & 0.023  & 8.428& 8.533 & 8.451 & 8.401 & 8.453 &0.033\\                 
 51& 16468-78&BHB  &  ...    & ...   &1.361 & 1.280 & 1.498 & 0.010  & ...  & ...   & 1.527 & 1.531 & 1.529 &0.003\\                 
 52& 30-16   &bhb  &  5.650 & 5.830 &5.369 & 5.298 & 5.531 & 0.104  & 5.832& 5.830 & 5.741 & 5.735 & 5.784 &0.018\\                 
 53& 16468-80&bhb  &  ...    & ...   &4.043 & 3.786 & 4.297 & 0.002  & ...  & ...   & 4.562 & 4.386 & 4.474 &0.124\\                 
 54& 30-28   &BHB  &  ...    & ...   &0.804 & 0.793 & 0.831 & 0.101  & ...  & ...   & 0.861 & 0.861 & 0.861 &0.000\\                 
 55& 16468-90&BHB  &  ...    & ...   &4.743 & 4.700 & 4.771 & 0.116  & ...  & ...   & 5.052 & 4.956 & 5.004 &0.068\\                 
 56& CHSS 608&BHB  &  ...    & 6.728 &6.237 & 6.057 & 6.593 & 0.068  & ...  & 6.728 & 6.765 & 6.798 & 6.764 &0.025\\                 
 57& 11424-28&BHB  &  ...    & ...   &5.470 & 5.345 & 5.727 & 0.081  & ...  & ...   & 5.899 & 5.917 & 5.908 &0.013\\                 
 58& 30-038  &BHB  &  5.570 & 5.655 &5.440 & 5.404 & 5.466 & 0.124  & 5.778& 5.655 & 5.782 & 5.685 & 5.725 &0.037\\                 
 59& 57-121  &BHB  &  8.220 & 8.381 &6.641 & 6.146 &  ...  &$-$0.020& 7.944& 8.381 & 7.622 & ...   & 7.982 &0.269\\                   
 60& AF-419  &BHB  &  7.930 & 8.416 &7.655 & 7.524 & 8.017 & 0.094  & 8.160& 8.416 & 8.214 & 8.300 & 8.272 &0.157\\                 
 61&11424-070&BHB  &  5.920 & 5.896 &5.483 & 5.280 & 5.776 & 0.052  & 5.986& 5.896 & 5.995 & 5.941 & 5.954 &0.026\\                 
 62&16927-22 &bhb  &  ...    & ...   &1.202 & 1.177 & 1.251 & 0.086  & ...  & ...   & 1.294 & 1.294 & 1.294 &0.000\\                 
 63&CHSS 663 &BHB  &  7.680 & 7.446 &7.140 & 6.822 & 7.469 & 0.038  & 7.709& 7.446 & 7.868 & 7.666 & 7.672 &0.100\\                 
 64&11424-82 &BHB  &  7.280 & 7.558 &7.122 & 7.060 & 7.338 & 0.117  & 7.540& 7.558 & 7.584 & 7.624 & 7.576 &0.021\\                 
 65&16940-45 &BHB  &  ...    & ...   &3.217 & 3.021 & 3.551 & 0.007  & ...  & ...   & 3.617 & 3.627 & 3.622 &0.007\\                 
 66&16927-55 &BHB  &  5.770 & 5.141 &4.902 & 4.563 & 5.659 &$-$0.009& 5.621& 5.141 & 5.577 & 5.767 & 5.526 &0.156\\                   
 67&16940-70 &bhb  &  7.160 & 6.864 &6.216 & 5.831 & 6.810 & 0.005  & 7.043& 6.864 & 6.998 & 6.954 & 6.965 &0.044\\                 
 68&16940.72 &BHB  &  ...    & ...   &4.433 & 4.356 & 4.586 & 0.093  & ...  & ...   & 4.758 & 4.747 & 4.752 &0.008\\                
\hline
\end{tabular}
\end{table*}
\end{center}

\section{RR Lyrae stars.}

 \subsection{The new RR Lyrae stars.} 
 Seven of the stars listed in  Table 5 (identified by a 7 in the Notes
 column) have not previously be identified as RR Lyrae stars. 
 All are of low amplitude and all but one are of Bailey type $c$.  
 Their $V$ and $B-V$ light
  curves are given in Figs. B1 and B2. The ephemerides and photometric data 
 for these variables are given in Table B1. The observations were made in
 the same way as those described by Kinman \& Brown (2010) but in general 
 there are fewer observations than for the stars observed in that paper. This is
paricularly true of AF-194, AF-197, AF-400 and AF-430. Consequently, the periods
 of these stars are less reliable than we would wish. In assigning periods and
 Bailey types, however, we took into account the mean $(B-V)$ colours which are
 well determined and which can be used to distinguish between Bailey type $c$
 and Bailey type $ab$. These Bailey types are therefore more certain than 
 would be inferred from the $V$ light-curves alone.

\begin{center}
\begin{table*}
\caption{Ephemerides and Photometric Data for the Seven New RR Lyrae Stars. \label{t:B1}
}
%\begin{flushleft}
\begin{tabular}{@{}lcccccc@{}}
\noalign{\smallskip} \hline
 Star  & Type & Period & JD$_{max}$&$\langle V\rangle$&$V_{amp}$&$\langle (B-V)
\rangle$ \\
       &      & (days) & 24000+ & (mag) & (mag) & (mag)  \\
\hline \hline
 AF-194 & $ab$ &0.8410:  & 49736.740  & 15.85 & 0.45 & 0.40       \\
 AF-197 &  $c$ &0.38802  & 49416.650  & 15.50 & 0.40 & 0.30      \\
 RR7-086 &$c$  &0.353878 & 47538.525  & 16.14 & 0.63 & 0.29       \\
 AF-316 & $c$  &0.3455178& 49010.793  & 16.13 & 0.46 & 0.28      \\
 RR7-101 &$c$  &0.329836 & 47538.685  & 16.15 & 0.60 & 0.22       \\
 AF-400 &$c$   &0.403114 & 50503.429  & 14.10 & 0.40 & 0.20       \\  
 AF-430 &$c$   &0.30517  & 50503.723  & 14.90 & 0.40 & 0.22       \\
\hline
\end{tabular}
\end{table*}
\end{center}

% Table B2
\begin{center}
\begin{table*}
\caption{Distance estimates of RR Lyrae stars. Adopted distance is D.
\label{t:B2}}
%\begin{flushleft}
\begin{tabular}{@{}lccccccccc@{}}
\noalign{\smallskip} \hline
 No. &Star&$D_{a}$&$D_{b}$&$D_{c}$&$d_{a}$&$d_{b}$&$d_{c}$& D & error      \\
     &    &(kpc) & (kpc)  & (kpc) & (kpc) & (kpc) & (kpc) & (kpc) & (kpc)   \\
\hline \hline
    1 & V385 Aur &22.163& ...  & ...  & 22.738&  ...  &  ...  & 22.74 & 1.14  \\
    2 & V386 Aur &16.362& ...  & ...  & 16.786&  ...  &  ...  & 16.79 & 0.84  \\
    3 & V387 Aur &17.131& ...  & ...  & 17.575&  ...  &  ...  & 17.58 & 0.88  \\
    4 & V389 Aur &23.288& ...  & ...  & 23.892&  ...  &  ...  & 23.89 & 1.19  \\
    5 & VX Lyn   &18.293& ...  & ...  & 18.767&  ...  &  ...  & 18.77 & 0.94  \\
    6 & VY Lyn   &10.157& 9.851& 9.972& 10.420&  9.875& 10.464& 10.25 & 0.23  \\
    7 & VZ Lyn   &12.527& ...  & ...  & 12.852&  ...  &  ...  & 12.85 & 0.64  \\
    8 & WX Lyn   &17.347& ...  & ...  & 17.797&  ...  &  ...  & 17.80 & 0.89  \\
    9 & AS Lyn   &33.035& ...  & ...  & 33.891&  ...  &  ...  & 33.89 & 1.69  \\
   10 & WZ Lyn   & 5.352& 5.674& 5.230&  5.491&  5.688&  5.488& 05.56 & 0.08  \\
   11 & XZ Lyn   &13.473& ...  & ...  & 13.822&  ...  &  ...  & 13.82 & 0.69  \\
   12 & TW Lyn   & 1.644& 1.651& 1.644&  1.687&  1.655&  1.725& 01.69 & 0.02  \\
   13 & YY Lyn   & 7.307& 7.566& 7.087&  7.496&  7.584&  7.437& 07.51 & 0.05  \\
   14 & YZ Lyn   &20.675& ...  & ...  & 21.211&  ...  &  ...  & 21.21 & 1.06  \\
   15 & AU Lyn   &27.497& ...  & ...  & 28.210&  ...  &  ...  & 28.21 & 1.41  \\
   16 & ZZ Lyn   &10.448&12.344&10.102& 10.719& 12.374& 10.600& 11.23 & 0.07  \\
   17 & RW Lyn   & 2.810& 2.598& 2.761&  2.883&  2.604&  2.897& 02.79 & 0.12  \\
   18 & AV Lyn   &15.668& ...  & ...  & 16.074&  ...  &  ...  & 16.07 & 0.80  \\
   19 & AC Lyn   &13.774& ...  & ...  & 14.131&  ...  &  ...  & 14.13 & 0.71  \\
   20 & AD Lyn   &10.536&12.494&10.144& 10.809& 12.524& 10.644& 11.33 & 0.74  \\
   21 & AW Lyn   &12.594&11.718&12.363& 12.921& 11.746& 12.973& 12.55 & 0.49  \\
   22 & AX Lyn   &37.560& ...  & ...  & 38.534&  ...  &  ...  & 38.53 & 1.92  \\
   23 & AY Lyn   &17.562& ...  & ...  & 18.017&  ...  &  ...  & 18.02 & 0.90  \\
   24 &P 54-13   & 7.953& 6.648& 7.797&  8.159&  6.664&  8.182& 07.67 & 0.63  \\
   25 & AZ Lyn   &15.465& ...  & ...  & 15.866&  ...  &  ...  & 15.87 & 0.79  \\
   26 & BB Lyn   &16.922& ...  & ...  & 17.361&  ...  &  ...  & 17.36 & 0.87  \\
   27 & BC Lyn   &18.481& ...  & ...  & 18.960&  ...  &  ...  & 18.96 & 0.95  \\
   28 & AF 194   &10.832&11.637&10.615& 11.113& 11.665& 11.139& 11.31 & 0.22  \\
   29 & AF 197   & 9.396& 9.177& 9.223&  9.640&  9.199&  9.678& 09.51 & 0.19  \\
   30 & DQ Lyn   & 1.416& 1.707& 1.387&  1.453&  1.711&  1.455& 01.54 & 0.10  \\
   31 & RR7 032  & 6.072& 5.963& 5.955&  6.229&  5.977&  6.249& 06.15 & 0.11  \\
   32 & RR7 034  & 8.636& 9.103& 8.314&  8.860&  9.125&  8.724& 08.90 & 0.14  \\
   33 &P 81 129  & 5.954& 6.030& 5.802&  6.108&  6.045&  6.088& 06.08 & 0.02  \\
   34 &AF Lyn    &12.433&12.253&12.209& 12.755& 12.282& 12.811& 12.62 & 0.21  \\
   35 &P 82 06   & 5.148& 5.223& 4.984&  5.281&  5.236&  5.230& 05.25 & 0.02  \\
   36 & AI Lyn   &19.799& ...  & ...  & 20.312&  ...  &  ...  & 20.31 & 1.02  \\
   37 & AK Lyn   &11.917&11.033&11.706& 12.226& 11.060& 12.283& 11.86 & 0.49  \\
   38 & RR7-079  & 3.809& 3.688& 3.735&  3.908&  3.697&  3.919& 03.84 & 0.09  \\
   39 & RR7-086  &12.779&15.001&12.274& 13.110& 15.037& 12.879& 13.68 & 0.84  \\
   40 & AL Lyn   &15.546&13.419&15.173& 15.949& 13.451& 15.921& 15.11 & 1.01  \\
   41 & AM Lyn   &21.678& ...  & ...  & 22.240&  ...  &  ...  & 22.24 & 1.11  \\
   42 &P 82-32   & 7.785& 8.543& 7.592&  7.987&  8.564&  7.966& 08.17 & 0.24  \\
   43 & AF 316   &12.757&12.218&12.519& 13.088& 12.247& 13.136& 12.82 & 0.35  \\
   44 & RR7-101  &12.967&12.239&12.723& 13.303& 12.268& 13.350& 12.97 & 0.43  \\
   45 &  TT Lyn  & 0.699& 0.694& 0.689&  0.717&  0.696&  0.723& 00.71 & 0.01  \\
   46 &AF 400    & 5.074& 5.915& 4.929&  5.206&  5.929&  5.172& 05.44 & 0.30  \\
   47 &AF 430    & 7.355& 7.697& 7.127&  7.546&  7.716&  7.478& 07.58 & 0.09  \\
   48&16927-123  & 3.326& 3.547& 3.248&  3.412&  3.556&  3.408& 03.46 & 0.06  \\
   49 & X LMi    & 2.224& 2.293& 2.188&  2.282&  2.299&  2.296& 02.29 & 0.01  \\
   50 &AG UMa    & 9.398& 9.453& 9.205&  9.642&  9.476&  9.659& 09.59 & 0.07  \\
   51 &BK UMa    & 2.869& 2.704& 2.825&  2.943&  2.711&  2.964& 02.87 & 0.10  \\
   52 &AK UMa    &12.736&13.915&12.503& 13.066& 13.948& 13.120& 13.38 & 0.35  \\
   53 &AO UMa    & 9.890&10.806& 9.692& 10.146& 10.832& 10.170& 10.38 & 0.28  \\
   54 &BN UMa    & 3.871& 4.136& 3.794&  3.971&  4.146&  3.981& 04.03 & 0.07  \\
   55 &CK UMa    & 4.984& 4.686& 4.886&  5.113&  4.697&  5.127& 04.98 & 0.17  \\
\hline
\end{tabular}
\end{table*}
\end{center}

% Figure B1
\begin{figure*}
\includegraphics[width=17.0cm, bb=20 325 575 650, clip=true]{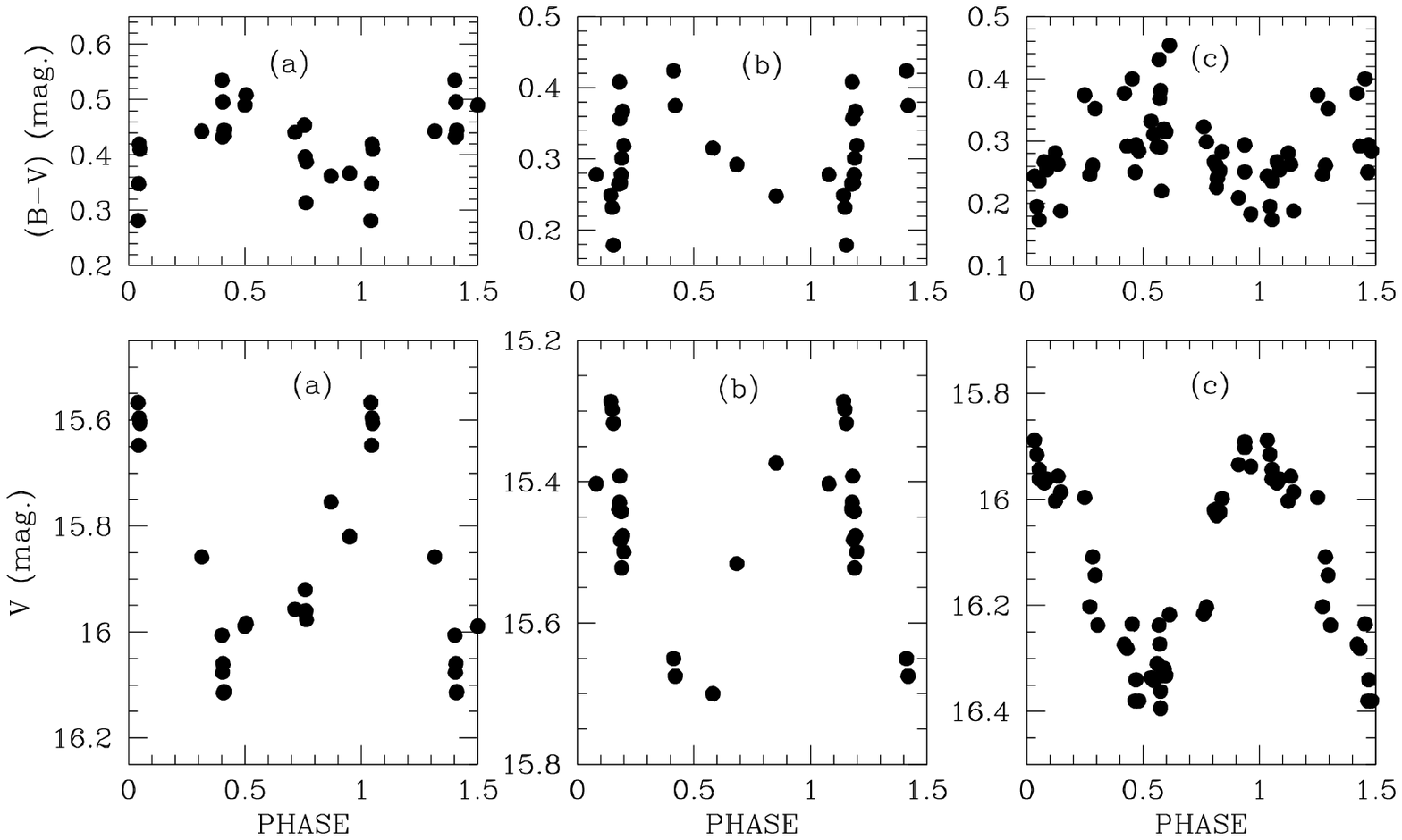}
\caption{ The $V$ and $(B-V)$ light curves for the RR Lyrae variable:     
     (a) AF-194, (b) AF-197 and (c) AF-316. Further information on these
      stars is given in Table B1.   
}
\label{f:B1}
\end{figure*}

\pagebreak

 \subsection{The Distances of the RR Lyrae stars.} 

   We use three methods to estimate the distances
  of our RR Lyrae stars; 
  details of these methods are given in Kinman et al. (2007b).

  (1) The absolute visual magnitude $M_{V}$ is derived 
  in terms of the metallicity [Fe/H] using coefficients given by 
  Clementini et al. (2003):
\begin{eqnarray}  
      M_V = 0.214[Fe/H] +0.86    \nonumber
\end{eqnarray}
  If [Fe/H] is not known, it is assumed to --1.6. An error of $\pm$0.5 dex in
  [Fe/H] leads to an error of $\pm$0.1 mag in the distance modulus and about
  5\% in the parallax.

  (2) The infrared absolute magnitude ($M_{K}$ is derived from [Fe/H] and the
      Period (P) (in days) in the form given by Nemec et al. (1994): 
\begin{eqnarray}  
      M_K = -2.40\log P + 0.06 [Fe/H] - 1.06    \nonumber
\end{eqnarray}
     The periods of the Bailey type $c$ stars must be ``fundamentalized". For 
 this purpose, we assumed that the ratio of the period of the first overtone 
 ($c$ type) to that of the fundamental ($ab$ type) is 0.745 (Clement et al.,
 2001). 

  (3) The infrared magnitude ($M_{K}$) can be derived from $(V-K)_{0}$ and 
     the metallicity [Fe/H]:  
\begin{eqnarray}  
      M_K = 1.166 + 0.18 [Fe/H] - 1.812 (V-K)_0  \nonumber    \\ 
            + 0.675(V-K)_0^2 - 0.183(V-K)_0^3   \nonumber
\end{eqnarray}
    If [Fe/H] is not known, it is assumed to be $-$1.6. In this case the 
    relation is the same as that used in method (5) for the BHB stars. 

  In a recent review, Feast (2011) has shown that the current calibration 
  of RR Lyrae absolute magnitudes is not satisfactory: there is a 
  significant spread in the coefficients and zero points derived from 
  trigonometric parallaxes, statistical parallaxes and pulsation parallaxes
  It is hoped that, in the future,
  new trigonometric parallaxes such as those recently given by
  Benedict et al. (2011) will eventually improve this situation. 
   The expressions for the RR Lyrae absolute magnitudes given above are 
  the best that we have available now but may well require some correction
  in the future. 

  Distances $D_{a}$, $D_{b}$ and $D_{c}$ were derived for our RR Lyrae
  stars using methods (1), (2) and (3) respectively and are given in 
 Table B2. There are 31 of these RR Lyrae stars that have both radial velocities
  and proper motions and that are closer than 17 kpc (the limit that we have 
taken for the proper motions to yield meaningful velocities). Of these 31 stars,
  19 have known [Fe/H] and 26 have $K$ magnitudes. As noted above, distances can
  be derived even if [Fe/H] is not known by assuming that it is --1.6, although
  this involves a loss of accuracy. We need the distances of the RR Lyrae stars to be
  as closely as possible on the same scale as that of the BHB stars. Now method (3)
  for the RR Lyraes (with [Fe/H] = --1.6) is the same as method (5) for the BHB stars
  (with $(B-V)_{0}$ = +0.18). We have therefore converted the distances D$_{a}$
  and D$_{b}$ to the scale of the distances D$_{c}$ by dividing them by the factors
  R$_{a}$ and R$_{b}$ respectively. R$_{a}$ is the mean value of D$_{a}$/D$_{c}$
  and equals 1.0228$\pm$0.0005. R$_{b}$ is the mean value of D$_{b}$/D$_{c}$ and
  equals 1.0468$\pm$0.0017. After we have divided the distances D$_{a}$ by 1.0228
  and the distances D$_{b}$ by 1.0468, they will be on the scale of D$_{c}$ which
  is the same as the BHB scale D$_{5}$. Now the BHB stars were all adjusted to be on
  the scale of BHB distance D$_{2}$. To get the RR Lyrae star's distances  on this 
  scale, they must further be divided by the factor F5 (see Sec. A3). This factor 
  (F5) must be evaluated at the blue edge of the instability gap ($(B-V)_{0}$ = 
  +0.18) where it has the value 0.953. 
  We call the final values of these RR Lyrae distances $d_{a}$, $d_{b}$ and
  $d_{c}$. All these distances and our adopted distance (D) which is the unweighted
  mean of d$_{a}$, d$_{b}$ and d$_{c}$ are given in Table B2. 

  For the 35 RR Lyrae stars where all three distances are available,
  our adopted distance is the arithmetic mean of the three distances and 
  $\sigma$ is the $rms$ scatter of a single distance. In Fig. B3  we have
  plotted $\sigma$/D against D for (a) the 9 stars for which [Fe/H] and
 $\langle K \rangle$ are best determined and (b) for the remaining 26 stars
  for which $\sigma$ is available. We see that $\sigma$/D is roughly 
 independent of distance and less than 0.06 except for the 7 numbered stars
 in Fig B3(b). It seems likely that the larger $\sigma$/D of these stars is
  produced by larger errors in $\langle K \rangle$. In the cases where only
  $d_{a}$ is available we therefore conservatively adopted $\sigma$ = 
  0.05 D for its error.
   Some allowance must be made for systematic errors in our
  distance scale and this can only be a rough estimate based on the spread
  amongst the various distance estimates that we have used. In 
  calculating the space motions we have included a systematic error of
  0.015 D in quadrature with the random errors given in Tables A1 and B2.

  Our distances for TT Lyn, TW Lyn and X LMi (0.71$\pm$0.01; 1.69$\pm$0.02;
  2.20$\pm$0.01 kpc) are in satisfactory agreement with those given in the 
recent compilation of bright RR Lyrae stars  by Maintz (2005) (0.71; 1.65; 
  2.20 kpc).

\pagebreak

%Figure B2
\begin{figure*}
\includegraphics[width=17.0cm, bb=20 325 575 650, clip=true]{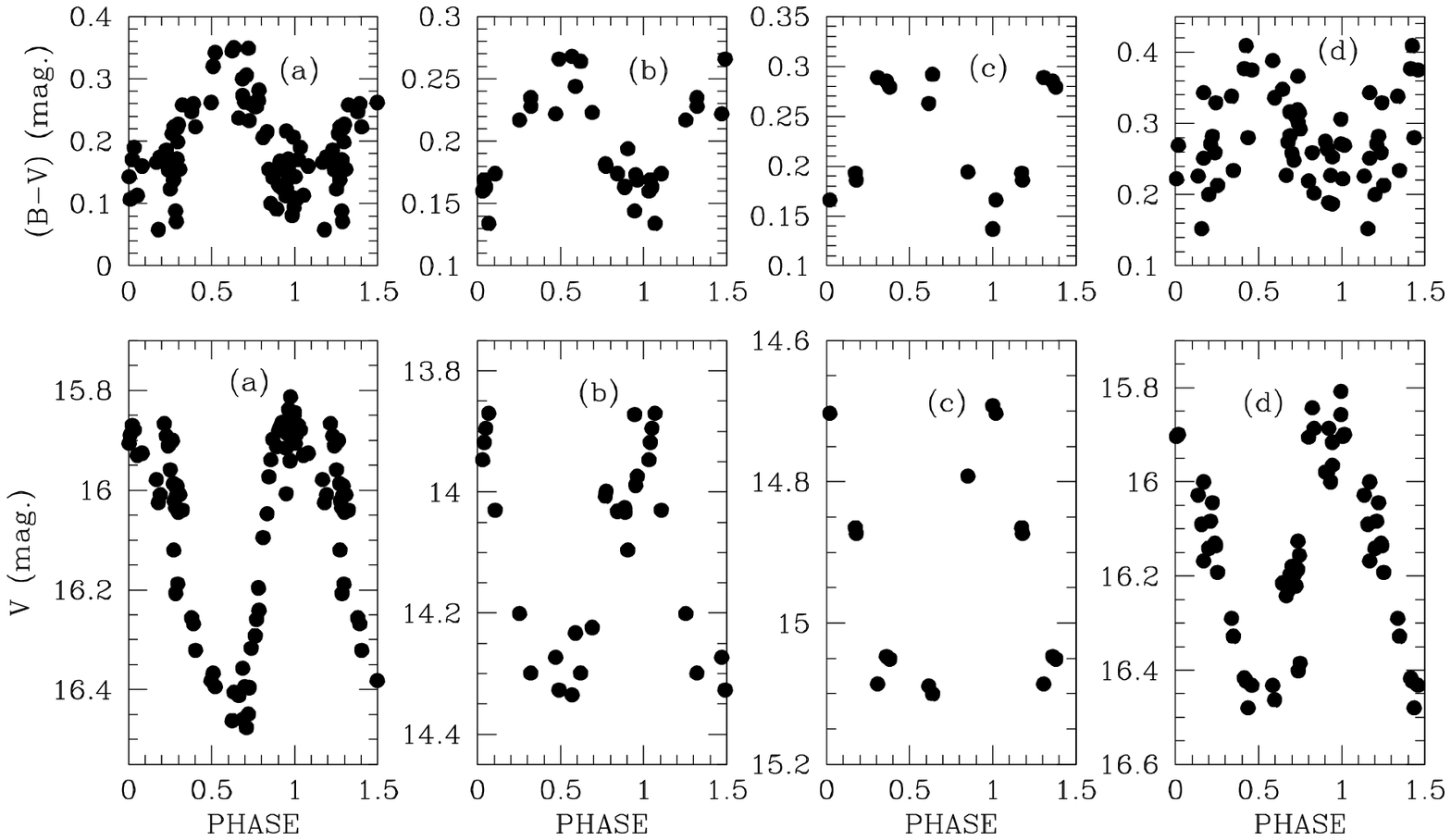}
\caption{ The $V$ and $(B-V)$ light curves for the RR Lyrae variable:     
     (a) RR7-101, (b) AF-400, (c) AF-430 and (d) RR7-086.                 
      Further information on these
      stars is given in Table B1.   
}
\label{f:B2}
\end{figure*}

% Figure B3
\begin{figure}
\includegraphics[width=8.5cm, bb=75 150 500 665, clip=true]{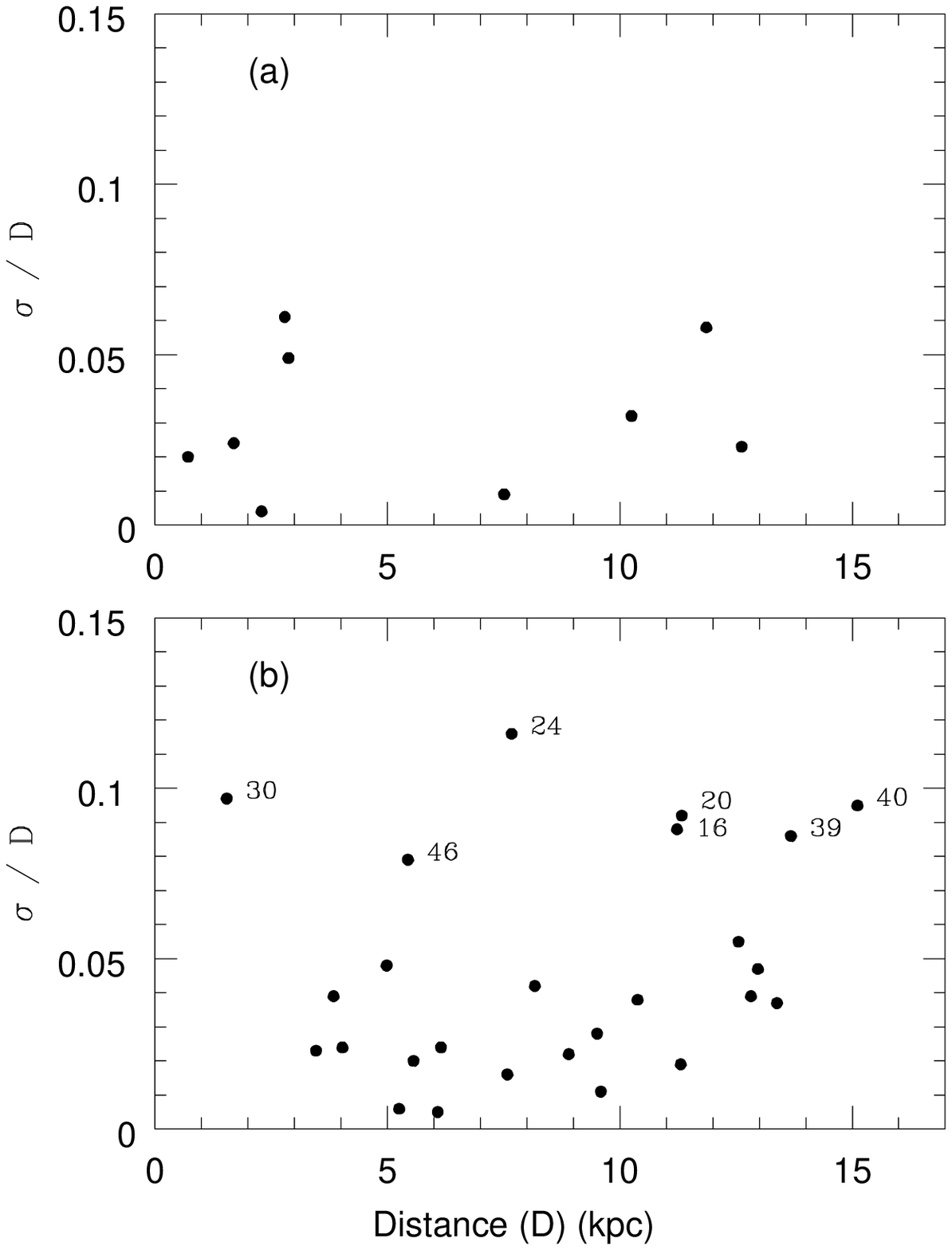}
\caption{ The $rms$ scatter ($\sigma$) in distance estimates divided
     by the adopted distance (D) in kpc (ordinate) $vs$ D (abscissa).
     (a) Nine stars with best determined [Fe/H] and $\langle K \rangle$.  
     (b) Twenty six remaining stars for which $\sigma$ is available. The numbers
     identify the stars in Table 2. 
      Further information on these
      stars see text.
}
\label{f:B3}
\end{figure}

\section{Proper Motions}

%Figure C1
\begin{figure*}
\includegraphics[width=17.0cm, bb=60 235 560 470, clip=true]{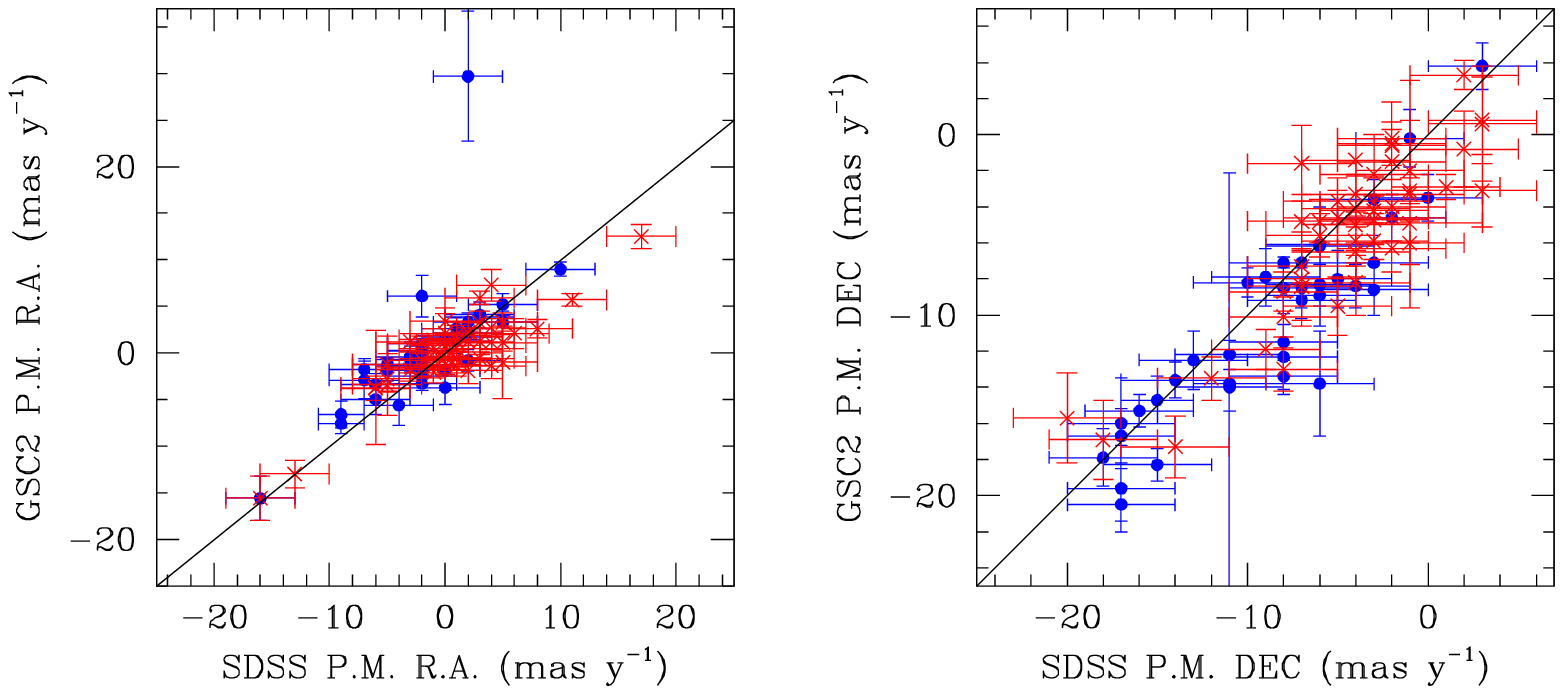}
\caption{ A comparison between the GSCII-SDSS proper motions (ordinate) 
  and those given in the SDSS DR7 catalogue (abscissa). The plot on the 
 left is for proper motions in R.A. and that on the right is for proper 
 motions in  Declination. The units are milliarcseconds per year. The 
 proper motions of the BHB stars are shown by blue filled circles and 
    those of the RR Lyrae stars by red crosses.
}
\label{f:C1}
\end{figure*}

The proper motions used in this paper come from astrometric data that were 
 assembled from the Second Guide Star Catalog (GSC-II) Lasker et al. (2008)
  and the Seventh Data Release of the Sloan Digital Sky Survey
SDSS DR7 Abazajian et al. 2009; Yanny et al. 2009). Absolute proper motions
were obtained by correcting the relative proper motions to a reference frame
provided by a Large Quasar Reference Frame assembled by Andrei et al. (2009).
Proper motions were computed for 77 million sources by combining SDSS second-epoch 
 positions with multi epoch positions derived from the GSC-II database and spanning
 a time baseline of 40 to 50 years. As described in Spagna et al. (2010), proper
 motion formal errors are typically in the range 2 to 3 mas y$^{-1}$ at 
 intermediate magnitudes ($16<r<18.5$). Comparisons against a sample of 80\,000 QSO
 indicate that the random errors of the two catalogues are, on average, comparable
 but that the reference system of the SDSS proper motions is affected by a global 
 systematic rotation $\Delta\mu\simeq --0.40$ mas y$^{-1}$, which is not 
 present in the GSC-II frame. 
  Further details concerning the data base from which our proper motions were 
  obtained may be found in Spagna et al. (2010a; 2010b).

  Overall, we consider these proper motions to be the most accurate available 
  for the magnitude range, $14\la V \la 17$, covered by the bulk of our objects.
  It is important, however, to compare our proper motions with those from 
  other catalogs because their proper motions may be preferable for the
  brightest stars or for a rare case where the GSC-II + SDSS DR7 error is 
  unusually large.

\subsection{Proper Motions for the Brighter or Anomalous Stars.}

 Table C1 gives the GSC-II---SDSS proper motions for our brighter program stars
 together with those given by the NOMAD catalogue (Zacharias et al., 2004),
 the UCAC3 catalogue (Zacharias et al., 2009) and the SDSS DR7 catalogue
 (Abazajian et al., 2009). 
Following this comparison, we have chosen to use the NOMAD proper motions for
 the brighter BHB stars RR7-15, RR7-64, BS 16473-09 and BS 16927-22 and for
 the brighter RR Lyrae stars TT Lyn, X LMi and DQ LYN 
 \footnote{The NOMAD catalogue was chosen 
 because the proper motion that it gives for a given object is the one which is 
 preferred from among a number of major catalogues (all of which are on the 
 International Celestial Reference System).}. 
 These are all stars whose $V$ magnitudes are brighter than 12.3

 Fig C1 gives separate plots for the proper motions in R.A. and Declination for
 the GSC-II---SDSS against those given by the SDSS DR7. Both catalogues are 
 based on quite similar material (i.e. first epochs from the POSS photographic
 plates and second epochs from the SDSS measurements), but they were processed 
 and calibratedly in different and independent ways. The plot shows good 
 agreement in general at the milliarc second per year level except for the 
 proper motion in R.A. for the BHB star P 30-38. We have preferred the
 SDSS DR7 proper motion because it roughly agrees with that given by the 
 NOMAD catalogue and because the errors in the GSC-II---SDSS catalogue for this
 star are unusually large. 

% Table C1
\begin{center}
\begin{table*}[h]
\caption{A Comparison of the proper motions from different sources for the 
        brightest of our program stars. \label{t:C1}
	}
%\begin{flushleft}
\begin{tabular}{@{}lccccc@{}}
\noalign{\smallskip} \hline
 Star  & $\langle V\rangle$& SDSS~(DR7)& NOMAD & UCAC3 &GSCII-SDSS   \\
       &       & $\mu_{\alpha}$~~~~~~~~$\mu_{dec}$& $\mu_{\alpha}$~~~~~~~~$\mu_{dec}$&
         $\mu_{\alpha}$~~~~~~~~$\mu_{dec}$& $\mu_{\alpha}$~~~~~~~~$\mu_{dec}$ \\
       & (mag) &  mas y$^{-1} $~~~~mas y$^{-1}$&mas y$^{-1}$~~~~mas y$^{-1}$&
       mas y$^{-1} $~~~~mas y$^{-1}$&mas y$^{-1}$    ~~~~mas y$^{-1}$\\
\hline \hline
 RR7-15 & 11.75 & ...~~~~~~~~... &--20.2$\pm$1.0~~~--35.0$\pm$0.7 &
                             --19.1$\pm$2.0~~~--36.0$\pm$0.8 &
                             --12.0$\pm$3.0~~~--43.5$\pm$1.7 \\
 RR7-23 & 12.63 & ...~~~~~~~~... & +07.7$\pm$1.6~~~--22.6$\pm$0.7 &
                              +07.6$\pm$1.3~~~--22.0$\pm$0.6 &
                              +09.6$\pm$1.7~~~--25.0$\pm$1.4 \\
 RR7-64 & 11.23 & ...~~~~~~~~...         &--05.1$\pm$0.7~~~--05.4$\pm$0.7 &
                             --06.0$\pm$1.6~~~--05.0$\pm$1.0 &
                             --02.4$\pm$2.9~~~--00.2$\pm$1.6 \\
11419-01&12.79 & --02$\pm$3~~~--15$\pm$3  &--08.0$\pm$1.0~~~--20.1$\pm$0.7 &
                             --08.3$\pm$3.1~~~--17.5$\pm$1.4 &
                              +06.1$\pm$2.2~~~--18.3$\pm$0.9 \\
16473-09& 10.91 & ...~~~~~~~~... &--01.7$\pm$0.6~~~--00.1$\pm$0.7 &
                             --02.2$\pm$0.9~~~--00.3$\pm$0.6 &
                              +10.7$\pm$3.1~~~--17.9$\pm$5.5 \\
 P 30-38& 14.39 &  +02$\pm$3~~~--11$\pm$3  & +04.1$\pm$5.9~~~--11.7$\pm$5.6 &
                                ...        ~~~~~~~~  ...         &
                              +29.7$\pm$7.0~~~--14.0$\pm$11.9 \\
16927-22& 11.18 & ...~~~~~~~~... & +06.8$\pm$0.8~~~--29.8$\pm$0.6 &
                              +06.0$\pm$0.9~~~--30.3$\pm$0.7 &
                              +09.2$\pm$0.6~~~--30.0$\pm$1.2 \\
RW LYN  &12.91 &  +03$\pm$3~~~--18$\pm$3  & +06.3$\pm$0.7~~~--23.4$\pm$1.2 &
                              +07.5$\pm$1.0~~~--23.6$\pm$0.8 &
                              +07.3$\pm$1.7~~~--15.7$\pm$2.5 \\
TT LYN  & 09.84 & ...~~~~~~~~... &--81.9$\pm$1.5~~~--41.8$\pm$0.9 &
                             --84.6$\pm$0.8~~~--42.4$\pm$0.7 &
                             --50.2$\pm$13.6~~~--25.6$\pm$9.2 \\
TW LYN  & 12.07 & ...~~~~~~~~... &--06.6$\pm$5.0~~~--01.1$\pm$3.2 &
                              +03.2$\pm$0.8~~~  +04.3$\pm$0.7 &
                              +00.6$\pm$3.8~~~  +02.6$\pm$2.9 \\
 X LMi  & 12.30 & ...~~~~~~~~... & +07.8$\pm$1.3~~~--17.3$\pm$0.7 &
                              +06.6$\pm$1.0~~~--16.7$\pm$1.0 &
                              +12.4$\pm$3.2~~~--14.0$\pm$5.1 \\
DQ LYN  & 11.46 & ...~~~~~~~~... &--01.9$\pm$0.8~~~--28.7$\pm$1.0 &
                             --01.0$\pm$1.0~~~--29.7$\pm$2.7 &
                              +02.9$\pm$1.8~~~--20.5$\pm$2.8 \\
RR7-079 &13.44 & --02$\pm$3~~~--12$\pm$3  &--15.5$\pm$3.8~~~--18.7$\pm$3.6 &
                             --17.8$\pm$4.3~~~--18.9$\pm$4.1 &
                              +01.0$\pm$1.1~~~--13.5$\pm$1.2 \\
\hline
\end{tabular}
\end{table*}
\end{center}

\section{Thick Disc}

  Fig. D1 shows the  L$_{\perp}$ $vs$ L$_{z}$ plot for the 
 thick disc stars within 2 kpc from Bensby et al. (2011) (crosses) 
 and the thick disc stars within 0.5 kpc (Ages 8 to 11 Gyr and --0.4 $<$
[Fe/H] $<$ --0.5) from the Geneva-Copenhagen Survey of the Solar 
 Neighbourhood III (Holmberg et al., 2009)(small black open circles);                       
  Most of these stars have L$_{\perp}$ $<$
 650 kpc km s$^{-1}$ and 1100 $<$ L$_{z}$ $<$ 2000 kpc km s$^{-1}$ ; we have used
 these limits  to define the location of stars that belong to the thick disc.   
\footnote{Only one disc RR Lyrae (TW Lyn) is known to lie outside this location}. 
 There is a strong concentration of local RR Lyrae stars (red circles) in 
 this location but rather few local BHB stars (Fig D1); 
  these stars are listed in Table D1.  
 Table D2 gives the mean properties of the 46 RR Lyrae stars in this thick
 disc location: 5 stars (11\%) have [Fe/H] $\leq$ --1.50 and have a high   
 probability of being 
 {\it halo} stars; 19 (41\%) have [Fe/H] $>$ --0.5 and can 
 definitely be called {\it disc} stars. The remaining 22 (48\%) have 
 intermediate [Fe/H] (--0.5 $>$ [Fe/H] $>$ --1.5) and mean values of their
 orbital eccentricity and maximum orbital 
 height above the galactic plane (z$_{max}$) and angular momentum
 L$_{\perp}$ that lie between those of the stars with [Fe/H] $>$ --0.5
 and those with [Fe/H] $<$ --1.5. 
 All of the 19  RR Lyrae stars with [Fe/H] $>$ --0.7 and 73\% of the 22
 with --0.7 $>$ [Fe/H] $>$ --1.5 have L$_{\perp}$ $<$ 325 kpc km s$^{-1}$ ;     
consequently $\sim$82\% of the likely thick disc stars, those with 
  [Fe/H] $>$ --1.5, should have L$_{\perp}$ $<$ 325 kpc km s$^{-1}$ 
  and 1100 $<$ L$_{z}$ $<$ 2000 kpc km s$^{-1}$. The isolation of
 a purer sample of disc stars requires additional chemical or kinematic 
 information. Only three local BHB stars (within 1 kpc) have L$_{\perp}$ $<$
 650 kpc km s$^{-1}$ and 1100 $<$ L$_{z}$ $<$ 2000 kpc km s$^{-1}$ and two of
 these have [Fe/H] $<$ --2.0 and so are almost certainly {\it halo} stars.
 Thus, unlike RR Lyrae stars, BHB stars do not have a strong {\it disc} 
 component in the solar neighbourhood as was found by Kinman et al. (2009).

% Figure D1
\begin{figure*}
\includegraphics[width=15.0cm, bb=30 390 500 680, clip=true]{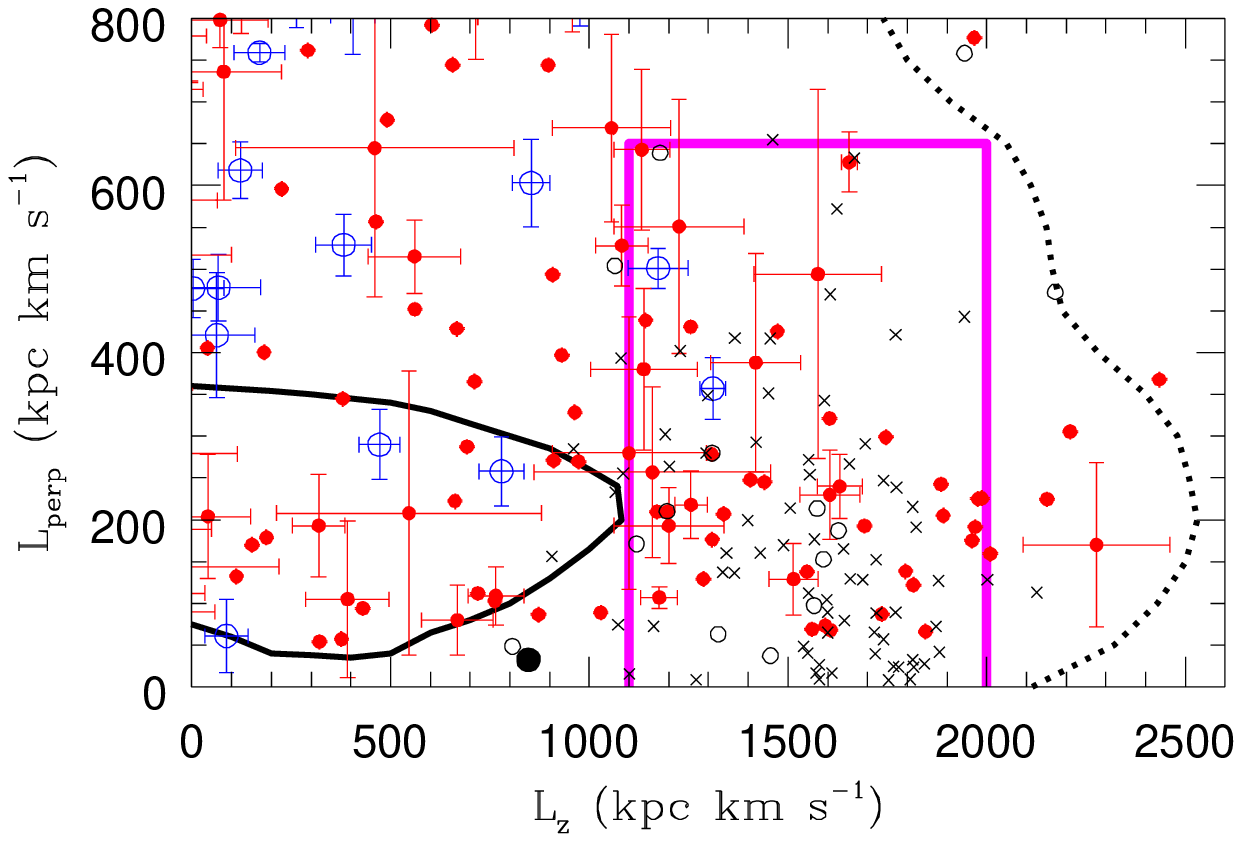} 
\caption{Angula Momenta L$_{\perp}$ $vs$ L$_{z}$. RR Lyrae stars within 1 kpc
are shown by red filled circles. Open circles are stars within 500 pc, Age 
8 to 11 Gyr, $-0.4 >$[Fe/H]$ > -0.5$ taken fromm the Geneva - Copenhagen Survey
(Holmberg et al., 2009). Crosses are thick disc stars within 2 kpc taken from
Bensby et al. (2011). The large black filled circle is the star Arcturus.
 The black full and dotted contours are taken from Fig 3 of Morrison et al.
 (2009). The magenta box shows the adopted location of the Thick Disc stars.
}
\label{f:D1}
\end{figure*}

% Table D1
\begin{center}
\begin{table*}
\caption{Candidates for Thick Disc RR Lyrae and BHB stars with distances (D) within 2 kpc.
\label{t:D1}} 
%\begin{flushleft}
\begin{tabular}{@{}lccccccclcccccc@{}}
\noalign{\smallskip} \hline
  D $<$ 1 kpc &  &  &  & &  &  &  &  1 $<$ D $<$ 2 kpc &   &   &  &  &  &  \\
 Star  & [Fe/H] & S$^{\dagger}$ & Ecc. & z$_{max}$ & L$_{\perp}$ & L$_{z}$&  &
 Star  & [Fe/H] & S$^{\dagger}$ & Ecc. & z$_{max}$ & L$_{\perp}$ & L$_{z}$ \\ 
\hline \hline
 RR LYRAE:&      &   &    &    &   &    & &RR LYRAE:    &   &   &   &   &   \\
   FW LUP &--0.20&(1)&0.06&0.13&068&1607& &CN LYR&--0.58&(1)&0.16&0.32& 66&1846\\
   KX LYR &--0.46&(2)&0.31&0.41&069&1561& &CG PEG&--0.50&(1)&0.03&0.39& 87&1736\\
   DX DEL &--0.50&(1)&0.22&0.23&073&1594& &HH PUP&--0.50&(1)&0.12&0.35&122&1815\\
   SW DRA &--1.12&(1)&0.42&0.71&107&1176& &ST OPH&--1.30&(2)&0.10&0.56&129&1514\\
   AV PEG &--0.36&(1)&0.27&0.39&129&1287& &TW HER&--0.88&(1)&0.13&0.61&139&1796\\
 v494 SCO &--1.01&(2)&0.12&0.21&138&1548& &SS CNC&--0.56&(1)&0.18&0.83&159&2009\\
   RS BOO &--0.62&(1)&0.17&0.94&191&1971& &TW LYN&--0.66&(1)&0.21&2.13&170&2276\\
   SW AND &--0.41&(1)&0.13&0.47&193&1692& &TZ AQR&--1.24&(1)&0.20&1.39&193&1200\\
   DH PEG &--1.24&(2)&0.27&0.52&207&1338& & Z MIC&--1.10&(1)&0.39&1.25&210&1196\\
   AR PER &--0.57&(1)&0.06&0.40&226&1987& &UW OCT&--0.49&(2)&0.47&1.58&210&1170\\
   UY CYG &--0.95&(1)&0.11&0.23&240&1630& &EZ LYR&--1.29&(1)&0.42&0.65&176&1309\\
   AN SER &--0.39&(1)&0.10&0.84&245&1440& &RR LEO&--1.51&(1)&0.38&1.04&218&1256\\
 v690 SCO &--1.16&(2)&0.19&0.48&248&1405& &RR GEM&--0.55&(1)&0.18&0.73&225&2152\\
   DM CYG &--0.57&(1)&0.13&0.48&205&1890& &AA CMI&--0.15&(1)&0.09&0.43&225&1978\\
   XZ DRA &--0.89&(1)&0.10&0.68&299&1746& &AV SER&--1.20&(3)&0.11&0.86&230&1605\\
 v445 OPH &--0.19&(2)&0.11&0.81&388&1419& &RW ARI&--1.16&(2)&0.26&1.44&238&1639\\
    T SEX &--1.33&(1)&0.30&1.18&431&1255& & U PIC&--0.72&(1)&0.09&1.01&242&1885\\
   UU VIR &--0.87&(2)&0.52&1.71&439&1142& &BS APS&--1.33&(2)&0.30&1.12&264&1852\\
   RX ERI &--1.33&(1)&0.00&1.67&628&1654& &CP AQR&--0.77&(1)&0.31&1.07&278&1308\\
    W CVN &--1.22&(1)&0.29&2.47&643&1132& &RW DRA&--1.55&(2)&0.60&1.11&280&1100\\
          &      &   &    &    &   &    & &RW TRA&--0.13&(1)&0.14&0.78&321&1605\\
   BHB:   &      &   &    &    &   &    & &AT SER&--2.03&(1)&0.16&1.47&380&1137\\
  HD 4850 &--1.18&(4)&... &... &357&1311& &BR AQR&--0.74&(1)&0.11&1.65&426&1473\\
  HD 8376 &--2.82&(4)&... &... &501&1173& &VV PEG&--1.88&(2)&0.26&2.55&495&1575\\
BD +01 0548&--2.23&(4)&...&... &299&1794& &RV SEX&--1.10&(2)&0.40&2.14&551&1226\\
           &      &   &    &    &   &    & &BO AQR&--1.80&(2)&0.46&1.82&257&1159\\
\hline
\end{tabular}
     \\
$^\dagger$ Sources of [Fe/H]: (1) Feast et al. (2008); (2) Beers et al. (2000);
   (3) Layden et al. (1994); (4) Kinman et al. (2009).\\  
     The orbital eccentricity (Ecc.) and the maximum height in kpc above the galactic plane\\ (z$_{max}$) are taken from Maintz \& de Boer (2005).~~~~
L$_{\perp}$ and L$_{z}$ are in units of kpc km s$^{-1}$. \\
\end{table*}
\end{center}

% Table D2
\begin{center}
\begin{table*}
\caption{Mean properties of RR Lyrae stars with 1100$<$L$_{z}$$<$2000 kpc km s$^{-1}$
 and L$_{\perp}$ kpc km s$^{-1}$ as a function of [Fe/H]. \label{t:D2}
 }
%\begin{flushleft}
\begin{tabular}{@{}cccccc@{}}
\noalign{\smallskip} \hline
Range in [Fe/H] & Number &$\langle$[Fe/H]$\rangle$ & $\langle$Ecc.$\rangle$ &
 $\langle$z$_{max}$$\rangle$ & $\langle$J$_{\perp}$$\rangle$  \\
       &of stars &    &     & kpc & km s$^{-1}$ kpc   \\
\hline \hline
[Fe/H]$\leq$--1.50  & 5&--1.75$\pm$0.11 &0.37$\pm$0.09&1.60$\pm$0.31&326$\pm$56  \\
--0.70$\leq$[Fe/H]$\geq$--1.50&22&--1.09$\pm$0.04&0.23$\pm$0.03&1.07$\pm$0.13 & 293$\pm$36 \\

[Fe/H]$\geq$--0.70& 19&--0.44$\pm$0.04 & 0.17$\pm$0.02&0.67$\pm$0.11& 177$\pm$21
 \\
\hline
\end{tabular}
     \\
    The orbital eccentricity (Ecc.) and the maximum height in kpc above the galactic plane (z$_{max}$) are taken from Maintz \& de Boer (2005)\\ 
\end{table*}
\end{center}

\section{The Angluar Momenta L$_{\perp}$ and L$_{z}$ for objects in Figures 
   4, 5 \& 6.}

% Tables E1, E2, E3 \& E4 
Tables \ref{t:E1}, \ref{t:E2}, \ref{t:E3} \& \ref{t:E4}
give the Anglar Momenta L$_{\perp}$ and L$_{z}$ for the 
RR Lyrae stars at the North Galactic Pole (Table 2 in Kinman et al., 2007b), 
the BHB  stars at the North Galactic Pole (Table 1 in Kinman et al., 2007b), 
Local BHB stars within 1 kpc (Kinman et al., 2000) using proper motions taken
from the NOMAD catalogue (Zacharias et al., 2009) and Globular Clusters within
10 kpc. We have only included RR Lyrae and BHB stars at the North Galactic Pole
whose distances are less than 10 kpc.

 Definitions of L$_{\perp}$ and L$_{z}$ are given in the Appendix of Kepley 
 et al. (2007). The calculations of these quantities and their errors 
 were made using a 
 program kindly made available by Heather Morrison and modified for our use
 by Carla Cacciari. 
 %Table E5 
 Table \ref{t:E5} compares the values of L$_{\perp}$ and L$_{z}$
 that we calculated with this program for a number of halo objects with those
 given by Re Fiorentin et al. (2005) and Morrison et al. (2009). The differences
 are generally less than the quoted errors and are probably largely produced
 by differences in the proper motions that were used. An exception is 
 HD 128279, where our values of L$_{\perp}$ and L$_{z}$ agree with those of
 Re Fiorentin et al. (2005) but not with those of Morrison et al. (2009).

\begin{center}
\begin{table*}
\caption{Angular Momenta L$_{\perp}$ and L$_{z}$ for RR Lyrae stars at the North
  Galactic Pole$^{\dagger}$. \label{t:E1}
  } 
%\begin{flushleft}
\begin{tabular}{@{}lccclccclcc@{}}
\noalign{\smallskip} \hline
 Star  &  L$_{\perp}$ & L$_{z}$&  &
 Star  &  L$_{\perp}$ & L$_{z}$&  &
 Star  &  L$_{\perp}$ & L$_{z}$\\           
\hline \hline
NSV5476 &1074$\pm$440&--1005$\pm$497& & 
SV-CVn  &1059$\pm$228&--0827$\pm$238& & 
IP-Com  &1878$\pm$780&--1860$\pm$780 \\
V-Com   &1001$\pm$454& +0902$\pm$1285& & 
FV-Com  &0786$\pm$369& +0611$\pm$497& & 
SA57-019&1041$\pm$389&--1089$\pm$438 \\
CD-Com  &6464$\pm$768&--4461$\pm$795& & 
U-Com   &0690$\pm$057&--0723$\pm$160& & 
EO-Com  &1743$\pm$483&--1889$\pm$528 \\
AF-031  &3935$\pm$710&--2075$\pm$1170& & 
SW-CVn  &0469$\pm$210& +0153$\pm$313& & 
TZ-CVn  &0903$\pm$368&--1010$\pm$595 \\
TU-Com  &0951$\pm$628&--1218$\pm$1459& & 
DV-Com  &1406$\pm$595&--0718$\pm$721& & 
SA57-047&2809$\pm$515&--3615$\pm$516 \\
CK-Com  &2207$\pm$407&--2985$\pm$459& & 
AF-791  &4599$\pm$544& +1355$\pm$442& & 
MQ-Com  &1572$\pm$393&--1895$\pm$501 \\
CL-Com  &0868$\pm$592& +0165$\pm$476& & 
TX-Com  &1101$\pm$405&--0993$\pm$395& & 
IS-Com  &1472$\pm$177&--1642$\pm$338 \\
AT-CVn  &1103$\pm$666&--0409$\pm$1106& & 
AP-CVn  &0875$\pm$419&--0254$\pm$359& & 
AF-882  &0636$\pm$315&--0445$\pm$390 \\
RR-CVn  &0465$\pm$118&--1375$\pm$336& & 
AF-155  &1078$\pm$548&--0310$\pm$739& & 
        &            &               \\
S-Com   &0560$\pm$083& +0147$\pm$260& & 
TY-CVn  &0802$\pm$263&--0747$\pm$471& & 
        &            &               \\
\hline
\end{tabular}
     \\
$^\dagger$ The stars are identified in Table 2 of Kinman et al. (2007b);~~
L$_{\perp}$ and L$_{z}$ are in units of kpc km s$^{-1}$.     \\
\end{table*}
\end{center}

\begin{center}
\begin{table*}
\caption{Angular Momenta L$_{\perp}$ and L$_{z}$ for BHB stars at the    
  North Galactic Pole$^{\dagger}$. \label{t:E2}
  } 
%\begin{flushleft}
\begin{tabular}{@{}lccclccclcc@{}}
\noalign{\smallskip} \hline
 Star  &  L$_{\perp}$ & L$_{z}$&  &
 Star  &  L$_{\perp}$ & L$_{z}$&  &
 Star  &  L$_{\perp}$ & L$_{z}$\\           
\hline \hline
16549-51&0605$\pm$363&--0197$\pm$234& & 
AF-078  &0990$\pm$209& +1159$\pm$314& & 
SA57-001&1430$\pm$342&--1742$\pm$503 \\
AF-006  &0499$\pm$164&--0973$\pm$349& & 
AF-769  &0646$\pm$189&--1423$\pm$408& & 
SA57-006&0404$\pm$191&--0280$\pm$412 \\
AF-727  &0916$\pm$159& +1812$\pm$323& & 
16026-67&0523$\pm$153&--0946$\pm$443& & 
SA57-007&0590$\pm$343&--0135$\pm$215 \\
AF-729  &1193$\pm$577&--1024$\pm$717& & 
AF-100  &0410$\pm$149&--0656$\pm$325& & 
SA57-017&0810$\pm$253&--1037$\pm$492 \\
AF-029  &1080$\pm$450& +1127$\pm$591& & 
16466-08&0742$\pm$480&--0298$\pm$555& & 
SA57-032&1621$\pm$604&--1514$\pm$670 \\
AF-030  &0800$\pm$392&--0902$\pm$514& & 
AF-108  &1279$\pm$224&--2185$\pm$370& & 
SA57-040&0794$\pm$320& +0719$\pm$358 \\
16022-26&0708$\pm$350&--0574$\pm$571& & 
AF-111  &1796$\pm$308& +1059$\pm$394& & 
SA57-045&1025$\pm$405& +0035$\pm$379 \\
AF-038  &2494$\pm$824&--1878$\pm$1130& & 
AF-112  &0509$\pm$091& +1126$\pm$180& & 
AF-825  &1082$\pm$407& +0263$\pm$247 \\
AF-041  &1668$\pm$511&--1928$\pm$635& & 
AF-113  &0284$\pm$067&--0940$\pm$159& & 
AF-841  &0303$\pm$157&---391$\pm$273 \\
AF-045  &0555$\pm$075& +1342$\pm$180& & 
AF-115  &1571$\pm$457&--1484$\pm$494& & 
AF-848  &0716$\pm$085& +0648$\pm$176 \\
AF-048  &0230$\pm$123& +0118$\pm$314& & 
16466-15&1220$\pm$154& +1194$\pm$453& & 
SA57-066&1856$\pm$420& +0428$\pm$601 \\
AF-052  &1583$\pm$776&--0621$\pm$785& & 
AF-131  &1329$\pm$612&--0759$\pm$664& & 
AF-854  &1820$\pm$584&--0275$\pm$456 \\
AF-053  &2091$\pm$768&--1609$\pm$828& & 
AF-134  &0688$\pm$117& +0211$\pm$155& & 
SA57-080&0555$\pm$229& +0244$\pm$366 \\
16026-28&0746$\pm$094& +1411$\pm$179& & 
16031-44&0913$\pm$182& +0163$\pm$457& & 
AF-866  &0458$\pm$281&--0132$\pm$396 \\
AF-754  &1161$\pm$733&--0175$\pm$1120& & 
15622-48&0502$\pm$276&--0436$\pm$507& & 
AF-900  &1052$\pm$378& +0069$\pm$203 \\
AF-755  &1582$\pm$762&--0676$\pm$1068& & 
15622-07&0224$\pm$105& +0417$\pm$322& & 
AF-909  &0534$\pm$249& +0144$\pm$330 \\
AF-068  &1160$\pm$405& +0257$\pm$460& & 
AF-138  &1113$\pm$299&--0837$\pm$494& & 
AF-914  &1054$\pm$394&--0232$\pm$304 \\
AF-070  &1037$\pm$400&--0050$\pm$570& & 
15622-09&0652$\pm$113& +0379$\pm$262& & 
AF-916  &0772$\pm$284& +0803$\pm$274 \\
AF-075  &2398$\pm$809&--1208$\pm$760& & 
AF-797  &1318$\pm$417&--1194$\pm$519& & 
AF-918  &0439$\pm$247& +0261$\pm$318 \\
AF-076  &1741$\pm$720&--0524$\pm$599& & 
AF-804  &0397$\pm$153&--0258$\pm$312& & 
        &            &               \\
\hline
\end{tabular}
     \\
$^\dagger$ The stars are identified in Table 1 of Kinman et al. (2007b);~~
L$_{\perp}$ and L$_{z}$ are in units of kpc km s$^{-1}$. \\
\end{table*}
\end{center}

\begin{center}
\begin{table*}
\caption{Angular Momenta L$_{\perp}$ and L$_{z}$ for Local BHB stars 
  $^{\dagger}$. \label{t:E3}
  } 
%\begin{flushleft}
\begin{tabular}{@{}lccclccclcc@{}}
\noalign{\smallskip} \hline
 Star  &  L$_{\perp}$ & L$_{z}$&  &
 Star  &  L$_{\perp}$ & L$_{z}$&  &
 Star  &  L$_{\perp}$ & L$_{z}$\\           
\hline \hline
HD 2857 &0421$\pm$075 & +0063$\pm$098& & 
HD 78913&0172$\pm$038 &--0769$\pm$040 & & 
HD117880&0319$\pm$108&--0701$\pm$168 \\
HD 4850 &0357$\pm$037 & +1311$\pm$032 & & 
HD 86968&0477$\pm$035 & +0003$\pm$078& & 
HD 128801&0828$\pm$037 & +0977$\pm$038  \\
HD 8376 &0501$\pm$024 & +1173$\pm$75 & & 
HD 87047&1124$\pm$037 &--0738$\pm$131& & 
HD 130095&0579$\pm$031 &--0184$\pm$167 \\
HD 13780&0258$\pm$041 & +0779$\pm$057 & & 
HD 87112&0399$\pm$48 &--0316$\pm$099& & 
HD 130201&0603$\pm$052 & +0854$\pm$47  \\
HD 14829&1215$\pm$045 & +0272$\pm$104& & 
HD 93329&0599$\pm$048 &--0814$\pm$078 & & 
HD 139961&0766$\pm$039 &--1211$\pm$162 \\
HD 31943&0290$\pm$042 & +0472$\pm$052 & & 
HD 106304&1305$\pm$064 &--0009$\pm$052 & & 
HD 161817&1033$\pm$018 &--0530$\pm$029 \\
HD 252940&0529$\pm$037 & +0382$\pm$070 & & 
BD +42$^{\circ}$2602&0835$\pm$046 & +0263$\pm$110& & 
HD 167105&0061$\pm$044 & +0087$\pm$054  \\
HD 60778&0888$\pm$051 & +0638$\pm$053  & & 
HD 109995&0759$\pm$011 & +0170$\pm$064 & & 
HD 180903&0618$\pm$034 & +0122$\pm$055  \\
HD 74721&0803$\pm$046 & +0405$\pm$055  & & 
BD +25$^{\circ}$ 2602&0478$\pm$040 & +0067$\pm$107& & 
HD 213468&1282$\pm$040&--0119$\pm$189 \\
\hline
\end{tabular}
     \\
$^\dagger$ The stars are taken from Table 1 of Kinman et al. (2000);~~
L$_{\perp}$ and L$_{z}$ are in units of kpc km s$^{-1}$. \\
\end{table*}
\end{center}

\begin{center}
\begin{table*}
\caption{Angular Momenta L$_{\perp}$ and L$_{z}$ for Galactic Globular Clusters 
  within 10 kpc$^{\dagger}$. \label{t:E4}
  } 
%\begin{flushleft}
\begin{tabular}{@{}lccclccclcc@{}}
\noalign{\smallskip} \hline
Cluster&  L$_{\perp}$ & L$_{z}$&  &
Cluster&  L$_{\perp}$ & L$_{z}$&  &
Cluster&  L$_{\perp}$ & L$_{z}$\\           
\hline \hline
NGC 104 &0618$\pm$064& +1169$\pm$051& & 
NBC 6093&0299$\pm$127&--0005$\pm$053& & 
NGC 6522&0199$\pm$079& +0057$\pm$215\\  
NGC 288 &0571$\pm$075&--0244$\pm$109& & 
NGC 6121&0073$\pm$024& +0136$\pm$074& & 
NGC 6553&0154$\pm$018& +0920$\pm$180 \\
NGC 362 &0377$\pm$146&--0225$\pm$132& & 
NGC 6144&0379$\pm$026&--0179$\pm$013& & 
NGC 6626&0145$\pm$031& +0480$\pm$113 \\
NGC 3201&1199$\pm$086&--2659$\pm$098& & 
NGC 6171&0358$\pm$065& +0367$\pm$111& & 
NGC 6656&0521$\pm$092& +0843$\pm$114 \\
NGC 4372&0596$\pm$095& +0813$\pm$054& & 
NGC 6205&2066$\pm$258&--0446$\pm$194& & 
NGC 6712&0383$\pm$054& +0109$\pm$125 \\
NGC 4833&0197$\pm$064& +0174$\pm$116& & 
NGC 6218&0334$\pm$034& +0758$\pm$074& & 
NGC 6723&0437$\pm$056&--0045$\pm$112 \\
NGC 5139&0114$\pm$023&--0439$\pm$020& & 
NGC 6254&0557$\pm$061& +0630$\pm$106& & 
NGC 6752&0336$\pm$018& +0971$\pm$068 \\ 
NGC 5272&1255$\pm$224& +0644$\pm$180& & 
NGC 6266&0178$\pm$022& +0231$\pm$102&  & 
NGC 6809&0759$\pm$040& +0134$\pm$075\\  
NGC 5904&1198$\pm$317& +0337$\pm$091 & & 
NGC 6304&0093$\pm$009& +0338$\pm$101& & 
NGC 6838&0097$\pm$047& +1161$\pm$020\\  
NGC 5927&0225$\pm$065& +1028$\pm$051& & 
NGC 6397&0764$\pm$022& +0503$\pm$043& & 
NGC 7009&0715$\pm$114&--0457$\pm$110\\   
\hline
\end{tabular}
     \\
$^\dagger$ 
L$_{\perp}$ and L$_{z}$ are in units of kpc km s$^{-1}$. They were derived
 from the data given by Vande Putte \& Cropper (2009).\\
\end{table*}
\end{center}

\begin{center}
\begin{table*}
\caption{Comparison of Angular Momenta L$_{\perp}$ and L$_{z}$ for Halo stars   
  derived from three sources$^{\dagger}$. \label{t:E5}
  } 
%\begin{flushleft}
\begin{tabular}{@{}cccccccccc@{}}
\noalign{\smallskip} \hline
     &\multicolumn{2}{c}{\it This Paper} & &\multicolumn{2}{c}{\it Morrison et al.(2009)}&  &\multicolumn{2}{c}{\it Re Fiorentin et al.
 (2005)}      \\
 Star  &  L$_{\perp}$ & L$_{z}$&  &
     L$_{\perp}$ & L$_{z}$&  &
     L$_{\perp}$ & L$_{z}$\\           
\hline \hline
UY CYG  &0113$\pm$048& +1670$\pm$061& & 
         0240$\pm$038& +1630$\pm$057& & 
            ...      &     ...       \\
VZ HER  &0196$\pm$072& +0316$\pm$075& & 
         0193$\pm$061& +0319$\pm$066& & 
              ...    &    ...        \\
RV SEX  &0421$\pm$131& +1279$\pm$199& & 
         0551$\pm$152& +1226$\pm$164& & 
             ...     &     ...       \\
SW BOO  &0686$\pm$097&--0022$\pm$155& & 
         0878$\pm$175& +0616$\pm$394& & 
             ...     &      ...      \\
AN LEO  &1232$\pm$123&--0201$\pm$250& & 
         1014$\pm$234& +0220$\pm$372& & 
             ...     &    ...        \\
V LMI   &0377$\pm$116& +0142$\pm$154& & 
         0208$\pm$170& +0546$\pm$334& & 
             ...     &     ...       \\
X LMI   &0223$\pm$064& +1005$\pm$080& & 
         0255$\pm$176& +0913$\pm$329& & 
            ...      &     ...       \\
UV VIR  &0356$\pm$130&--0510$\pm$196 & & 
         0144$\pm$126&--0070$\pm$290& & 
            ...      &     ...       \\
TT CNC  &2027$\pm$238& +0797$\pm$128& & 
         2123$\pm$225& +0718$\pm$119& & 
         2483$\pm$...& +0562$\pm$... \\
AR SER  &2082$\pm$135& +0656$\pm$196& & 
         2017$\pm$108& +0637$\pm$170& & 
         1934$\pm$...& +0322$\pm$... \\
TT LYN  &1959$\pm$129& +0838$\pm$074& & 
         2032$\pm$136& +0847$\pm$072& & 
         2124$\pm$...& +0748$\pm$... \\
HD 128279&2075$\pm$320& +1046$\pm$188& & 
          0475$\pm$050& +2220$\pm$030 & &
         1844$\pm$...&  +1194$\pm$...\\
HD 237846&1546$\pm$032& +0931$\pm$038& & 
          1568$\pm$026& +0948$\pm$033 &  &
         1443$\pm$...&  +0774$\pm$...\\
HD 214161&2181$\pm$075& +0941$\pm$064& & 
         2125$\pm$087&  +0836$\pm$134 &  &
         2173$\pm$...&  +0952$\pm$...\\
\hline
\end{tabular}
     \\
$^\dagger$ 
L$_{\perp}$ and L$_{z}$ are in units of kpc km s$^{-1}$. \\
\end{table*}
\end{center}

\newpage

\section{Groups of $outliers$.}

 This section summarizes additional data about the two groups of outliers called
 H99 and K07.
\subsection{The H99 group.}
   This group was discovered by Helmi et al. (1999) and further investigated
  by Re Fiorentin et al. (2005). Roederer et al. (2010) made a detailed 
  chemical analysis of these stars 
  (excluding the RR Lyrae stars) and showed that they had a metallicity range 
  of --3.4$<$[Fe/H]$<$--1.5 but were otherwise chemically homogeneous. They 
  concluded that the wide metallicity range precluded the progenitor being
  a single globular cluster. We note that the globular cluster NGC 6205 (M13)
 ([Fe/H] = --1.57) has a rather similar L$_{\perp}$ to that of this group but
  a different L$_{z}$. The 4 RR Lyrae stars that belong to the H99 group
  RZ CEP, TT CNC, AR SER and TT LYN have [Fe/H] = --1.77, --1.57, --1.78 and
  --1.56 respectively (Re Fiorentin et al. 2005),  
  and  XZ CYG and CS ERI  have ([Fe/H] =
 --1.44 and --1.41 respectively. The mean [Fe/H] of these six RR Lyrae stars
 is --1.59 and the $rms$ scatter in their [Fe/H] about this mean is 0.16 which
 is comparable with the likely error in their metallicity. The period 
 distribution of the six RR Lyrae stars correspond to Oosterhoff type I; NGC 
 6205 has too few RR Lyrae stars to have a reliable Oosterhoff type. The
 RR Lyrae members of H99 therefore are more homogeneous than the later-type 
 stars in the group {\it and could possibly have originated from a single 
 globular cluster}. 

\subsection{The K07 group.}
 Kepley et al. (2007) identified six low metallicity {\it outliers} in their
  Table 5. Two of these {\it outliers}  (RV CAP and HD 214925) that have
    similar location on the L$_{\perp}$ $vs$ L$_{z}$ plot and 13
  other {\it outliers} are also close to this location. 
    The assumed boundaries of what we call the K07 group are 
 --2300 $<$ L$_{z}$ $<$ --1500 and 
 +1300 $<$ L$_{\perp}$ $<$ +2200 and are shown by the green rectangle in Figs.
 \ref{f:4}, \ref{f:5} \& \ref{f:6}. 
 The  15 low metallicity stars within these boundaries are listed in Table
 \ref{t:k07}.
  These stars cover a wide range of distances and it seems unlikely that they
  all belong to the same group. It is more likely that they belong
 to several groups that have similar L$_{\perp}$ and L$_{z}$. Thus, 
 AT VIR and RV CAP (at $\sim$1 kpc) have similar properties and IP COM and 
 EO COM (at $\sim$7 kpc) also have similar properties. Further analysis 
 requires more accurate data and is beyond the scope of this paper.

%\newpage 

\twocolumn

\label{lastpage}


\clearpage

\begin{thebibliography}{99}

\bibitem[Abazajianet al.(2009)]{AAM4} Abazajian, K.,
  Adelman-McCarthy, J., Ag\'{u}eros, M. et al., 2009, ApJS, 182, 543 
   (SDSS Data Release 7)
\bibitem[An et al.(2008)]{an08} AN, D., Johnson J., Clem, J., Yanny, B., 
   Rockosi, C., et al. 2008, ApJS, 179, 326
\bibitem[Andrei et al.(2011)]{and11} Andrei, A., Souchay, J., Zacharias, N.,
 Smart, R., et al., 2009, A\&A, 505, 385 
\bibitem[Beers et al.(1996)]{b96} Beers, T.C., Wilhelm, R., Dionidis, S.P. \& 
     Mattson, C.J. 1996, ApJ Suppl., 103, 433     
\bibitem[Beers et al.(2000)]{b00} Beers, T.C., Chiba, M., Yoshii, Y. et al.  
     2000, AJ, 119, 2866 
\bibitem[Beers et al.(2012)]{b12} Beers, T., Carollo, D., Ivezic, D., An, D.,
   et al. 2012,ApJ, 746, 34   
\bibitem[Behr(2003)]{beh03} Behr, B., 2003, ApJS, 149, 101  
\bibitem[Bell et al.(2010)]{bel10} Bell. E., Xue, X., Rix, H-W., Ruhland, C.,
     Hogg, D. 2010, AJ, 140, 1850 
\bibitem[Benedict et al.(2011)]{ben011} Benedict, G.F., McArthur, B.,
  Feast, M., Barnes, T., Harrison, T. et al. 2011, AJ, 142, 187  
\bibitem[Bensby et al.(2011)]{ben11}R Bensby, T., Alves-Brito, A., Oey, M.,
  Yong, D., Mel\'{e}ndez, J., 2011, ApJL, 735, 46  
\bibitem[Brown et al.(2003)]{br03} Brown,W.R., Geller, M.J., Kenyon, S.J. 
     et al. 2003, AJ, 126, 1362     
\bibitem[Brunthaler et al.(2011)]{bru11} Brunthaler, A., Reid, M., Menten, K.,
  Zheng, X.-W., Bartkiewicz, A., Choi, Y. et al., 2011, AN, 332, 461
\bibitem[Cacciari et al.(2005)]{cac05} Cacciari, C., Corwin, T., Carney, B.,
 2005, AJ, 129, 267  
\bibitem[Cardelli, Clayton \& Mathis(1989)]{ccm89} Cardelli, J.A., Clayton, 
     G.C. \& Mathis, J.S. 1989, ApJ, 345, 245     
\bibitem[Carollo et al.(2007)]{car07} Carollo, D., Beers, T., Lee, Y., Chiba, M., et al.,
       2007. Nature, 450, 1020 
\bibitem[Carollo et al.(2010)]{car10} Carollo, D. Beers, T., Chiba, M., 
 Norris, J. et al., 
 2010, ApJ, 712, 692 
\bibitem[Carollo et al.(2012)]{car12} Carollo, D. Beers, T., Bovy, J.,
  Sivarani, T. et al., 2012, ApJ, 744, 195 
\bibitem[Chiba \& Beers(2000)]{chi00}  
	Chiba, M.,  Beers, T. C. 2000, AJ, 119, 2843	
\bibitem[Clement et al.(2001)]{cle01} Clement, C., Muzzin, A., Dufton, Q.,
 Ponnampalam, T., Wang, J., Burford, J., et al., 2001, AJ, 122, 2587 
\bibitem[Clementini et al.(2003)]{cle03} Clementini, G., Gratton, R., 
     Bragaglia, A., Carretta, E., Di Fabrizio, L. \& Maio, M. 2003, AJ, 125, 
     1309      
\bibitem[De Jong et al.(2010)]{dej10} De Jong, J., Yanny, B., Rix, H.-W., 
   Dolphin, A. et al., 2010, ApJ, 714, 663
\bibitem[De Lee(2008)]{del08} De Lee, N., 2008, PhD Thesis, Michigan State
          University, Lansing, Michigan, U.S.A. 
\bibitem[Deason et al.(2011)]{dea11} Deason, A., Belokurov, V., Evans, N.,
  2011, MNRAS, 416, 2903 
\bibitem[Dehnen \& Binney(1998)]{db98} 
	Dehnen, W.,  Binney, J. J. 1998, MNRAS, 298, 387
\bibitem[Dorman(1993)]{dor93} Dorman, B., Rood, R., O'Connell, R., 1993,
    ApJ, 419, 596  
\bibitem[Dotter et al.(2010)]{dot10} Dotter, A., Sarajedini, A., Anderson, J.,
 Aparicio, A., et al. 2010, ApJ, 708, 698 
\bibitem[Feast et al.(2008)]{fea08} Feast, M., Laney, C., Kinman, T., van     
  Leeuwen, F., Whitelock, P., 2008, MNRAS, 386, 2115 
\bibitem[Feast(2011)]{fea11} Feast, M. 2011, Carnegie Observatories Astrophysics
  Series, Vol. 5, ed. A. McWilliam (Pasadena: Carnegie Observatories) 
\bibitem[Fernley \& Barnes(1997)]{fer97} Fernley, J., Barnes, T. 1997, A\&AS,
   125, 313 
\bibitem[Ferraro et al.(1997)]{fer97} Ferraro, F.R., Carretta, E., Corsi, C.E., 
  Fusi Pecci, F., Cacciari, C., Buonanno, R., Paltrinieri, B. \& Hamilton, D.
     1997, A \& A, 320, 757 
\bibitem[Font et al.(2011)]{fon11} Font, A., McCarthy, I., Crain, R., Theuns, T.,
    et al., 2011, MNRAS, 416, 2802 
\bibitem[Hattori \& Yoshii(2011)]{hat11} Hattori, K., Yoshii, Y., 2011,
 MNRAS, 418, 2418   
\bibitem[Helmi et al.(1999)]{hel99} Helmi, A., White, S.D., de Zeeuw, P.T. 
     \& Zhao, H. 1999, Nature, 402, 53 
\bibitem[Helmi(2008)]{hel08} Helmi, A. 2008, Astron. Astrophys. Rev. 15, 145
\bibitem[Holmberg et al.(2009)]{hol11} Holmberg, J., Nordstr\"{o}m, B., 
  Andersen, J., 2009, A\&A, 501, 941 
\bibitem[Jeffery et al.(2007)]{jef07} Jeffery, E., Barnes, T., Skillen, I.,
 Montemayor, T., 2007, ApJS, 171, 512 
\bibitem[Johnson \& Soderblom(1987)]{john87}  
	Johnson, D. R. H,,  Soderblom, D. R. 1987,  AJ, 93, 864	
\bibitem[Jones \& Walker(1988)]{jw88} Jones, B.F. \& Walker, M.F. 1988, AJ, 95, 1755
\bibitem[Jurcsik \& Kovacs(1996)]{jur96} Jurcsik, J., Kovacs, G., 1996, 
      A\&A, 312, 111 
\bibitem[Jurcsik et al.(2006)]{jur06} Jurcsik, J., Sodor, A., Varadi, M.,
  Vida, K., Posztobanyi, K., Szing, A., et al., IBVS, 5709 
\bibitem[Kemper(1982)]{kem82} Kemper, E., AJ, 87, 1395  
\bibitem[Kepley \& al.(2007)]{kep07} Kepley, A. Morrison, H. Helmi, A., 
         Kinman, T. et al. 2007, AJ, 134, 1579
\bibitem[Kholopov et al.(1985)]{kho85} Kholopov, P. et al. 1985 Gen. Cat. of
       Variable Stars (http://www.sai.msu.ru/gcvs/gcvs/) 
\bibitem[Kinemuchi et al.(2006)]{kin06} Kinemuchi, K., Smith, H., Wo\'{z}niak,
   P., McKay, T., 2006, AJ, 132, 1202 
\bibitem[Kinman et al.(1966)]{kin65} Kinman, T., Wirtanen, C., Janes, K.,
         1966, ApJS, 13, 379  
\bibitem[Kinman et al.(1982)]{kin82} Kinman, T., Mahaffey, C.,  Wirtanen, C., 
     1982, AJ, 87, 314 
\bibitem[Kinman, Suntzeff \& Kraft(1994)]{ksk94} Kinman, T.D., Suntzeff, N.B.,
   Kraft, R.P., 1994, AJ, 108, 1722
\bibitem[Kinman et al.(2000)]{kin00}  
 Kinman, T. D., Castelli, F.,  Cacciari, C., Bragaglia,  A., Harmer, D.,
        Valdes, F.,  2000, A\&A, 364, 102  
\bibitem[Kinman, Saha \& Pier(2004)]{kin04}  Kinman, T., Saha, A., Pier,
  J., 2004, ApJL, 605, 25  
\bibitem[Kinman et al.(2007a)]{kin07a} Kinman, T., Salim, S., Clewley, L.,
      2007a, ApJL, 662, 111 
\bibitem[Kinman et al.(2007b)]{kin07b} Kinman, T., Cacciari, C., Bragaglia, A.,
        Buzzoni, A., Spagna, A.  2007b, MNRAS, 375, 1381 
\bibitem[Kinman et al.(2009)]{kin09} Kinman, T., Morrison, H., Brown, W.,
     2009, AJ, 137, 3198 
\bibitem[Kinman \& Brown(2010)]{kin10} Kinman T.. Brown, W., 2010, IBVS 5935
\bibitem[Kinman \& Brown(2011)]{kin11} Kinman T.. Brown, W., 2011, AJ, 141, 168
\bibitem[Klement(2010)]{kle10} Klement, R.J., 2010, Astron. Astrophys.
    Rev., 18, 567 
\bibitem[Lasker et al.(2008)]{las08} Lasker, B.M., Lattanzi, M., McLean, B., 
     et al. 2008, AJ, 136, 735 
\bibitem[Layden(1994)]{l94} Layden, A. 1994, AJ, 108, 1016 
\bibitem[Lee \& Carney(1999)]{lc99} Lee, J-W., Carney, B.W. 1999,
           AJ, 118, 1373  
\bibitem[Liu \& Janes(1990)]{liu90} Liu, T., Janes, K., 1990, ApJ, 354, 273
\bibitem[Maintz(2005)]{mai05} Maintz, G., 2005, A\&A, 442, 381
\bibitem[Maintz \& de Boer(2005)]{mdb05} Maintz, G. \& de Boer, K.S. 2005, 
     A\& A, 442, 229 
\bibitem[Mar\'{i}n-Franch et al.(2009)]{mar09} Mar\'{i}n-Franch, A.,
 Aparicio, A., Piotto, G., Rosenberg, A., et al., 2009, ApJ, 694, 1498 
\bibitem[McCarthy et al.(2012)]{mcc12} McCarthy, I., Font, A., Crain, R., 
   Schaye, J. et al., 2012, MNRAS, in press 
\bibitem[McClusky(2008)]{mcc08} McClusky, J. 2008, IBVS 5825 
\bibitem[Miceli et al.(2008)]{mic08} Miceli, A., Rest, A., Stubbs, C.,
 Hawley, S., et al., 2008, ApJ, 678, 865  
\bibitem[Morrison et al.(2009)]{mor09} Morrison, H., Helmi, A., Sun, J.,
 Liu, P., et al. 2009, ApJ, 694, 130   
\bibitem[Nemec et al.(1994)]{nem94} Nemec, J., Nemec, A., Lutz, T., 
   1994, AJ, 108, 222 
\bibitem[Oser et al.(2010)]{ose10} Oser, L., Ostriker, J., Naab, T., 
     Johansson, P. et al., 2010, ApJ., 725, 2312  
\bibitem[Paltrinieri et al.(1998)]{pal98} Paltrinieri, B., Ferraro, F.R., Fusi
     Pecci, F. \& Carretta, E. 1998, MNRAS, 293, 434       
\bibitem[Pesch \& Sanduleak(1989)]{pes89} Pesch, P., Sanduleak, N. 1989, ApJS, 71, 549
\bibitem[Pier, Saha \& Kinman(2003)]{pie03} Pier, J., Saha, A., Kinman, T.,
     2003, IBVS 5459  
\bibitem[Preston, Shectman \& Beers(1991)]{pre91}  
	Preston, G. W., Shectman, S. A., Beers, T. C. 1991, ApJ, 375, 121
\bibitem[Re Fiorentin et al.(2005)]{fio05} Re Fiorentin, P., Helmi, A., 
  Lattanzi, M., Spagna, A., 2005, A\&A, 439, 551 
\bibitem[Reid \& Brunthaler(2004)]{rei04} Reid, M., Brunthaler, 2004,
      ApJ, 616, 872  
\bibitem[Roederer et al.(2010)]{roe10} Roederer, I., Sneden, C., Thompson, I.,
 Preston, G., Shectman, S., 2010, ApJ, 711, 573  
\bibitem[Ruhland et al.(2011)]{ruh11} Ruhland, C., Bell, E., Rix, H-W., Xue, 
 X.,    2011, ApJ, 731, 119
\bibitem[Saha \& Oke(1964)]{sah64} Saha, A., Oke, B., 1984, ApJ, 285, 688
\bibitem[Schlegel, Finkbeiner \& Davies(1998)]{schle98}  
	Schlegel, D. J.,  Finkbeiner, D. P., Davis, M. 1998, ApJ, 500, 525
\bibitem[Schmidt(1956)]{sch56}Schmidt, M., 1956, B.A.N. 13, 15
\bibitem[Schmidt(2002)]{sch02}Schmidt, E., 2002, AJ, 123, 965 
\bibitem[Schmidt et al.(1995)]{sch95} Schmidt, E., Chab, J. Reiswig, D.,
   109, 1239 
\bibitem[Schmidt \& Seth(1996)]{sch96} Schmidt, E., Seth., A., 1995, AJ,
        112, 2769 
\bibitem[Sch\"{o}nrich, Asplund \& Casagrande(2011)]{sch11} Sch\"{o}nrich, R.,
    Asplund, M., Casagrande, L.. 2011, MNRAS, 415, 3807 
\bibitem[Sesar et al.(2010)]{ses10} Sesar, B., Ivezi\'{c}, Z., Grammer, S.,
         Morgan, D., et al., 2010, ApJ, 708, 717 
\bibitem[Sesar et al.(2011)]{ses11} Sesar, B., Juri\'{c}, M., Ivezi\'{c},
  \u{Z}., 2011, ApJ, 731, 4 
\bibitem[Sirko et al.(2004)]{sir04} Sirko, E., Goodman, J., Knapp, G.R. et al. 
     2004, AJ, 127, 899 
\bibitem[Smith et al.(2009)]{smi09} Smith, M., Evans, N., Belokurov, V.,     
          Hewitt, P., et al., 2009, MNRAS, 399, 1223 
\bibitem[Smith et al.(2010)]{smi10} Smith, K., Bailer-Jones, C., Klement, R.,
       Xue, X. 2010, A\&A, 522, 88
\bibitem[Smith et al.(2011)]{smi11} Smith, H., Catelan, M., Kuehn, C.
 2011, Carnegie Obs. Astrophysics Ser., Vol. 5 (ed. A. McWilliam)
 (Pasadena: Carnegie Obs.) 
\bibitem[Sodor, Jurcsik \& Szeidl(2009)]{sod09} Sodor, A., Jurcsik, J., Szeidl, B.,
     2009, MNRAS, 394, 261 
\bibitem[Spagna et al.(2010a)]{spa10a}  
 Spagna, A., Bucciarelli, B., Lattanzi, M., Re Fiorentin, P., Smart, R., 2010a,
        Mem. S.A. It., 14, 67  
\bibitem[Spagna et al.(2010b)]{spa10b}  
 Spagna, A., Lattanzi, M., Re Fiorentin, P., Smart, R., 2010b, A\&A, 510, L4 
\bibitem[Usher \& Mitchell(1982)]{usm82} Usher, P., Mitchell. K. 1982, ApJS, 49, 27
\bibitem[Valenti et al.(2004)]{val04} Valenti, E., Ferraro, F.R., Perina, S. 
     \& Origlia, L. 2004, A\&A, 419, 139  
\bibitem[van den Bergh(1993)]{vdb93} van den Bergh, S. 1993, AJ, 105, 971
\bibitem[Vande Putte \& Cropper(2009)]{van09} Vande Putte, D., Cropper, M.,
     2009,  MNRAS, 392, 113 
\bibitem[Watkins et al.(2009)]{wat09} Watkins, L., Evans, N., Belokurov, V.,
  Smith, M., et al., 2009, MNRAS, 398, 1757 
\bibitem[Xue et al.(2011)]{xue11} Xue,X.X., Rix, H.-W., Yanny, B., Beers, T. 
       et al. 2011. ApJ, 738, 79 
\bibitem[Yanny et al.(2009)]{ya09} Yanny, B., Rockosi, C., Newberg, et al. 
         2009, AJ, 137, 4377  
\bibitem[Zacharias et al.(2004)]{zac04} Zacharias, N., Monet, D., Levine, S., et
  al. 2004, BAAS, 36, 1418 (NOMAD Catalogue)
\bibitem[Zacharias et al.(2009)]{zac09} Zacharias, N., Finch, C., Girard. T., 
   Hambly, N. et al., 2009, AJ, 139, 2184 (UCAC3 Catalogue) 
\bibitem[Zinn(1993)]{z93} Zinn, R.  1993, in {\it The Globular Cluster $-$    
   Galaxy Connection }, eds. Smith, G \& Brodie, J., ASP 
	Conf.\ Ser.\ Vol.\ 48, p.\ 38
\bibitem[Zolotov et al.(2009)]{zol09} Zolotov, A., Willman, B., Brooks, A., 
   Governato, F. et al. 2009, ApJ, 721, 738 
\bibitem[Zolotov et al.(2010)]{zol10} Zolotov, A., Willman, B., Brooks, A., 
   Governato, F. et al. 2010, ApJ, 738, 79 
\bibitem[Zolotov(2011)]{zol11} Zolotov, A., 2011, BAAS, 43, 2011.

\end{thebibliography}
\end{document}